\documentclass{aa}
\usepackage{graphicx}
\usepackage{txfonts}
\usepackage{subcaption} 
\usepackage{lscape} 
\usepackage{placeins} 
\usepackage{hyperref}
\usepackage{multirow}

\usepackage{tikz}
\usetikzlibrary{arrows.meta, positioning, shapes.geometric, fit, calc}

\usepackage{CJKutf8} 

\begin{document}
    \title{A pilot study on the CSST astrometric capability}
    \subtitle{Detecting astrometric binaries with \emph{Gaia} synergy via simulated data}

    \author{
        Shangyu Wen
        (\begin{CJK*}{UTF8}{gbsn}文尚宇\end{CJK*})\inst{1,2}
    \and
        Shilong Liao
        (\begin{CJK*}{UTF8}{gbsn}廖石龙\end{CJK*})\inst{1,2}
        \thanks{Corresponding author: \email{shilongliao@shao.ac.cn}}
    \and
        Zhaoxiang Qi
        (\begin{CJK*}{UTF8}{gbsn}齐朝祥\end{CJK*})\inst{1,2}
    \and
        Zhensen Fu
        (\begin{CJK*}{UTF8}{gbsn}傅震森\end{CJK*})\inst{1}
    \and
        Xiyan Peng
        (\begin{CJK*}{UTF8}{gbsn}彭喜衍\end{CJK*})\inst{1}
    \and
        Ye Ding
        (\begin{CJK*}{UTF8}{gbsn}丁页\end{CJK*})\inst{1,2}
    \and
        Qiqi Wu
        (\begin{CJK*}{UTF8}{gbsn}武琦琦\end{CJK*})\inst{1}
    \and
        Qi Xu
        (\begin{CJK*}{UTF8}{gbsn}徐琦\end{CJK*})\inst{1,2}
    \and
        Xun Sun
        (\begin{CJK*}{UTF8}{gbsn}孙迅\end{CJK*})\inst{1,2}
    \and
        Keyu Zhu
        (\begin{CJK*}{UTF8}{gbsn}朱科宇\end{CJK*})\inst{1,2}
    \and
        Yong Yu
        (\begin{CJK*}{UTF8}{gbsn}于涌\end{CJK*})\inst{1,2}
    }

    \institute{
        Shanghai Astronomical Observatory, Chinese Academy of Sciences, 80 Nandan Road, Shanghai 200030, PR China.
    \and
        School of Astronomy and Space Sciences, University of Chinese Academy of Sciences, No. 19(A) Yuquan Road, Beijing 100049, PR China.
    }

    \date{Received XXX; accepted XXX}

    \abstract
    {The China Space-station Survey Telescope (CSST) will provide deep, wide-field epoch astrometry over a 10-year mission. Astrometric binary orbital solutions can provide constraints on the masses of stellar and compact-object components. Recovery of orbital parameters depends on both astrometric precision and temporal coverage. Combining CSST and \emph{Gaia} epoch astrometry extends the temporal baseline and therefore improves binary detection.}
    {We evaluate CSST, \emph{Gaia}, and joint epoch astrometry for binary-candidate selection and 12-parameter (12p) orbit fitting at faint magnitudes ($g>17.8$). We also test how a more regular CSST cadence affects the number of 12p fits satisfying our criteria.}
    {We constructed a mock catalog, simulated CSST and \emph{Gaia} epoch astrometry, and fitted five-parameter (5p) single-star models to derive astrometric diagnostics, proper-motion anomaly features, and observational-sampling features. These features were used in a four-stage histogram-based gradient-boosting classifier. Selected candidates were then fitted with a 12p orbital model and assessed using two nested selection tiers.}
    {On the independent test set, the classifier reaches a precision of $0.802$ and a recall of $0.181$ among eligible true binaries. In the scenario-specific fitted samples, joint astrometry raises the fiducial fraction from $6.76\%$ for \emph{Gaia} alone to $10.46\%$; for fitted binaries with $P_{\rm true}>15\,{\rm yr}$, it rises from $2.37\%$ to $6.78\%$. The current CSST schedule yields few fiducial fits, while idealized regular cadences increase the yield mainly at $g\lesssim21$.}
    {In the simulation, joint CSST and \emph{Gaia} epoch astrometry yields higher fractions of fitted unresolved binaries satisfying the stated criteria than the \emph{Gaia}-only solution. A practical strategy is to select binary candidates from 5p diagnostics and astrometric anomalies, obtain more regular CSST follow-up observations for promising binaries, and then fit 12p orbital models and apply the selection criteria.}

    \keywords{astrometry -- binaries: general -- methods: numerical}
\maketitle
\nolinenumbers

\section{Introduction\label{sect:intro}}
    China Space-station Survey Telescope (CSST), a major science project of the China Manned Space Program, is planned for launch in the near future. CSST has an expected operational lifetime of 10 years. It will operate in the same orbit as the China Manned Space Station, allowing in-orbit maintenance and upgrades through docking with the space station \citep{2026SCPMA..6939501C,2011SSPMA..41.1441Z,2014RAA....14.1055S}. The CSST wide-field survey will cover about $17\,500\,{\rm deg}^2$ and reach a magnitude limit of $g<26.3\,{\rm mag}$ (Table~\ref{table:csst_mag_lim}) \citep{2026SCPMA..6939501C,zhan2021wide}. With deep repeated observations over a decade, CSST will provide a new opportunity for high-precision optical astrometry beyond the current \emph{Gaia} magnitude range.

    \begin{table*}[h!]
        \caption{Magnitude limits of the CSST bands \citep{2026SCPMA..6939501C,zhan2021wide}.}
        \label{table:csst_mag_lim}
        \centering
        \begin{tabular}{c|ccccccccc}
            \hline\hline
            \multirow{2}{*}{Sky area (deg$^{2}$)} & \multicolumn{9}{c}{Magnitude limit for different bands (point source, $5\sigma$, AB mag)} \\
             & $NUV$ & $u$ & $g$ & $r$ & $i$ & $z$ & $y$ & $J^{\prime}$\tablefootmark{a} & $H^{\prime}$\tablefootmark{b} \\ \hline
            17\,500\tablefootmark{c} & $25.4$ & $25.4$ & $26.3$ & $26.0$ & $25.9$ & $25.2$ & $24.4$ & $23.5$ & $23.0$ \\
            400\tablefootmark{d} & $26.7$ & $26.7$ & $27.5$ & $27.2$ & $27.0$ & $26.4$ & $25.7$ & & \\ \hline
        \end{tabular}
        \tablefoot{\tablefoottext{a}{0.9--1.3\textmu m.}\tablefoottext{b}{1.3--1.7\textmu m.}\tablefoottext{c}{Wide-field survey.}\tablefoottext{d}{Selected deep fields.}}
    \end{table*}

    Binary and higher-order multiple systems are common, and binaries constitute the dominant form of stellar multiplicity \citep{2013ARA&A..51..269D,2022A&A...657A...7K}. Stellar mass is a primary parameter governing stellar evolution and shapes observable properties such as spectra and luminosities \citep{2012Sci...337..444S}. Mass measurements provide key constraints on stellar evolution and binary formation. Precise mass measurements are especially important for addressing outstanding astrophysical problems, such as the nature of the mass gap between neutron stars and black holes \citep{2012ApJ...757...36K,2016A&A...587A..61C}. Determining whether the observed mass gap is astrophysical or produced by selection effects requires larger and more complete binary samples. Low-mass stars also dominate the stellar population according to the initial mass function \citep{2019NatAs...3..482K}. By pushing to fainter magnitudes, CSST can access more distant and lower-mass binary populations.

    Spectroscopic measurements generally yield minimum companion masses unless the orbital inclination is constrained independently. Astrometric epoch data provide an efficient way to detect binaries through the positional wobble induced by their companions. With astrometric epoch data, the relative precision of mass determinations can reach below $3\%$ \citep{2021A&ARv..29....4S}.

    Binaries also provide useful tests of parallax systematics. The nearly identical true parallaxes of their components enable an internal consistency check \citep{2024A&A...691A..81D,2021MNRAS.506.2269E}, and an orbital parallax derived from a combined astrometric and radial-velocity orbital solution provides an independent parallax reference \citep{2025AJ....169..211D}. When a binary is processed with a five-parameter (5p) single-star model, part of the orbital motion can be absorbed into the fitted parallax and proper motion, leading to systematic biases in the astrometric parameters \citep{2020MNRAS.495..321P}.

    \emph{Gaia} extends space-based optical astrometry to $G\lesssim 21$ \citep{2023A&A...674A...1G}, and its third data release (DR3) \texttt{nss\_two\_body\_orbit} catalog provides a large sample of robust astrometric binary solutions \citep{2023A&A...674A...9H}. Recent studies of \emph{Gaia} DR3 binary detectability further show that the astrometric binary selection function depends strongly on orbital period, semi-major axis, and observational cadence \citep{2024OJAp....7E.100E,2024A&A...688A...1C}. However, the present \emph{Gaia} DR3 orbital solutions remain limited for long-period binaries and faint sources. Most \emph{Gaia} DR3 binaries with orbital solutions are brighter than $19\,{\rm mag}$ and lie within $5\,{\rm kpc}$ \citep{2023A&A...674A...9H}. Future \emph{Gaia} data releases will extend the time baseline, with DR4 covering $66$ months and DR5 reaching $10.5$ years, and are expected to increase the orbital solution yield substantially \citep{2024OJAp....7E.100E}.

    For binaries with periods longer than the \emph{Gaia} observing baseline, the observed orbital arc can still reveal binarity through acceleration solutions, elevated renormalized unit weight error (RUWE), or astrometric excess noise, while the limited phase coverage makes a full Keplerian solution and precise recovery of the orbital elements more difficult. Combining data from different missions achieves a longer baseline and therefore increases the covered orbital fraction. For example, \citet{2023A&A...672A..82L} derived orbital solutions for three relatively long-period binaries by combining Hipparcos and \emph{Gaia} DR3 data, demonstrating the value of jointly analyzing data from multiple missions to detect long-period binary candidates (e.g. $P>15\,{\rm yr}$). Within the overlapping magnitude range of CSST and \emph{Gaia}, joint epoch astrometry can extend the temporal baseline. This extension is especially valuable for long-period binaries, for which only a fraction of the orbit is covered by a single mission.

    The recovery of orbital parameters cannot be inferred from single-epoch astrometric precision alone. It also depends strongly on the observation schedule, including the time baseline and the observational cadence. Even under regularly sampled, high signal-to-noise observations, sufficient orbital coverage is required to constrain the orbital period and the semi-major axis \citep{2014A&A...563A.126L}. Together with the parallax, these fitted orbital parameters determine the dynamical mass. The importance of sampling is also seen in \emph{Gaia} unresolved-binary detectability, where the sensitivity window depends on the observing baseline and scanning law \citep{2024A&A...688A...1C,2024OJAp....7E.100E}.

    CSST also has important faint-end potential. Ground-based optical surveys face a trade-off among sky coverage, depth, and astrometric precision. The Legacy Survey of Space and Time (LSST) will reach $g\leq 25$ over $18\,000\,{\rm deg}^2$, with astrometric precision limited to tens of milliarcseconds \citep{2019ApJ...873..111I}. The Pan-STARRS1 $3\pi$ survey achieved wider sky coverage ($\delta>-30\degr$) and astrometric precision better than $5\,{\rm mas}$, but with a brighter single-observation limit of $g<22\,{\rm mag}$ \citep{2016arXiv161205560C}. Other deep space missions, such as \emph{Euclid} and the Nancy Grace Roman Space Telescope, reach faint magnitudes but have shorter nominal lifetimes or smaller wide-survey sky coverage for this specific astrometric-binary application (\citealp{2022A&A...662A.112E}; Roman Observations Time Allocation Committee final report, \citeyear{2025arXiv250510574O}).

    CSST combines depth, wide sky coverage, repeated observations, and a 10-year mission duration. Recent simulations also suggest that CSST observations can increase samples of substellar companions and brown-dwarf binaries around low-mass objects \citep{2026AJ....171..121X}. These properties could extend astrometric binary searches to fainter candidates. At the faint end, the recovery of orbital parameters remains sensitive to cadence and signal-to-noise ratio. In the overlap regime with \emph{Gaia} epoch astrometry, joint fitting remains particularly important.

    In this work, we use mock data to assess how CSST can contribute to faint-end binary detection. We first select binary candidates using diagnostics from 5p fits, proper motion anomaly (PMa) features, and observational-sampling features. We then fit the selected candidates with a 12-parameter (12p) orbital model and apply the selection criteria to the orbital solutions. Finally, we test how replacing the current CSST survey schedule with more regular observing schedules would affect the yields of CSST and joint fits satisfying the selection criteria.

    This paper proceeds as follows. Sect.~\ref{sect:mock_epoch_astrometry} describes the CSST, \emph{Gaia}, and joint observing schedules. Sect.~\ref{sect:input_catalog} describes the construction of the input catalog. Sect.~\ref{sect:models_and_mock_obs} presents the astrometric models and the mock observation procedure. Sect.~\ref{sect:select_binary_candidates} describes the features and classifier used to select binary candidates. Sect.~\ref{sect:bin_orb_sol} presents the orbital solution results. Sect.~\ref{sect:csst_schedule_opt} discusses the effect of a more regular CSST observing cadence. Finally, we summarize the findings in Sect.~\ref{sect:concl_and_disc}. Appx.~\ref{appx:detailed_ast_models} presents the forward models for single stars and binaries. Appx.~\ref{appx:pma_calc} gives the calculation of the proper motion anomaly between CSST and \emph{Gaia} and derives its significance and uncertainty. Appx.~\ref{appx:selection} illustrates the pipeline used to select binary candidates from the simulated single-star-model solutions. Appx.~\ref{appx:orb_sol} briefly describes the orbital fitting algorithm. Appx.~\ref{appx:resolved_binary} shows a toy injection--recovery experiment for resolved binaries.

\section{Mock epoch astrometry\label{sect:mock_epoch_astrometry}}

\subsection{CSST observations\label{subsect:csst_obs_schedule}}
    CSST is designed to conduct a deep and wide-field survey. The initial 10-year CSST schedule spans approximately July 2027 to July 2037 \citep{2023FrASS..1046603F}\footnote{This is only an initial schedule, which may be optimized and/or rearranged in the future.}. We set the CSST reference epoch to $t_{\rm C}={\rm J}2032.5$. The schedule provides the telescope pointing direction, velocity, and barycentric position at each observing epoch $t_j$.

\begin{figure*}[h!]
        \begin{subfigure}[h!]{0.45\linewidth}
            \centering
            \includegraphics[width=\linewidth]{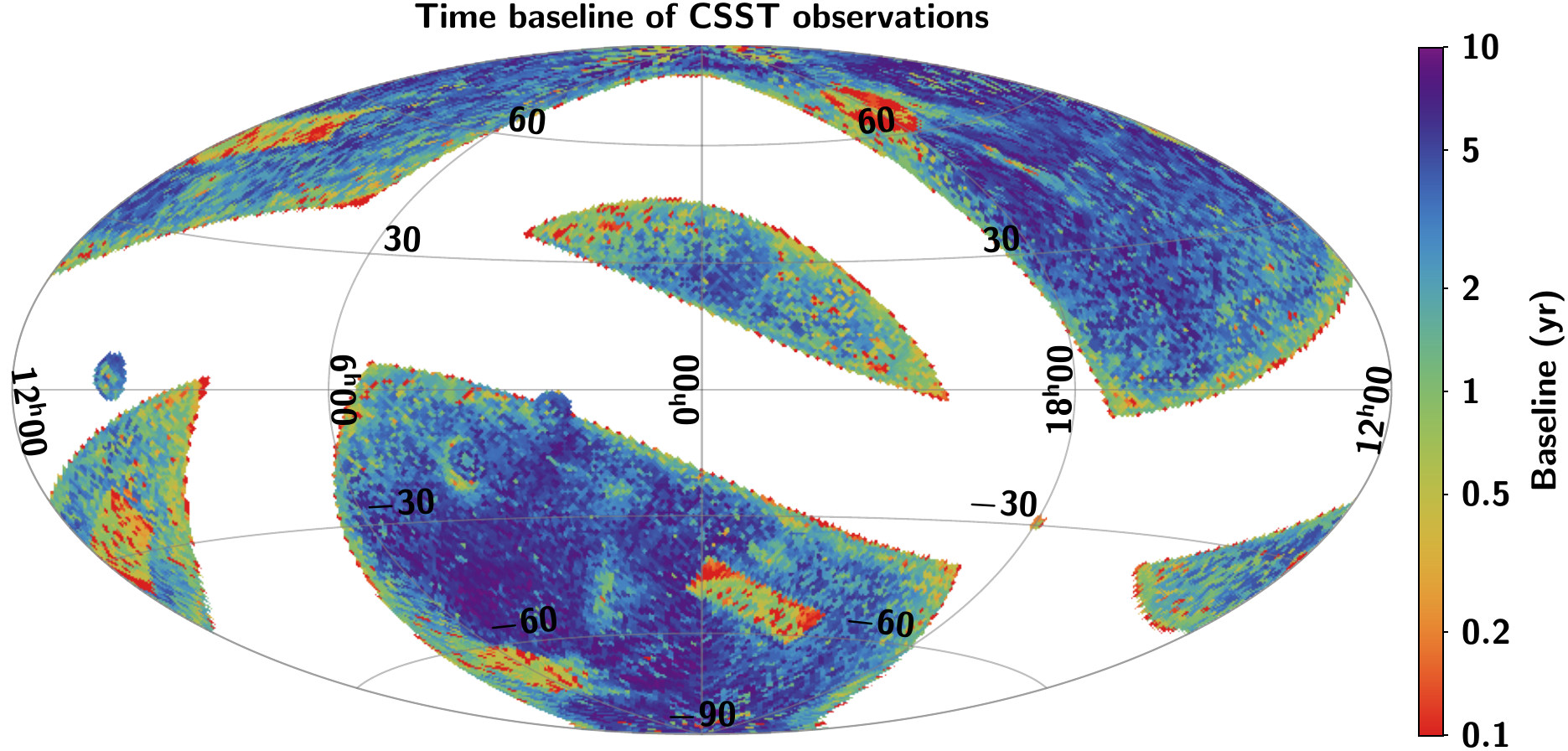}
            \includegraphics[width=\linewidth]{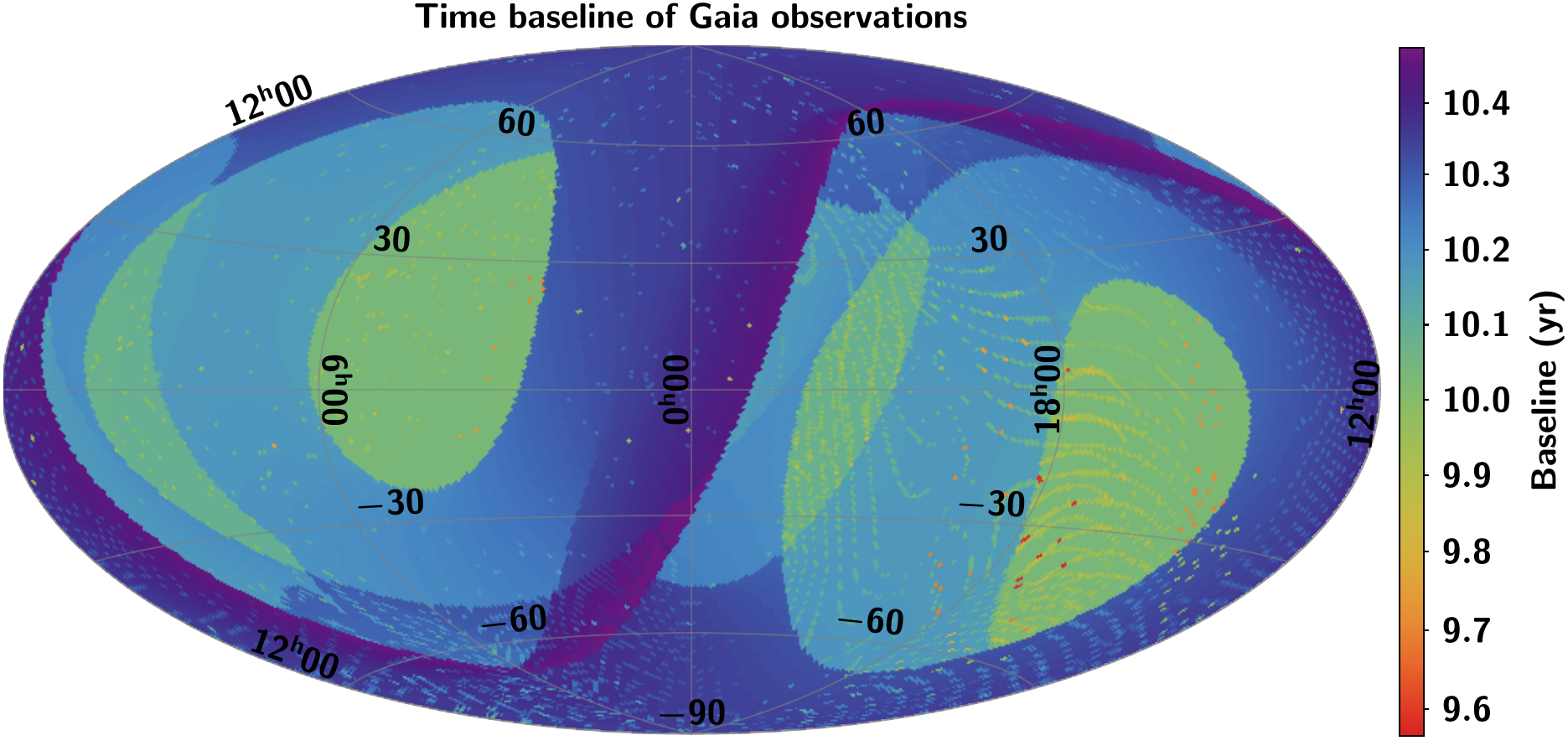}
            \includegraphics[width=\linewidth]{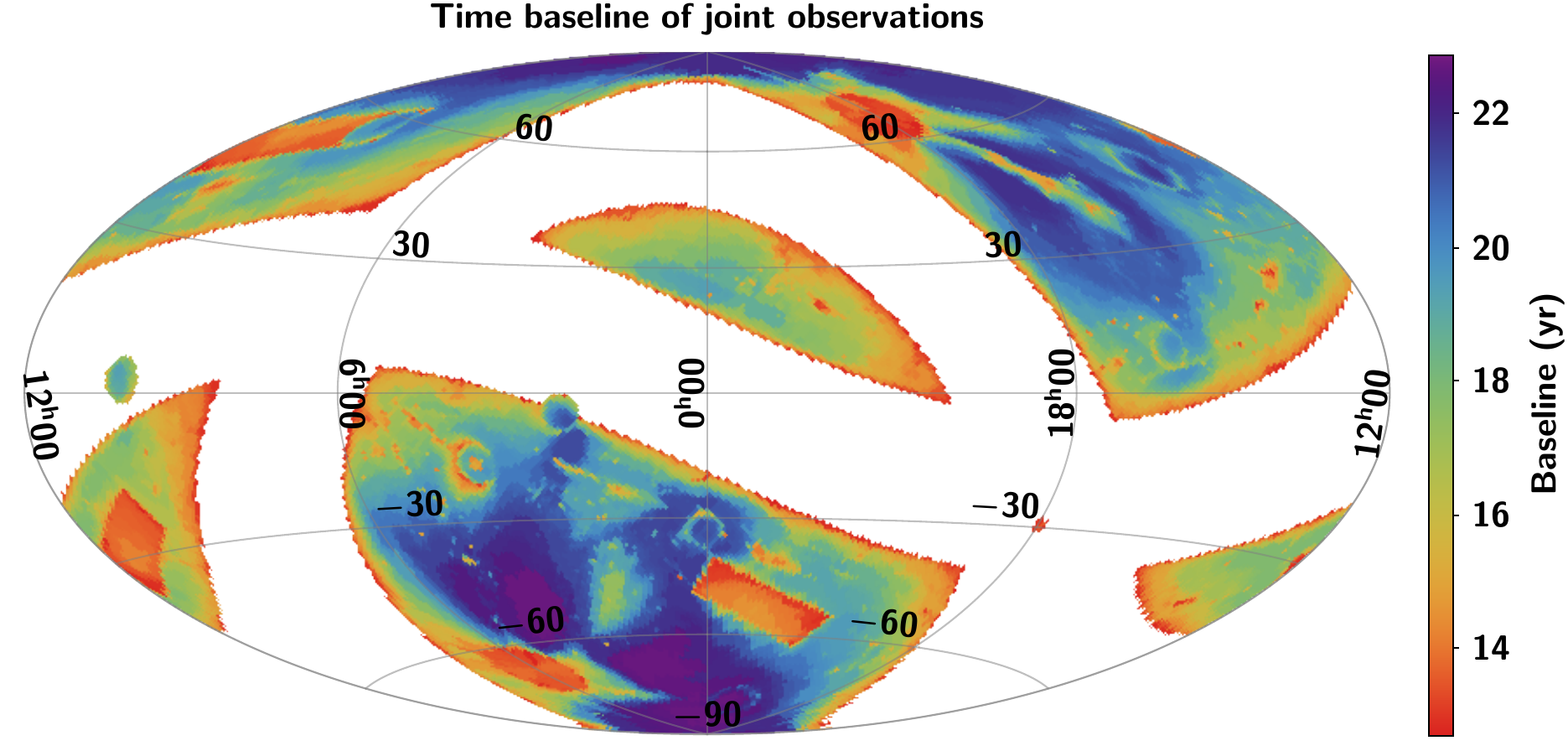}
            \caption{Time baselines}
            \label{subfig:obs_temporal_dist}
        \end{subfigure}
        \hfill
        \begin{subfigure}[h!]{0.45\linewidth}
            \centering
            \includegraphics[width=\linewidth]{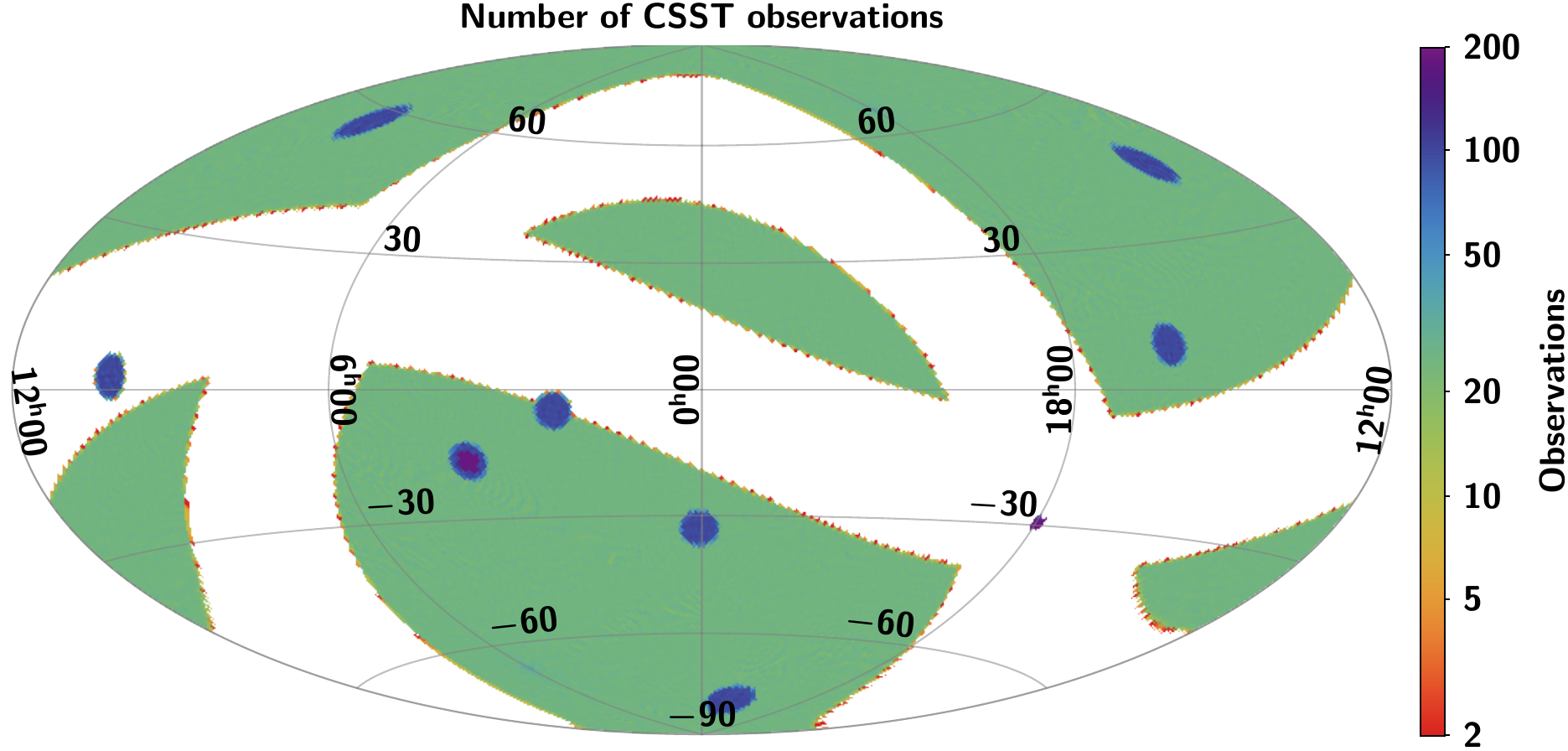}
            \includegraphics[width=\linewidth]{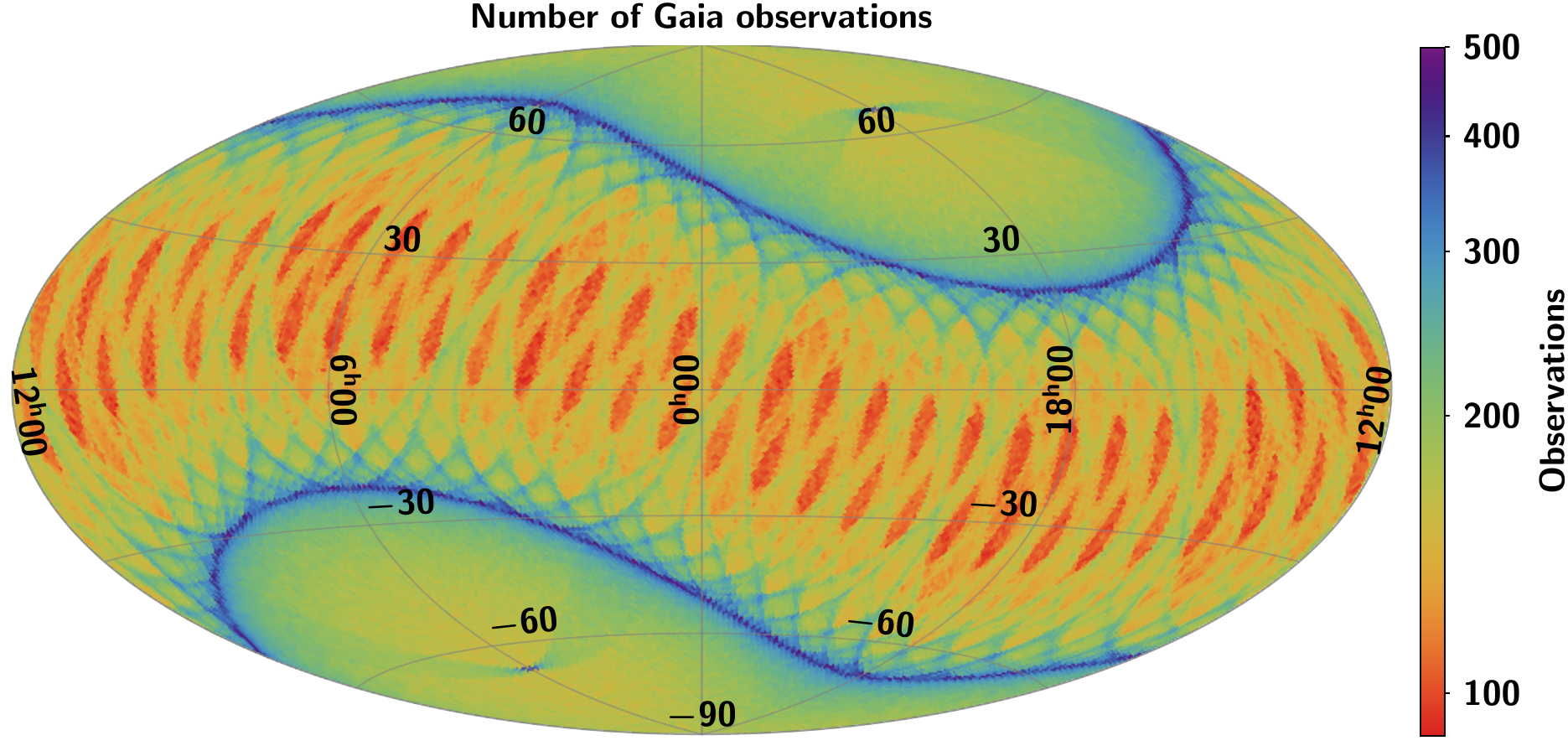}
            \includegraphics[width=\linewidth]{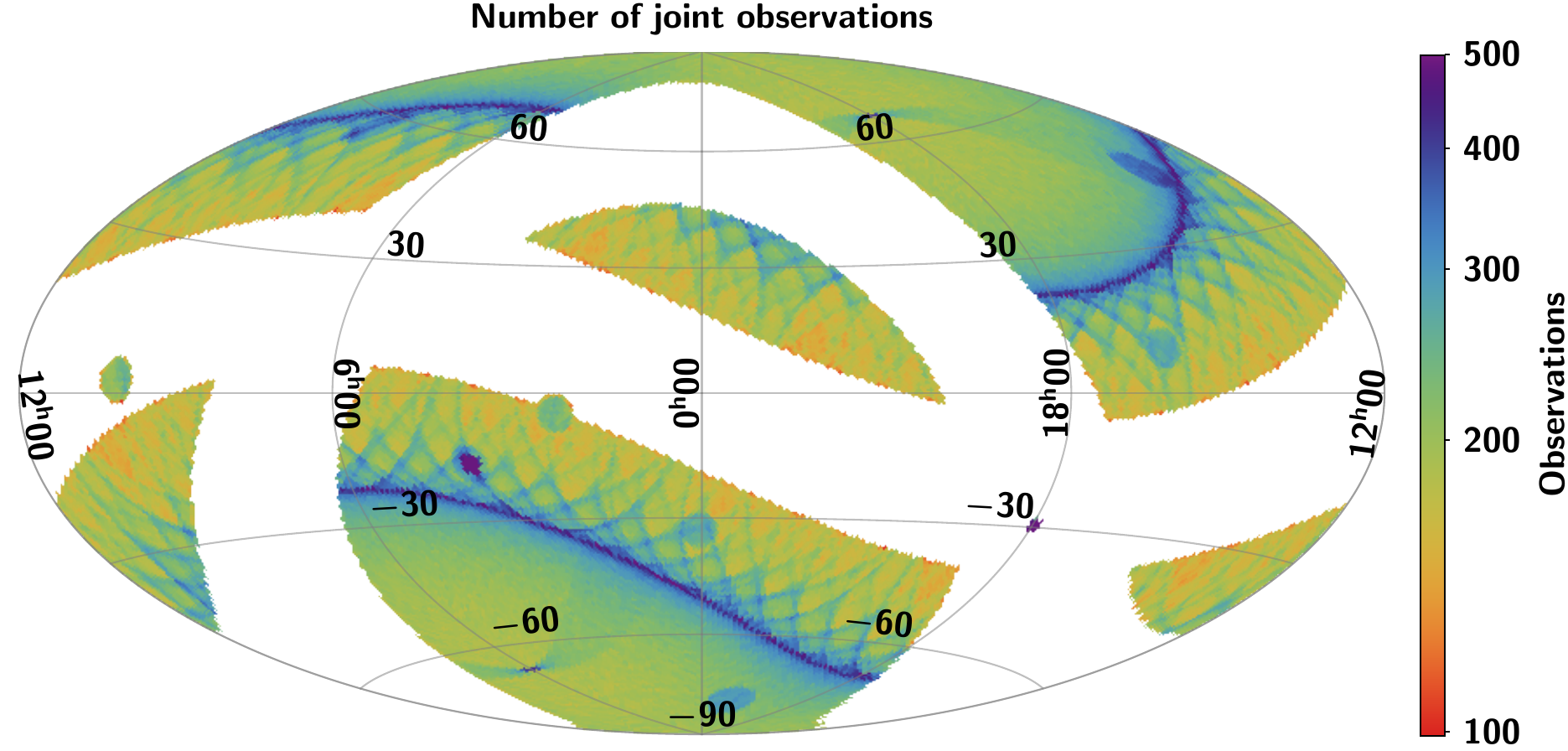}
            \caption{Number of observations}
            \label{subfig:obs_spatial_dist}
        \end{subfigure}
        \caption{Sky distributions of the time baselines (\emph{left}) and numbers of observations (\emph{right}) for the CSST (\emph{top}), \emph{Gaia} (\emph{middle}), and the joint (\emph{bottom}) mock observations, shown in Aitoff projection of equatorial coordinates on a HEALPix level 6 grid. White regions in the CSST and joint panels are outside the current CSST survey schedule and are therefore excluded from the joint sample, although they are covered by \emph{Gaia}. Right ascension increases from right to left.}
        \label{fig:obs_temporal_and_spatial_dist}
    \end{figure*}

    Fig.~\ref{fig:obs_temporal_and_spatial_dist} presents the temporal and spatial observation distributions of the CSST survey schedule. The spatial distribution shows an intentional focus on nine selected deep fields. The main survey avoids regions at low Galactic and ecliptic latitudes, where the long exposures ($\geq 150\,{\rm s}$) would be affected by bright sky background and an increased risk of detector damage \citep{2026SCPMA..6939501C}.

    Although CSST will observe in multiple bands, we use the $g$-band single-epoch observation error model for all CSST observations for simplicity\footnote{The survey schedule and the single-epoch observation errors in the relevant bands are similar to those in the $g$ band.}. The single-epoch observation errors in right ascension (RA) and declination (Dec) are modeled as functions of $g$-band magnitude following \citet{2023FrASS..1046603F}:
    \begin{align}
        \sigma_{\alpha^{\star},0}&=2.914\times 10^{-4}\mathrm{e}^{0.4014g}+1.120\times 10^{-18}\mathrm{e}^{1.694g}\ \text{mas},\\
        \sigma_{\delta,0}&=2.682\times 10^{-4}\mathrm{e}^{0.4112g}+6.925\times 10^{-19}\mathrm{e}^{1.703g}\ \text{mas}.
    \end{align}
    The simulated single-epoch observation errors did not include the attitude jitter. We therefore added an attitude noise term of $\sigma_{\rm att}=0.1\,{\rm mas}$ in quadrature to the single-epoch observation errors. This value is comparable to \emph{Gaia}'s attitude noise of $0.076\,{\rm mas}$ \citep{2021A&A...649A...2L}. The adopted single-epoch observation errors are therefore
    \begin{align}
        \sigma_{\alpha^{\star}}=\sqrt{\sigma_{\alpha^{\star},0}^{2}+\frac{\sigma_{\mathrm{att}}^{2}}{2}},\quad
        \sigma_{\delta}=\sqrt{\sigma_{\delta,0}^{2}+\frac{\sigma_{\mathrm{att}}^{2}}{2}}.
        \label{eq:csst_err}
    \end{align}
    The adopted CSST single-epoch observation errors are shown in Fig.~\ref{fig:csst_gaia_obs_err}.

    To assign scheduled CSST visits to mock sources, we discretized the celestial sphere with \texttt{healpy} \citep{2019JOSS....4.1298Z}. For each scheduled pointing, we identified the corresponding sky pixel and assigned its observing epoch to all mock sources in that sky pixel. This approximation introduces an epoch-assignment mismatch that is primarily set by the difference between the sky-pixel area and the telescope FoV. To limit this mismatch, we used HEALPix level 6\footnote{This choice gives a sky-pixel area of approximately $0.839\,{\rm deg}^{2}$, comparable to the CSST FoV of approximately $1.1\,{\rm deg}^{2}$.}.

    The survey schedule is shown in the \emph{top row} of Fig.~\ref{fig:obs_temporal_and_spatial_dist}. The adopted CSST schedule covers $23\,770$ HEALPix level 6 pixels. Across these covered pixels, the mean time baseline is $3.32\,{\rm yr}$, and $90\%$ have a baseline shorter than $6.47\,{\rm yr}$. By comparison, the adopted \emph{Gaia} scanning law has a mean baseline of $10.22\,{\rm yr}$ and a 90th-percentile baseline of $10.38\,{\rm yr}$ across all level 6 pixels. We quantify the temporal unevenness using the normalized metric $\sigma_{t}/t_{\rm len}$, where $\sigma_{t}$ denotes the standard deviation of the gaps between adjacent observations and $t_{\rm len}$ is the time baseline of each scenario in each pixel. Among pixels with at least three observations, the median normalized unevenness is $0.093_{-0.024}^{+0.038}$ for CSST, compared with $0.010_{-0.005}^{+0.002}$ for \emph{Gaia}. Therefore, the current CSST wide-field survey schedule may reduce orbital-phase coverage and thereby weaken binary detection and orbit recovery.

\subsection{\emph{Gaia} observations\label{subsect:gaia_obs_schedule}}
    \emph{Gaia} observes with a scanning law. Each transit provides astrometric information primarily in the along-scan (AL) direction, while the across-scan (AC) direction is much less precise. We adopted a Gaussian observation error for each AL CCD transit, with the standard deviation given by \citet{LL:LL-136}
    \begin{align}
        \sigma_{\eta,\text{CCD}}=\max\left(0.135,\sqrt{0.085^{2}+0.055x+0.0017x^{2}}\right)\ \text{mas},\label{eq:gaia_err_ccd}
    \end{align}
    where $x=10^{0.4(G-15)}$ and $G$ is the \emph{Gaia} $G$-band Vega magnitude. The AC observation error is about 10 times the AL observation error, and the number of AL observations is about 10 times that of AC observations \citep{2012A&A...538A..78L}. We therefore ignored AC measurements in the mock \emph{Gaia} astrometry.

    Following \citet{2024OJAp....7E.100E}, we combined successive CCD measurements into one FoV transit and adopted the assumed per-FoV-transit AL observation error
    \begin{align}
        \sigma_{\eta}=\frac{1}{\sqrt{8}}\sigma_{\eta,\text{CCD}}.\label{eq:gaia_err_fov}
    \end{align}
    The adopted FoV-level AL prescription approximates the per-transit observation error and omits CCD-level correlations, attitude correlations, and basic-angle systematics. Hereafter, $\sigma_{\eta}$ denotes this single-epoch FoV-level AL observation error.

    To simulate \emph{Gaia} observations, we made use of the \emph{Gaia} Observation Forecast Tool (GOST)\footnote{\href{https://gaia.esac.esa.int/gost/}{https://gaia.esac.esa.int/gost/}}, which provides the observation epochs $t_j$, scan angles $\psi_j$, and AL parallax factors $\Pi_j$ for a given direction. We queried the GOST scanning law on a HEALPix level 6 grid for the entire mission, from 2014-07-25T10:31:26 TCB to 2025-01-15T06:16:32 TCB. Each source was then assigned the GOST scan sequence of the HEALPix level 6 pixel containing its sky position. We selected $t_{\rm G}={\rm J}2019.75$ to be the reference epoch of \emph{Gaia}.

    The \emph{middle row} of Fig.~\ref{fig:obs_temporal_and_spatial_dist} shows the \emph{Gaia} observing time span and the number of FoV transits. The adopted full-mission scanning law provides approximately 180 FoV transits per level 6 sky pixel on average and an average time baseline of about 10.2 years, yielding regular temporal sampling for astrometric solutions.

\subsection{Joint CSST and \emph{Gaia} observations\label{subsect:joint_obs_schedule}}
    For sources within the overlapping magnitude range, we combined CSST observations and \emph{Gaia} AL transits at the epoch level. The resulting data set provides a longer time baseline for astrometric motion than either mission alone. Because the combined observations span from late 2014 to July 2037, we set the joint reference epoch to $t_{\rm J}=t_0={\rm J}2026.0$, where $t_0$ denotes the input catalog reference epoch.

\begin{figure}[!h]
        \centering
        \includegraphics[width=\linewidth]{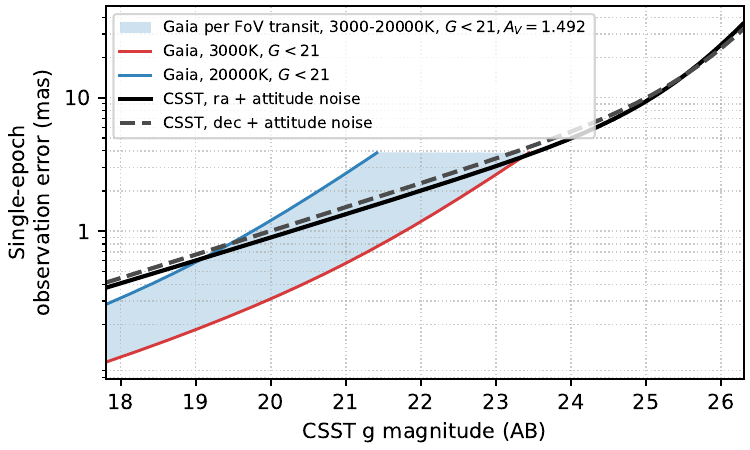}
        \caption{Single-epoch observation errors adopted for the simulated CSST and \emph{Gaia} observations. The black and grey curves show the CSST RA and Dec observation errors after adding an attitude-noise term of $0.1\,{\rm mas}$ in quadrature. The blue shaded region shows the effective \emph{Gaia} single-epoch FoV-level AL observation error mapped onto the CSST $g$ magnitude axis. The mapping includes the blackbody color, the offset between the AB and Vega zero points, and the adopted CSST and \emph{Gaia} extinctions evaluated at the median $A_V=1.492$ of the input sample. The shaded region spans $3000\,{\rm K}<T_{\rm eff}<20000\,{\rm K}$ and is truncated at $G<21$.}
        \label{fig:csst_gaia_obs_err}
    \end{figure}

    Fig.~\ref{fig:csst_gaia_obs_err} compares the single-epoch observation errors adopted for CSST and \emph{Gaia}. The CSST curves show the RA and Dec observation errors after including the attitude noise in Eq.~\eqref{eq:csst_err}, while the \emph{Gaia} curve shows the effective single-epoch FoV-level AL observation error in Eq.~\eqref{eq:gaia_err_fov}. For this display, we mapped the \emph{Gaia} single-epoch observation error model onto the CSST $g$ magnitude axis using blackbody spectra, so that the two models can be shown on a common magnitude axis. The mapping includes the blackbody color and the offset between the AB and Vega zero points introduced in Sect.~\ref{subsect:phot}. We adopted the extinctions $A_{G}=0.835A_{V}$ \citep{2019MNRAS.482.4570G} and $A_{g}=1.197A_{V}$ \citep{mwmsc}.

    We used $A_V=1.492$, the median value of the input MWMSC sample, for this display. The shaded region spans $3000\,{\rm K}<T_{\mathrm{eff}}<20000\,{\rm K}$ and is truncated at $G<21$. The approximation $A_{G}=0.835A_{V}$ was used only in this figure. The mock catalog uses the \emph{Gaia} magnitudes calculated by \texttt{isochrones} \citep{2015ascl.soft03010M} with the MWMSC $A_V$ of each source, as described in Sect.~\ref{subsect:phot}.

    Combining the CSST and \emph{Gaia} observations extends the time baseline to 18.2 years on average and increases the average number of observations to 220 per HEALPix level 6 pixel.

\section{The mock input catalog\label{sect:input_catalog}}
\subsection{The input samples\label{subsect:input_samples}}
    The mock input catalog was constructed from the Milky Way stellar mock catalog (MWMSC)\footnote{Online query and descriptions can be found at \href{https://nadc.china-vo.org/data/data/general/csst-trilegal/mwmsc/f}{https://nadc.china-vo.org/data/data/general/csst-trilegal/mwmsc/f}.} \citep{mwmsc}, which was developed for the CSST Survey Camera and is based on \texttt{TRILEGAL} \citep{2005A&A...436..895G,2016AN....337..871G}, a stellar population synthesis code. We adopted MWMSC because current catalogs face a trade-off between depth and sky coverage. For example, \emph{Gaia} provides a full-sky catalog but is limited to $G\simeq21$ \citep{2023A&A...674A...1G}, whereas deeper surveys, such as Pan-STARRS1 $3\pi$ and the DESI Legacy Imaging Surveys, reach $g\simeq 23$--$24$ over large areas but not the full sky \citep{2016arXiv161205560C,2019AJ....157..168D}. No existing real optical stellar catalog simultaneously reaches the CSST faint end, covers the full sky, and provides the astrometric information required for this work. MWMSC covers the full sky and provides astrophysical parameters, photometry, and basic astrometric parameters for about $12.6$ billion simulated stars down to $g=27.5$ \citep{mwmsc}, which is fainter than the adopted CSST limit of $g=26.3$.

    We used MWMSC as the underlying Galactic stellar population and then generated the evolved masses, orbital elements, and photometry required for the subsequent mock observations and orbital solutions. This construction preserves the sky distribution, proper motions, distances, ages, metallicities, extinctions, and initial masses from a physically motivated Milky Way simulation, while allowing the binary evolution results and orbital elements to be generated in a controlled way.

    We first applied a set of cuts before drawing the sample. Sources were required to have $\varpi>0.03\,{\rm mas}$ to remove extremely distant ($\gtrsim 30\,{\rm kpc}$) sources. We also required the sources to be initially detectable in the CSST $g$ band with $17.8<g<26.3$, where the bright limit was set to avoid saturation \citep{2026AJ....171..121X}. As described in Sect.~\ref{subsect:csst_obs_schedule}, we consider only the CSST $g$ band in this work. The initial magnitude requirement limits the size of the input sample. After photometric processing, we reapply the CSST detectable-magnitude cut to the final total magnitude in Sect.~\ref{subsect:phot}.

    We further restricted the input parameters to the range supported by the evolution and photometric generation process in Sects. \ref{subsect:evolution} and \ref{subsect:phot}, requiring $-4<[{\rm M/H}]<0.5$ and $0.1M_{\sun}<m_{1,\rm init}<150M_{\sun}$, where the upper limit is adopted from \citet{2005Natur.434..192F}. For binaries, the secondary initial mass $m_{2,\rm init}=m_{1,\rm init}q_{\rm init}$ was required to satisfy the same mass range. Pre-main-sequence sources were excluded to keep the input catalog consistent with \texttt{COSMIC} \citep{2020ApJ...898...71B}, which starts from the zero-age main sequence.

    From all MWMSC sources passing these cuts, we drew a random sample of $20\,000\,000$ sources and retained their positions, distances, proper motions, ages, metallicities, initial masses and mass ratios, and extinctions. The sampled catalog contains $13\,130\,084$ single stars and $6\,869\,916$ binaries. Binaries were evolved as described in Sect.~\ref{subsect:evolution} and then passed to the photometric calculation in Sect.~\ref{subsect:phot}. Single stars were passed directly to the same photometric calculation.

\subsection{Binary evolution\label{subsect:evolution}}
    MWMSC provides only the basic astrometric parameters for single stars, but not the orbital elements or the evolved binary properties required for our astrometric simulations. We therefore evolved the binaries with \texttt{COSMIC} \citep{2020ApJ...898...71B}. The initial periods and eccentricities were generated following \citet{LL:LL-136,1991A&A...248..485D}. The orbital periods were sampled from $\lg (P/{\rm d}) \sim N(4.8,2.3^2)$, truncated to $50\,{\rm d}<P<10000\,{\rm d}$, where $N(\cdot,\cdot)$ denotes a normal distribution.

    The lower limit was chosen because shorter-period binaries have smaller photocentric semi-major axes and are therefore more difficult to detect. The upper limit was chosen so that the \emph{Gaia} or CSST baseline covers a substantial fraction of the orbit. Specifically, \citet{2014A&A...563A.126L} found that even for uniformly sampled, high signal-to-noise observations, about $40\%$ orbital coverage is needed for an unbiased mass estimate. This period range therefore targets binaries whose orbital parameters may be recovered.

    The eccentricities followed a truncated normal distribution $N(0.31,0.15^2)$ for $P\leq1000\,{\rm d}$, and were set to $e=0.95\sqrt{U(0,1)}$ for longer periods, where $U(\cdot,\cdot)$ denotes a uniform distribution.

    We kept only binaries with well-defined final component masses and orbital parameters that could be used in the astrometric simulations. The filtering was applied sequentially. Binaries were removed
    \begin{enumerate}
        \item if \texttt{COSMIC} returned a merged or disrupted state,
        \item or if Roche-lobe overflow occurred either at the endpoint or at any time in the evolution history,
        \item or if both components were compact or massless remnants,
        \item or if the final period, eccentricity, separation, or component masses were invalid, for example infinite, NaN, or non-positive,
        \item or if one component had become massless while the other remained luminous.
            These binaries were not recycled into the sample of single stars, because they had undergone binary evolution but no longer provided a clean binary orbit.
    \end{enumerate}
    These filters removed $516\,558$ of the $6\,869\,916$ input binaries, corresponding to $7.52\%$, and left $6\,353\,358$ binaries with finite orbital elements and component masses.

    We assigned random orbital orientations and phases. The cosine of the inclination, $\cos i$, was drawn uniformly from $[-1,1]$. The argument of periastron $\omega$, the longitude of the ascending node $\Omega$, and the mean anomaly at the reference epoch $M_0$ were drawn uniformly from $[0,2\pi)$.

\subsection{Photometry\label{subsect:phot}}
    The effective temperatures and \emph{Gaia} photometry were calculated with \texttt{isochrones} \citep{2015ascl.soft03010M} based on the MIST evolution tracks \citep{2016ApJS..222....8D,2016ApJ...823..102C}. Although MWMSC provides apparent magnitudes, we recalculated the photometry by evaluating the same model with the source-specific MWMSC $A_V$ and again with $A_V=0$. This procedure yielded extincted and extinction-free component magnitudes in the \emph{Gaia} bands with a consistent extinction treatment.

    We constructed the CSST $g$ magnitudes from the extinction-free \emph{Gaia} $G$ photometry using the adopted \emph{Gaia} $G$ and CSST $g$ response curves \citep{2021A&A...649A...3R,2026SCPMA..6939501C}, blackbody spectra at the effective temperatures returned by \texttt{isochrones}, and the source-specific CSST extinction.\footnote{Throughout this work, $G$ denotes the \emph{Gaia} $G$-band Vega magnitude \citep{2021A&A...649A...3R}, while $g$ denotes the CSST $g$-band AB magnitude \citep{2026SCPMA..6939501C}.} For each unresolved binary, we calculated the two component $g$ magnitudes separately and obtained the total magnitude by summing their fluxes.

    The binary photometric calculation first classified binaries according to whether modeled magnitudes were available for the two stars. Among the $6\,353\,358$ binaries entering the photometric calculation, $5\,666\,162$ had modeled magnitudes for both components and $295\,869$ had a modeled magnitude only for the initially less massive component; we retained both groups. For the latter systems, the luminous component was relabeled as component 1 and the other component as component 2. A magnitude of $99$ was assigned to component 2 so that its optical flux was negligible. We discarded the $92$ binaries with a modeled magnitude only for the initially more massive component and the $391\,235$ binaries with no modeled magnitude for either component. The first discarded group consists mainly of rare model-boundary cases, while the second group does not provide an optically detectable source.

    We define the flux ratios as \citep{2002A&A...391..647G}
    \begin{align}
        f_{\rm G}=10^{-0.4(G_2-G_1)},\qquad
        f_{\rm C}=10^{-0.4(g_2-g_1)},
    \end{align}
    where the subscripts G and C denote the \emph{Gaia} $G$ band and the CSST $g$ band, respectively. We then computed the corresponding photocentric semi-major axis \citep{2002A&A...391..647G}
    \begin{align}
        a_{0,x} = a\left(\frac{q}{1+q}-\frac{f_{x}}{1+f_{x}}\right),\label{eq:a0}
    \end{align}
    where $x=\{{\rm G},{\rm C}\}$, $a$ is the angular relative semi-major axis of the binary, and $q=m_2/m_1$ is the mass ratio. The simulated $a_{0,x}$ is signed because the photocenter can lie on either side of the barycenter depending on the mass and flux ratios. In the orbital fitting procedure, this information is carried by the signs of the Thiele--Innes elements \citep{2009ApJS..182..205W} and therefore all fitted photocentric semi-major axes are positive.

    We then applied a catalog-level cut to flag binaries for which the unresolved photocenter approximation may become inappropriate. This cut is based on the photocentric semi-major axis. Similar criteria have been adopted in recent CSST simulations, where the photocentric semi-major axis is compared with the diffraction limit to separate unresolved and resolvable binaries \citep{2026AJ....171..121X}. In practice, whether a binary is resolved, marginally resolved, or unresolved depends on several factors beyond the photocentric semi-major axis, including the instantaneous separation projected onto the detector and the component magnitude difference \citep{LL:LL-136,2024OJAp....7E.100E}. The epoch-level effects of marginally resolved observations are treated later in the mock observation model through the binary observation bias prescription and an additional blending-noise term in Sect.~\ref{sect:models_and_mock_obs}. Here we use $|a_{0,x}|$ only as a catalog-level filter.

    For \emph{Gaia}, we adopted a commonly used angular resolution $u_{\rm G}=90\,{\rm mas}$ \citep{LL:LL-136,2023A&A...674A..25H,2024OJAp....7E.100E} as an approximation to distinguish between unresolved and marginally resolved AL astrometry. For CSST, we estimated the corresponding angular resolution from the diffraction limit \citep{2026AJ....171..121X}. Using a $6000\,{\rm K}$ blackbody spectrum and the CSST $g$-band throughput \citep{2026SCPMA..6939501C}, we obtained a $g$-band diffraction limit of $\simeq60\,{\rm mas}$, and adopted $u_{\rm C}=60\,{\rm mas}$.

    For binaries with two luminous components, we flagged a system as potentially resolved if its absolute photocentric semi-major axis exceeded the adopted angular resolution in either mission,
    \begin{align}
        |a_{0,\rm C}|>u_{\rm C}
        \quad {\rm or} \quad
        |a_{0,\rm G}|>u_{\rm G}.
    \end{align}
    After all preceding cuts, only three binaries satisfied this criterion, which is insufficient for statistically meaningful resolved-binary population results. We therefore excluded these three sources from the main sample. Appx.~\ref{appx:resolved_binary} presents an independent experiment on a toy population of resolved binaries. Finally, we reapplied the CSST magnitude cut to the total $g$ magnitude calculated after obtaining the two component magnitudes and summing their fluxes. We retained binaries with $17.8<g_{\rm tot}<26.3$. After these filters, the input sample contained $4\,761\,319$ binaries.

    Among the $4\,761\,319$ input binaries, $1\,211\,353$ have total magnitudes satisfying both $G_{\rm tot}<21$ and $17.8<g_{\rm tot}<26.3$. In this $1.21$ million-source subset, the median absolute difference $|a_{0,\rm C}-a_{0,\rm G}|$ is $7.10\,$\textmu as, and $90\%$ have an absolute difference below $47.4\,$\textmu as.

    Single stars were processed with the same photometric procedure, except that no binary evolution or orbital parameters were assigned. Of the $13\,130\,084$ single-star inputs, $1\,167\,688$ did not obtain a magnitude from \texttt{isochrones}, primarily because they lay outside the model boundaries. The final CSST detectable magnitude range cut retained $9\,351\,614$ single stars. The final mock input catalog therefore contains $14\,112\,933$ sources in total.

\section{Astrometric models and mock observations\label{sect:models_and_mock_obs}}
\subsection{Astrometric models\label{subsect:models}}
    For each source, we generated the astrometric motion on a local tangent plane centered on the catalog position at $t_0={\rm J}2026.0$. The \emph{Gaia} observables are along-scan projections evaluated with the GOST scan angles and parallax factors, whereas CSST provides two-dimensional tangent-plane positions. The complete single-star forward models are given in Appx. \ref{appx:detailed_ast_models}.

    For binaries, we added the passband-dependent orbital displacement of the photocenter. We used the prescription of \citet{LL:LL-136} and \citet{2024OJAp....7E.100E} to describe the transition from unresolved photocenter motion to the marginally resolved regime, adopting $u_{\rm G}=90\,{\rm mas}$ and $u_{\rm C}=60\,{\rm mas}$. The full binary models are presented in Appx. \ref{appx:detailed_ast_models}.

\subsection{Generating the mock observations\label{subsect:mock_obs}}
    After evaluating the astrometric models, we added noise using the single-epoch observation errors described in Sect.~\ref{sect:mock_epoch_astrometry}. For \emph{Gaia}, we used the precomputed GOST scanning law, combined successive CCD transits into FoV transits, and randomly rejected each FoV transit with a 10\% probability following \citet{2021A&A...649A...4L} and \citet{2024OJAp....7E.100E}. For CSST, we added Gaussian noise using the adopted $g$-band single-epoch observation-error model, including the attitude-noise term.

    For binaries with two luminous components, we additionally included a Gaussian blending noise term of $0.5\,{\rm mas}$ when the instantaneous projected separation exceeds half of the corresponding resolution limit \citep{2024OJAp....7E.100E}, i.e. $|\Delta\eta_j|>u_{\rm G}/2$ for \emph{Gaia} and $\rho_j>u_{\rm C}/2$ for CSST. For \emph{Gaia}, this term is added in the AL direction; for CSST, it is added in quadrature to each of the RA and Dec errors.

    According to \citet{2018A&A...616A...2L}, the number of visibility periods $N_{\rm vis}$ is a more informative indicator of temporal sampling than the number of observations. In the \emph{Gaia} astrometric solution, a visibility period is a group of observations separated from other groups by a gap of at least four days. This definition downweights observations concentrated within a few short time intervals, because such observations are less useful for astrometry than observations spread across many distinct epochs.

    Following the same idea, we introduced the number of effective observations, $N_{\rm effobs}$. For \emph{Gaia}, we defined $N_{\rm effobs,G}$ to be identical to $N_{\rm vis,G}$.
    \begin{align}
        N_{\rm effobs,G}=N_{\rm vis,G}.
    \end{align}
    For CSST, each observation provides two observables, so we count
    \begin{align}
        N_{\rm effobs, C}=2N_{\rm vis, C}.
    \end{align}
    For the joint scenario, the number of effective observations was taken to be the sum of the two contributing scenarios,
    \begin{align}
        N_{\rm effobs, J}=N_{\rm effobs, C}+N_{\rm effobs, G}=2N_{\rm vis, C}+N_{\rm vis, G}.
    \end{align}

    We generated mock observations only for sources that are, in principle, sufficiently observed for a 12p binary orbital solution. We required $N_{\rm effobs}\geq 13$ for all scenarios, corresponding to at least one degree of freedom beyond the 12 fitted parameters for evaluating parameter uncertainties. Sources with $G>21$ were treated as unobserved by \emph{Gaia} and assigned $N_{\rm effobs, G}=0$. In the joint scenario, we additionally required both missions to contribute data, $N_{\rm effobs, G}>0$ and $N_{\rm effobs, C}>0$, and required the combined number of effective observations to satisfy $N_{\rm effobs, J}\geq13$.

    After applying the $N_{\rm effobs}$ requirements, the numbers of sources for which observations were generated in the three scenarios are listed in Table~\ref{table:num_mock_sample}.
    \begin{table}[!h]
        \caption{Numbers of sources for which observations were generated after applying the effective-observation requirements. More details are given in Sect.~\ref{subsect:mock_obs}.}
        \label{table:num_mock_sample}
        \centering
        \begin{tabular}{c|c|c|c}
            \hline
            Scenario & Total & Single stars & Binaries\\
            \hline
            \emph{Gaia} & $3\,294\,321$ & $2\,082\,968$ & $1\,211\,353$\\
            CSST & $871\,963$ & $582\,677$ & $289\,286$\\
            Joint & $283\,306$ & $180\,995$ & $102\,311$\\
            \hline
        \end{tabular}
    \end{table}

\section{Selecting binary candidates\label{sect:select_binary_candidates}}
    The 5p single-star solution is linear in the astrometric parameters and was therefore computed for all sources. For each scenario, we fitted the 5p model to the simulated epoch astrometry with weighted least squares, and recorded the 5p solution, covariance matrix, and diagnostics.

    The 12p binary orbital model contains nonlinear orbital parameters and is computationally expensive, making it impractical to apply the 12p solution to all simulated sources. We therefore selected binary candidates using diagnostics derived from the 5p solutions, supplemented by auxiliary observational-sampling features. In future applications, the 12p fitting stage can also be applied to binary candidates selected independently, for example from acceleration solutions or overluminosity in future \emph{Gaia} or CSST data releases.

    The 5p diagnostics (Sect.~\ref{subsubsect:5p_diag}) summarize the goodness of fit, excess noise, parameter significances, and covariance of the 5p solution. The PMa features (Sect.~\ref{subsubsect:pma}) measure the difference between catalog proper motions from the 5p solutions and the long-baseline \emph{Gaia}--CSST proper motion. The auxiliary observational-sampling features (Sect.~\ref{subsubsect:obs_features}) include the number of effective observations, time baseline, and cadence statistics. These features describe the time sampling under which the 5p and PMa features are produced and the sensitivity with which these features respond to orbital motion. These diagnostics have been extensively used to search for signals related to orbital motion, e.g. \citet{2019A&A...632L...9K,2022MNRAS.510.3885G,2025A&A...704A.150A}.

\subsection{Diagnostic and auxiliary features\label{subsect:select_binary_candidates_features}}
    The 5p model is inadequate for binaries because it does not model orbital motion. The unmodeled orbital motion can appear as increased 5p residuals or inflated uncertainties, bias the fitted 5p parameters, or contribute to astrometric excess noise. For clarity, we describe the features used for binary candidate selection in terms of their physical origin: 5p diagnostics, proper motion anomaly (PMa) features, and auxiliary observational-sampling features. This organization is consistent with a machine-learning search for substellar companions \citep{2025A&A...704A.150A}.

\subsubsection{The 5p diagnostics\label{subsubsect:5p_diag}}
    The $F_2$ statistic transforms the $\chi^2$ statistic into an approximately standard normal variable \citep{wilson1931distribution}, making it more comparable between sources with different numbers of observations and between scenarios. This statistic has been extensively used to analyze Hipparcos and \emph{Gaia} data \citep{The_Hipparcos_and_Tycho_catalogues,2023A&A...674A...9H,2023A&A...674A..10H}. We therefore included $F_{2,\rm 5p}$, the $F_2$ statistic of the 5p solution, as a goodness-of-fit feature:
    \begin{align}
        F_{2,\rm 5p}=\sqrt{\frac{9\nu_{\rm 5p}}{2}}\left[\left(\frac{\chi^{2}_{\rm 5p}}{\nu_{\rm 5p}}\right)^{1/3}+\frac{2}{9\nu_{\rm 5p}}-1\right],
    \end{align}
    where $\nu_{\rm 5p}$ is the number of degrees of freedom of the 5p fit. In our three scenarios,
    \begin{align}
        \begin{split}
            \nu_{\rm 5p, G}&=N_{\rm epoch, G}-5,\\
            \nu_{\rm 5p, C}&=2N_{\rm epoch, C}-5,\\
            \nu_{\rm 5p, J}&=N_{\rm epoch, G}+2N_{\rm epoch, C}-5.
        \end{split}
    \end{align}
    Here $N_{\rm epoch,G}$ is the number of \emph{Gaia} FoV transits and $N_{\rm epoch,C}$ is the number of CSST observations.

    We also used the astrometric excess noise $\epsilon$ and its significance $s_{\epsilon}$. The astrometric excess noise is the additional noise introduced to make the residuals of a 5p fit statistically consistent with the adopted observation errors. A significant astrometric excess noise can therefore indicate unresolved astrometric signals, including orbital motion. This quantity has been used to search for binary-related astrometric signals \citep{2022MNRAS.510.3885G} and to constrain binary parameters \citep{2019A&A...632L...9K,2021A&A...645A...7K,2026A&A...710A.398L}.

    We also included information from the fitted 5p parameters and their covariance. Binary orbital motion can bias a 5p solution or be partly absorbed into the fitted five astrometric parameters \citep{2020MNRAS.495..321P}. We therefore included parameter significances as part of the 5p diagnostic feature group.

    For one-dimensional quantities, such as parallax, the significances were defined as the ratio of the fitted value to its uncertainty. For two-dimensional quantities, such as the position offset and proper motion, we defined the significance as the Mahalanobis distance \citep{mahalanobis} from the corresponding null vector. Here the null vector corresponds to zero position offset or zero proper motion.
    \begin{align}
        \begin{split}
            s_{\rm 5p, offset}&=\sqrt{\Delta\vec{P}_{\rm 5p}^{\rm T}\vec{C}_{\Delta\alpha^{\ast}\Delta\delta,\rm 5p}^{-1}\Delta\vec{P}_{\rm 5p}},\\
            s_{\rm 5p, \mu}&=\sqrt{\vec{\mu}_{\rm 5p}^{\rm T}\vec{C}_{\mu_{\alpha^{\ast}}\mu_{\delta},\rm 5p}^{-1}\vec{\mu}_{\rm 5p}},\\
            s_{\rm 5p, \varpi}&=\frac{\varpi_{\rm 5p}}{\sigma\left(\varpi_{\rm 5p}\right)}.
        \end{split}
    \end{align}
    where $\Delta\vec{P}_{\rm 5p}=\left[\Delta\alpha^{\ast}_{\rm 5p},\Delta\delta_{\rm 5p}\right]^{\rm T}$ and $\vec{\mu}_{\rm 5p}=\left[\mu_{\alpha^\ast, \rm 5p},\mu_{\delta, \rm 5p}\right]^{\rm T}$, and $\vec{C}_{\Delta\alpha^{\ast}\Delta\delta,\rm 5p}$ and $\vec{C}_{\mu_{\alpha^{\ast}}\mu_{\delta},\rm 5p}$ are the covariance matrices of the position and the proper motion, extracted from the 5p covariance matrix $\vec{C}_{\rm 5p}$.

    To summarize whether the 5p solution is poorly constrained along any direction in the parameter space, we also included $\sigma_{\rm 5D,max}$, defined as the square root of the largest singular value of the scaled 5p covariance matrix $\vec{S}\vec{C}_{\rm 5p}\vec{S}$, where $\vec{S}$ is the scaling matrix. For \emph{Gaia}, $\vec{S}={\rm diag}(1,1,\sin 45\degr,T/2,T/2)$, where $45\degr$ is the solar-aspect angle in the nominal scanning law and the factor $\sin45\degr$ accounts for the maximum AL parallax factor \citep{2018A&A...616A...2L}. For CSST and the joint 5p solution, we used the same covariance-compression idea but adopted $\vec{S}={\rm diag}(1,1,1,T/2,T/2)$. This choice reflects that, for two-dimensional astrometry, the parallactic displacement is measured as a tangent-plane vector, whose parallax factor has a scale of order unity for a near-Earth observer.

    The above features, including
    \begin{itemize}
        \item $F_{2,\rm 5p}$, which quantifies the overall goodness of fit of the 5p solution;
        \item $\epsilon$ and $s_{\epsilon}$, which measure the astrometric excess noise;
        \item the parameter significances $(s_{\rm 5p, offset}$, $s_{\rm 5p, \mu}$, $s_{\rm 5p, \varpi})$, which characterize the signal-to-noise level;
        \item $\sigma_{\rm 5D,max}$, which includes information about the uncertainties of the 5p solution;
    \end{itemize}
    were collected into the 5p diagnostic feature group for the binary classifier in Appx.~\ref{appx:selection}.

\subsubsection{Proper motion anomaly\label{subsubsect:pma}}
    The proper motion anomaly (PMa) measures the difference between a catalog-level proper motion and a long-baseline proper motion inferred from positions at different epochs. For a single star, these two proper motions should be statistically consistent. For an unresolved binary, the photocenter orbit can perturb the fitted short-baseline proper motion and generate a measurable PMa. This method has been used to identify stellar and substellar companions with \emph{Gaia} data \citep{2019A&A...623A..72K,2022A&A...657A...7K}, and related Hipparcos--\emph{Gaia} analyses compare proper motions measured over different time spans to capture the potential orbital motion \citep{2021ApJS..254...42B,2026AJ....172...53T}.

    In our implementation, the PMa feature group contains the total PMa, the two PMa components, their component uncertainties, and the PMa significance, for both \emph{Gaia} and CSST. These PMa features are only available for sources with successful 5p solutions in both \emph{Gaia} and CSST. The calculation is described in Appx.~\ref{appx:pma_calc}.

\subsubsection{Auxiliary observational-sampling features\label{subsubsect:obs_features}}
    The response of the 5p and PMa diagnostics to a given photocentric orbit depends on the time sampling of the observations. Diagnostics based on the 5p fit residuals depend on the number of observations, the time baseline, and the cadence \citep{2024A&A...688A...1C}. This dependence is also explicit in recent work that uses \emph{Gaia} DR3 astrometric excess noise to constrain binary-orbit inclinations \citep{2026A&A...710A.398L}. For PMa, the sensitivity depends on the observational coverage of the individual catalogs and on the baseline used to derive long-term proper motion \citep{2019A&A...623A..72K,2022A&A...657A...7K}. These considerations motivate including observational-sampling information when interpreting the 5p and PMa diagnostics.

    We therefore included auxiliary observational-sampling features as complementary inputs to the classifier. For the individual \emph{Gaia} and CSST stages, these features include the number of effective observations, the time baseline, and the median, standard deviation, and maximum gap between consecutive observing epochs. For the joint stages, we used the corresponding summaries for each mission, together with the inter-mission gap and the total time span.

\subsection{Binary candidates\label{sect:bin_candidates}}
    We selected binary candidates with a four-stage hierarchical cascade classifier based on the features described in Sect.~\ref{subsect:select_binary_candidates_features}. The stages were ordered by the available information: joint 5p diagnostics with PMa features, joint 5p diagnostics without PMa, \emph{Gaia} diagnostics, and CSST diagnostics. Each stage used a histogram-based gradient-boosting classifier implemented with \texttt{HistGradientBoostingClassifier} in \texttt{scikit-learn} \citep{2011JMLR...12.2825P} and trained on sources eligible for that stage. Sources accepted by an earlier stage were not passed to later stages, and the final candidate set was the union of the four accepted subsets. The definition of an eligible source and details of the classifier training, threshold selection, and cascade evaluation can be found in Appx.~\ref{appx:selection}.

    On the independent test set, the hierarchical cascade classifier selected $72\,554$ binary candidates out of the $896\,258$ sources that were eligible for at least one classifier. These candidates include $58\,199$ true binaries and $14\,355$ single stars in the simulation. The resulting precision is $0.802$. The selected true binaries correspond to $18.1\%$ of the eligible binaries.

    \begin{table}[!h]
        \caption{Test set performance of the hierarchical candidate classifier.}
        \label{table:candidate_selection_summary}
        \centering
        \begin{tabular}{l|r}
            \hline\hline
            Quantity & Value \\
            \hline
            Total sources in the test set & $7\,054\,416$ \\
            \quad Binaries & $2\,380\,323$ \\
            \quad Single stars & $4\,674\,093$\\ \hline
            Sources eligible for at least one classifier & $896\,258$ \\
            \quad Binaries & $322\,204$ \\
            \quad Single stars & $574\,054$ \\ \hline
            Selected binary candidates & $72\,554$ \\
            \quad True positives (binaries) & $58\,199$ \\
            \quad False positives (single stars) & $14\,355$ \\ \hline
            Precision & $0.802$ \\
            Recall among eligible binaries & $0.181$ \\
            \hline
        \end{tabular}
    \end{table}

    The four score thresholds were optimized jointly by applying the complete cascade to the validation set. The final candidate sample is the union of the candidates selected by the four stages from the test set. Sources accepted by an earlier stage are not passed to the later stages. Therefore, later stages act only on sources that were either not eligible for, or not accepted by, the preceding stages. The earlier stages select candidates using more complete feature sets, leaving the later stages with a more difficult remaining sample. Although the remaining samples are more difficult, the later stages still add true binaries that would otherwise be absent from the final candidate list.

\begin{figure*}[!h]
        \centering
        \includegraphics[width=0.9\textwidth]{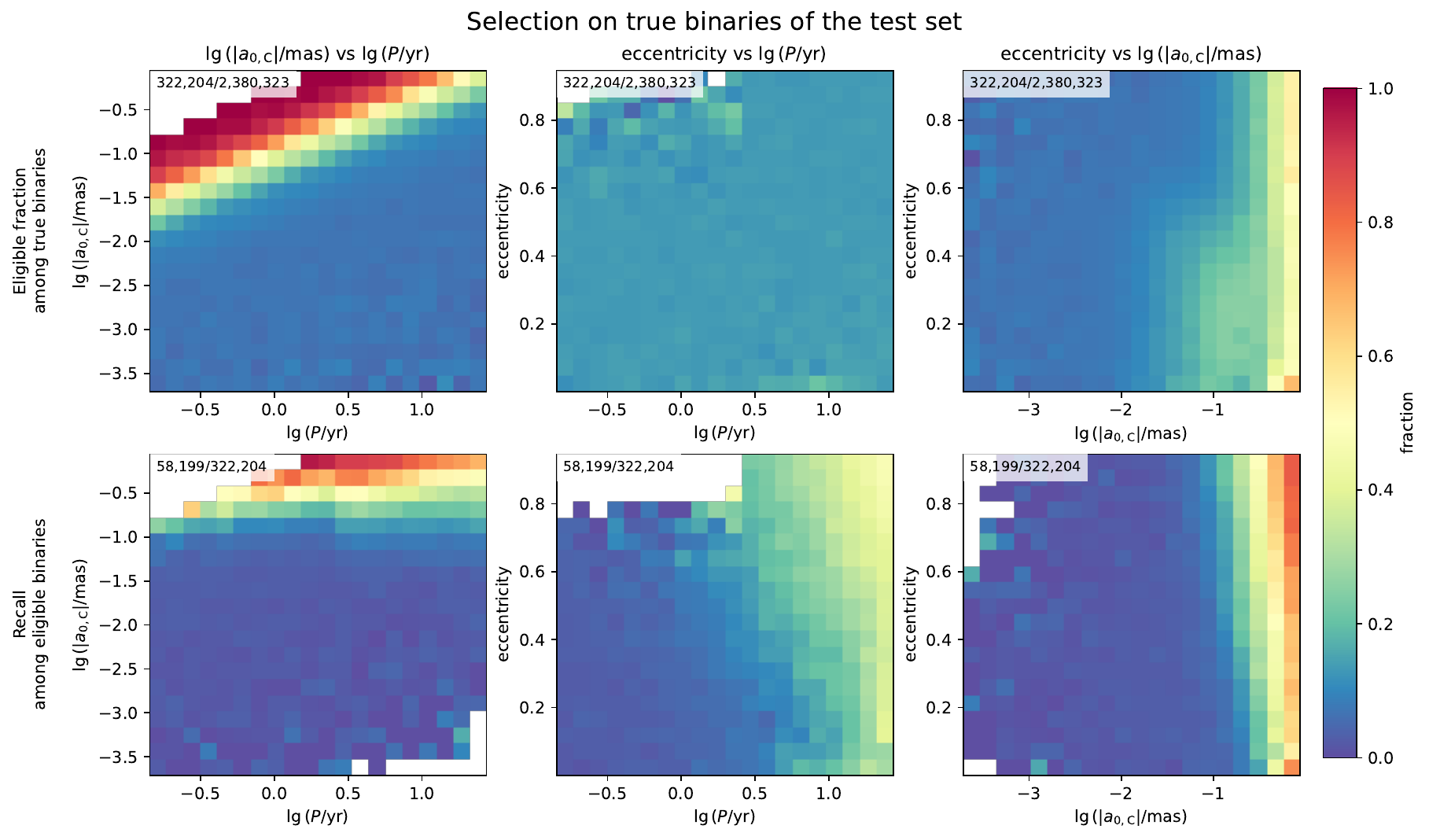}
        \caption{Selection of true binaries in the test set by the hierarchical candidate classifier. The \emph{top} row shows the eligible fraction, namely the fraction of true binaries that pass the pre-selection of at least one stage (see Appx.~\ref{appx:selection}). The \emph{bottom} row shows the recall of eligible binaries. The photocentric semi-major axis shown in all panels is $|a_{0,\rm C}|$. Because $a_0$ depends on the passband through the band-specific flux ratio, we consistently use the CSST value: the adopted CSST magnitude range, $17.8<g<26.3$, encompasses the full sample analyzed here, whereas the \emph{Gaia} sample is limited to $G<21$. In each panel, the plotted range along each axis is set by the 0.5th and 99.5th percentiles of the true binaries in the test set. The numbers annotated in the upper-left corner of the panels give eligible/total true binaries in the test set in the first row and selected/eligible true binaries in the second row. White bins contain too few data for a stable estimate of the fraction.}
        \label{fig:bin_candidates}
    \end{figure*}

    Fig.~\ref{fig:bin_candidates} shows the selection of true binaries in the test set. The top row shows the fraction of true binaries that are eligible for at least one stage, and the bottom row shows the recall among those eligible binaries. In each panel, the plotted range in each coordinate is set by the 0.5th and 99.5th percentiles of the corresponding distribution of true binaries in the test set. The numbers annotated in the upper-left corner of the panels give eligible/total true binaries in the test set in the first row and selected/eligible true binaries in the second row.

    The dominant trend is with the photocentric semi-major axis. Binaries with larger values of the photocentric semi-major axis $|a_{0,\rm C}|$\footnote{We use the CSST photocentric semi-major axis because $a_{0}$ is mission dependent through the passband-integrated flux ratio, and \emph{Gaia} has no observations for sources fainter than $G\gtrsim21$.} are both more likely to be eligible and more likely to be recalled once eligible for at least one stage. A period dependence is also visible. At fixed photocentric semi-major axis, the recall changes with orbital period because the 5p residuals, PMa, and observational coverage respond to different orbital periods. The eccentricity dependence is weaker but still visible. Higher eccentricity binaries produce faster photocenter motion near periastron. When the motion close to periastron is sampled, the 5p model absorbs the orbital motion less efficiently. This mainly increases the positional offset significances, which modestly improves the recall once the source is eligible for at least one stage.

\section{Orbital solution results\label{sect:bin_orb_sol}}

\subsection{Definition of selection tiers\label{subsect:orbfit_selection_tiers}}
    We applied the 12p orbital fit to the candidates selected in Sect.~\ref{sect:bin_candidates}. This sample contains $72\,554$ sources, of which $58\,199$ are true binaries and $14\,355$ are single stars. The fitting algorithm is described in Appx.~\ref{appx:orb_sol}.

    We define two nested selection tiers for the fitted solutions. The \emph{12p model preference} tier requires
    \begin{align}
        \Delta{\rm BIC}={\rm BIC}_{5\rm p}-{\rm BIC}_{12\rm p}\geq 10.
    \end{align}
    Here, BIC denotes the Bayesian information criterion, and the subscripts identify the 5p and 12p models. A positive $\Delta{\rm BIC}$ favors the 12p model, and the adopted threshold of 10 requires strong evidence for the 12p model \citep{2007MNRAS.377L..74L}.

    The \emph{fiducial} tier is a more restrictive subset of the 12p model preference tier. It additionally requires $F_{2,12{\rm p}}<25$, where $F_{2,12{\rm p}}$ is the $F_2$ statistic of the 12p solution, and $s_{a_0}=a_0/\sigma_{a_0}>1$, where $s_{a_0}$ is the photocentric semi-major-axis significance. We also exclude solutions that converge at the upper eccentricity boundary $e=0.99$ \citep{2019MNRAS.489..738H}.

\subsection{Summary of the orbital solution}
    \begin{table*}
        \caption{Orbital solution results for the candidate sample \tablefootmark{a}.}
        \label{tab:orbfit_results}
        \centering
        \begin{tabular}{l|c|c|c}
            \hline\hline
            Scenario & 12p-fitted sources\tablefootmark{b} & \multicolumn{2}{c}{Selected sources (binary/single; percentage\tablefootmark{c})} \\
            \cline{3-4}
            & binary/single & 12p model preference & Fiducial \\
            \hline
            \emph{Gaia} & $58\,147$/$14\,282$ & $7\,307$/$0$; $12.57\%$ & $3\,930$/$0$; $6.76\%$ \\
            CSST & $6\,715$/$2\,804$ & $35$/$0$; $0.52\%$ & $12$/$0$; $0.18\%$ \\
            Joint & $9\,403$/$3\,241$ & $2\,010$/$0$; $21.38\%$ & $984$/$0$; $10.46\%$ \\
            \hline
        \end{tabular}
        \tablefoot{\tablefoottext{a}{The candidate sample contains $72\,554$ sources, of which $58\,199$ are true binaries and $14\,355$ are single stars.}\tablefoottext{b}{A source is counted only if it satisfies the mock requirements for that scenario; the joint scenario additionally requires observations from both missions.}\tablefoottext{c}{Each percentage uses the number of 12p-fitted true binaries in that row as its denominator.}}
    \end{table*}

    Table~\ref{tab:orbfit_results} summarizes the two selection tiers for the 12p solutions. The 12p-fitted samples still contain single stars because the nonlinear 12p model can formally converge for single-star epoch data. In all three scenarios, none of the fitted simulated single stars satisfies the 12p model preference criterion $\Delta{\rm BIC}\geq10$. Within this simulation, the adopted $\Delta{\rm BIC}$ threshold therefore removes all single-star false positives that formally converge under the 12p model.

    The numbers in the two selection tiers differ among the scenarios. The joint 12p-fitted sample is smaller than the \emph{Gaia} 12p-fitted sample because it is drawn from the smaller subset of candidates that have observations from both missions and satisfy the joint effective observation requirement in Sect.~\ref{subsect:mock_obs}. CSST alone yields only $35$ solutions in the 12p model preference tier and $12$ fiducial solutions. This small yield is consistent with the adopted CSST observing schedule. Although CSST provides high-precision astrometric measurements, the temporal sampling is often too uneven for 12p solutions to satisfy the fiducial criteria. The joint fit yields $2\,010$ 12p model preference and $984$ fiducial solutions, corresponding to $21.38\%$ and $10.46\%$ of the fitted true binaries. In the scenario-specific fitted samples, the joint solution raises the 12p model preference fraction from $12.57\%$ for the \emph{Gaia} fit to $21.38\%$ for the joint fit, and the fiducial fraction from $6.76\%$ to $10.46\%$.

    The joint fit uses a single photocentric semi-major axis for both bands. We characterize the intrinsic shared-$a_0$ mismatch as $\Delta_{\rm shared}=\frac{1}{2}\left|a_{0,\rm C}-a_{0,\rm G}\right|$, where $a_{0,\rm C}$ and $a_{0,\rm G}$ are the true photocentric semi-major axes for CSST and \emph{Gaia}, respectively. For a common value adopted midway between the two band-specific true values, $\Delta_{\rm shared}$ is the absolute mismatch in each band; any other fitted $a_0$ produces a larger mismatch in one of the two bands. For the $984$ fiducial joint fits, $\Delta_{\rm shared}/|a_{0,\rm C}|$ has a median of $0.077$ and a 99th percentile of $0.106$; $\Delta_{\rm shared}/|a_{0,\rm G}|$ has a median of $0.091$ and a 99th percentile of $0.134$. For these fiducial joint fits, the formal fractional $a_0$ uncertainty $\sigma_{a_0}/a_{0}$ has a median of $0.337$. At the population level, the shared-$a_0$ mismatch is therefore smaller than the typical reported formal $a_0$ uncertainty in this simulation.

    \begin{figure*}[!h]
        \centering
        \includegraphics[width=0.7\textwidth]{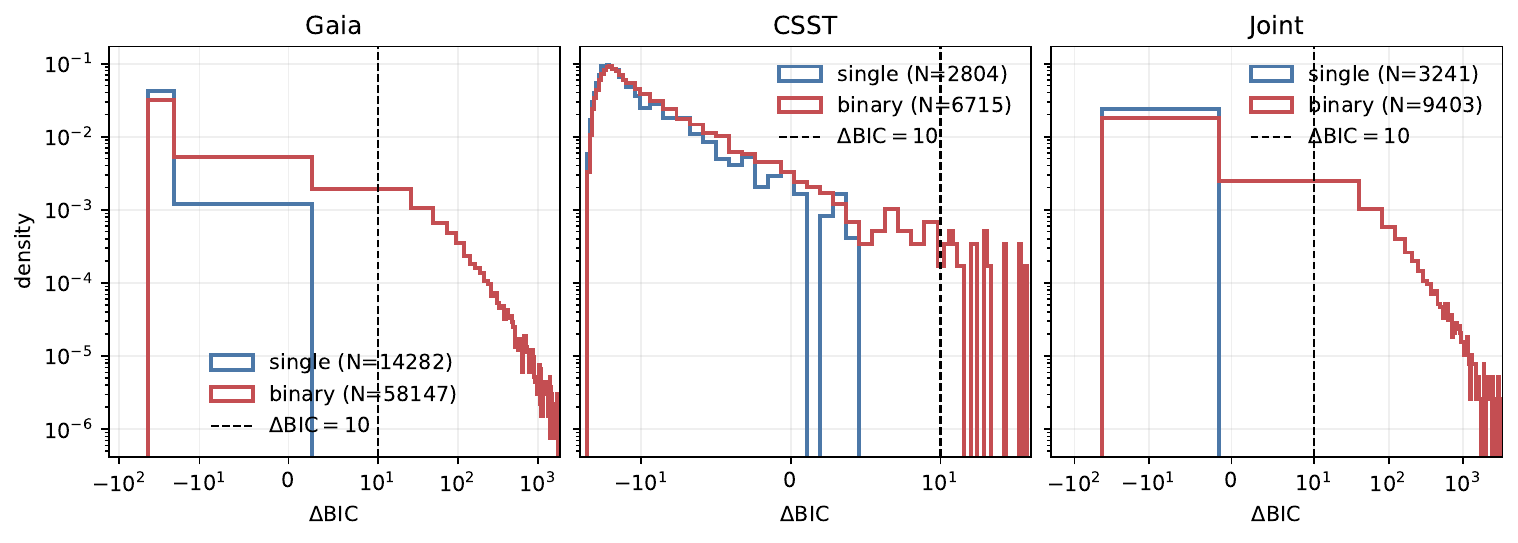}
        \caption{Histogram of the $\Delta{\rm BIC}$ values for 12p-fitted sources. The panels show the \emph{Gaia}, CSST, and joint solutions from left to right. Red and blue histograms represent binaries and single stars in the candidate set, respectively. The vertical dashed line marks $\Delta{\rm BIC}=10$, which defines the 12p model preference criterion. Only the range between the 0.1st and 99.9th percentiles of $\Delta{\rm BIC}$ is shown.}
        \label{fig:delta_bic_hist}
    \end{figure*}

    Fig.~\ref{fig:delta_bic_hist} illustrates the 12p model preference criterion. In the \emph{Gaia} and joint panels, binaries develop a positive-$\Delta{\rm BIC}$ tail, while all 12p-fitted single stars remain below the $\Delta{\rm BIC}=10$ threshold. The CSST panel shows much stronger overlap between the binary and single-star $\Delta{\rm BIC}$ distributions, and only a small number of sources pass the threshold. This explains the small number of CSST solutions in the 12p model preference tier, whereas the joint fit yields many more such solutions.

\subsection{Properties of the selected 12p solution subsets}
\begin{figure*}[!h]
        \centering
        \includegraphics[width=0.85\textwidth]{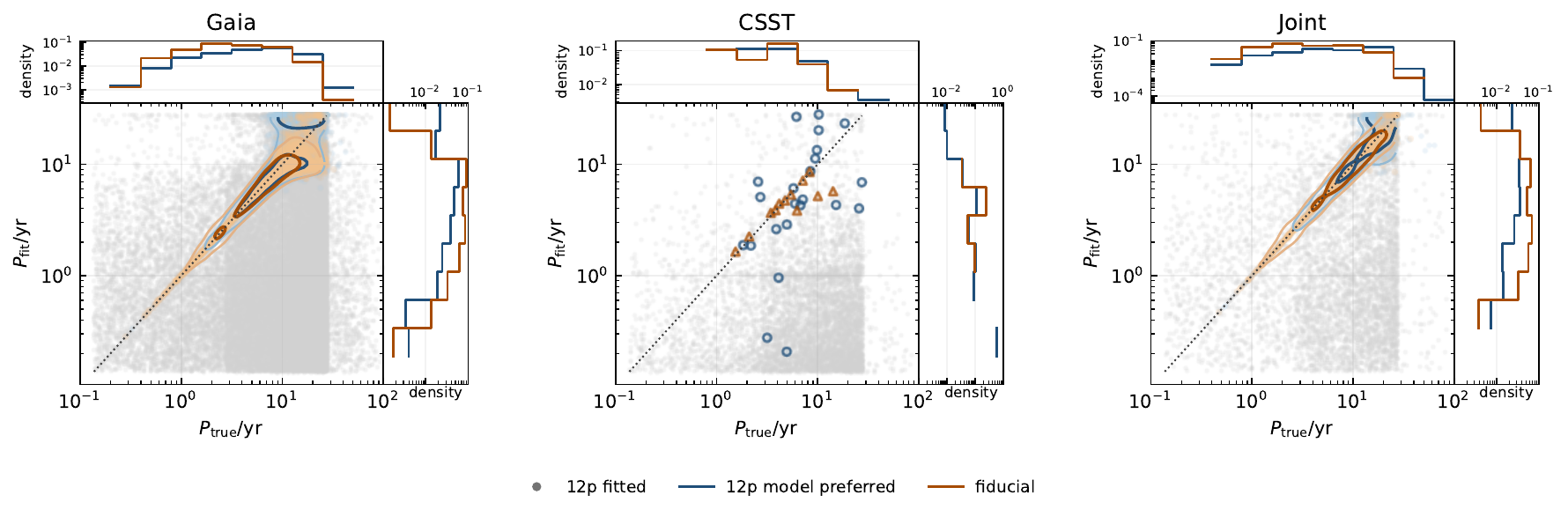}
        \caption{Fitted orbital period $P_{\rm fit}$ versus true orbital period $P_{\rm true}$ for 12p-fitted true binaries in the \emph{Gaia}, CSST, and joint scenarios. Light grey points show the 12p fits satisfying neither selection tier. In the \emph{Gaia} and joint panels, blue and orange contours show solutions that satisfy only the 12p model preference criterion and solutions in the fiducial tier, respectively. The thin outer and thick inner contours enclose approximately $90\%$ and $50\%$ of the probability mass of each selected group, respectively. The two selected CSST groups are plotted individually as blue circles and orange triangles because they contain too few solutions for stable density contours. The dotted diagonal marks perfect period recovery, $P_{\rm fit}=P_{\rm true}$. The top and right histograms show the separately normalized marginal densities of $P_{\rm true}$ and $P_{\rm fit}$ for the two selected groups.}
        \label{fig:porb_solved_vs_porb_true}
    \end{figure*}

    Fig.~\ref{fig:porb_solved_vs_porb_true} compares the fitted and true orbital periods of 12p-fitted true binaries. The light grey background points show the 12p fits satisfying neither selection tier. Their broad distribution indicates that obtaining a converged 12p fit does not guarantee accurate recovery of the orbital period. In the \emph{Gaia} and joint panels, the blue contours represent solutions that satisfy only the 12p model preference criterion, and the orange contours represent fiducial solutions. Both selected groups are more concentrated around $P_{\rm fit}=P_{\rm true}$ than the light-grey background points, and the fiducial solutions are more tightly concentrated than the solutions satisfying only the 12p model preference criterion. The two selected CSST groups contain too few solutions for stable density contours, so their sources are plotted individually.

\begin{figure*}[!h]
        \centering
        \includegraphics[width=0.85\textwidth]{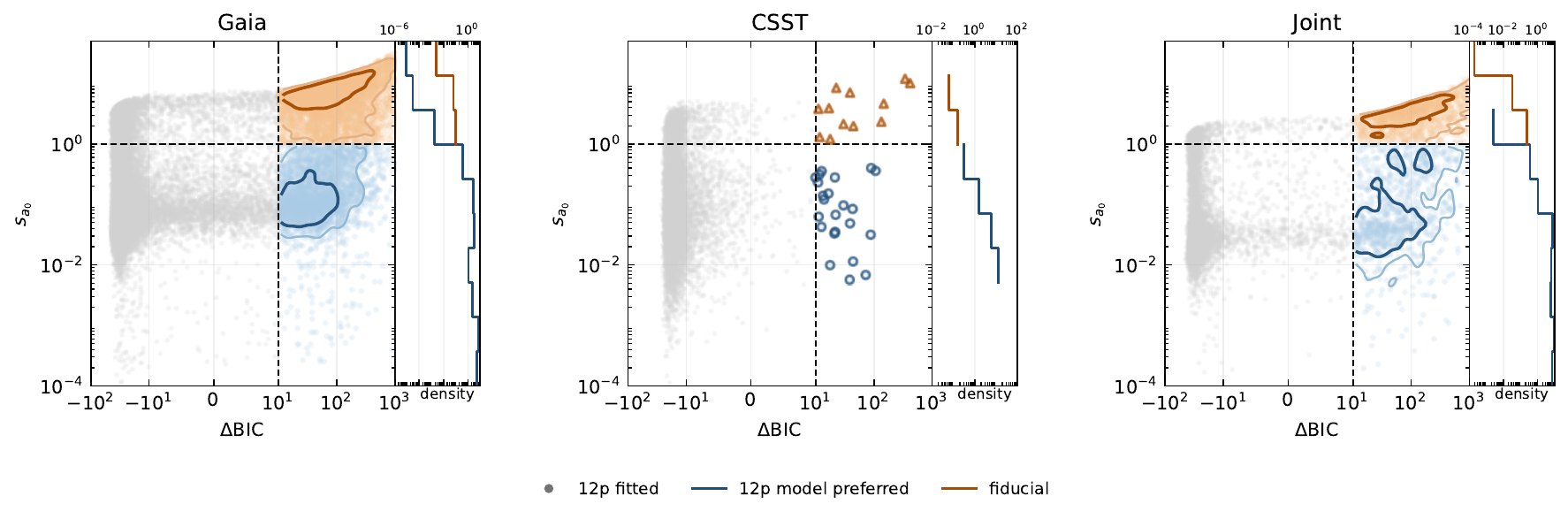}
        \caption{Photocentric semi-major-axis significance $s_{a_0}=a_0/\sigma_{a_0}$ versus $\Delta{\rm BIC}$ for 12p-fitted sources in the \emph{Gaia}, CSST, and joint scenarios. The color code, contour levels, and markers are the same as in Fig.~\ref{fig:porb_solved_vs_porb_true}. The vertical and horizontal black dashed lines mark $\Delta{\rm BIC}=10$ and $s_{a_0}=1$, respectively. The right histograms show the separately normalized marginal densities of $s_{a_0}$ for the two selected groups.}
        \label{fig:delta_bic_a0sig}
    \end{figure*}

    Fig.~\ref{fig:delta_bic_a0sig} shows the 12p-fitted sources in the $\Delta{\rm BIC}$--$s_{a_0}$ plane, where $s_{a_0}=a_0/\sigma_{a_0}$. The uncertainty $\sigma_{a_0}$ is returned by the 12p orbital fit and therefore depends on the number of observations, observational distribution and precision. A large $\Delta{\rm BIC}$ selects fits for which the 12p model is preferred over the 5p model, but it does not establish accurate recovery of the orbital parameters. In the \emph{Gaia} and joint panels, many orbital solutions satisfying only the 12p model preference criterion have $s_{a_0}\leq1$. The fiducial tier requires $s_{a_0}>1$ and therefore excludes these fits. The fiducial subset accounts for a larger fraction of the 12p-fitted true binaries in the joint scenario than in the \emph{Gaia} scenario. Fig.~\ref{fig:porb_solved_vs_porb_true} shows that the fiducial joint subset is more concentrated around $P_{\rm fit}=P_{\rm true}$ in this simulation.

\begin{figure*}[!h]
        \centering
        \includegraphics[width=0.7\textwidth]{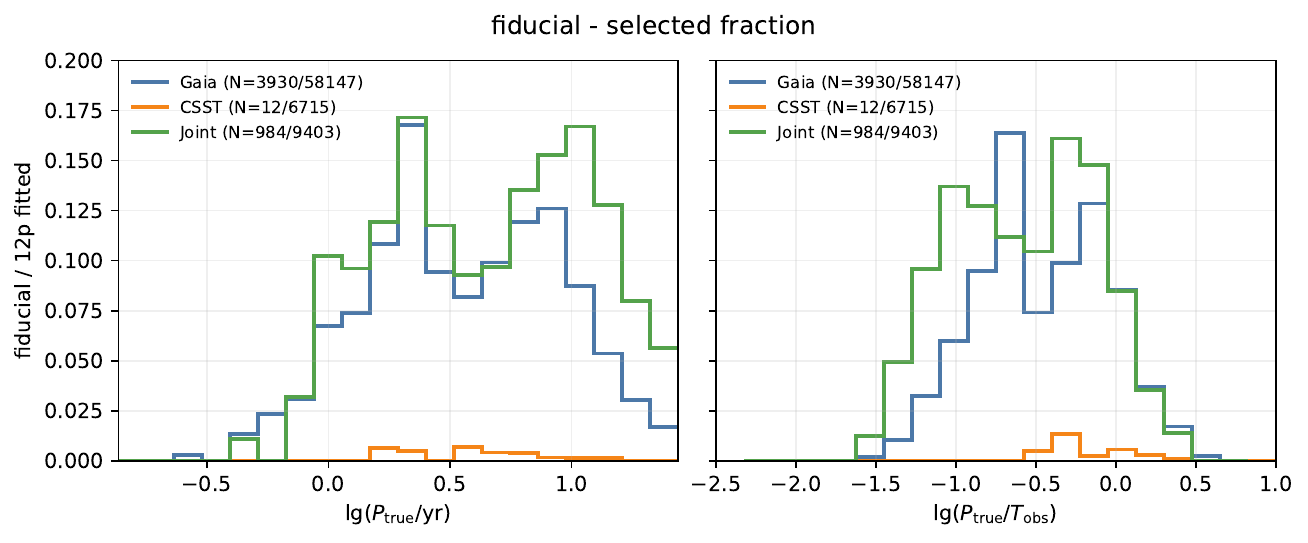}
        \caption{Fiducial fractions among 12p-fitted true binaries versus the true orbital period $P_{\rm true}$ (\emph{left}) and the ratio of the true orbital period to the observational time baseline of each source $P_{\rm true}/T_{\rm obs}$ (\emph{right}), for the \emph{Gaia}, CSST, and joint solutions. In each bin, the numerator is the number of true binaries satisfying the fiducial criteria, and the denominator is the number of 12p-fitted true binaries in the same scenario and bin. The legend gives the total fiducial/12p-fitted counts for each scenario.}
        \label{fig:fiducial_frac_period_coverage}
    \end{figure*}

    Fig.~\ref{fig:fiducial_frac_period_coverage} shows the fraction of 12p-fitted true binaries satisfying the fiducial criteria in bins of $P_{\rm true}$ and $P_{\rm true}/T_{\rm obs}$, where $T_{\rm obs}$ is the time baseline for each source. At a fixed $P_{\rm true}$, a shorter $T_{\rm obs}$ yields a larger $P_{\rm true}/T_{\rm obs}$ and a smaller observed orbital fraction. This ratio therefore measures the temporal coverage available for detecting the orbit and recovering its period.

    In the left panel, the joint fiducial fraction is higher than the \emph{Gaia} fraction toward the long-period end. At $P_{\rm true}<1\,{\rm yr}$, the fiducial fractions decrease rapidly in the \emph{Gaia} and joint scenarios. At fixed component masses, flux ratio, and parallax, shorter-period binaries have smaller photocentric semi-major axes and are therefore more difficult to detect. \emph{Gaia} scan-angle-dependent signals can additionally introduce period aliases in this regime \citep{2023A&A...674A..25H}. This behavior is consistent with Fig.~\ref{fig:porb_solved_vs_porb_true}, where fiducial solutions and solutions satisfying only the 12p model preference criterion are both rare at $P_{\rm true}<1\,{\rm yr}$.
    In the right panel, the fiducial fractions for both the \emph{Gaia} and joint solutions decrease once the true orbital period exceeds the observational time baseline. This pattern is consistent with temporal coverage contributing to the higher joint fiducial fraction at long periods. The CSST fiducial fraction remains low under the current survey schedule.

    For binaries with $P_{\rm true}>15\,{\rm yr}$, $482$ of $20\,330$ successfully fitted binaries in the \emph{Gaia} scenario and $202$ of $2\,978$ in the joint scenario satisfy the fiducial criteria. These counts correspond to $2.37\%$ and $6.78\%$, respectively. Joint fitting therefore raises the long-period fiducial fraction from $2.37\%$ to $6.78\%$, a factor of $2.86$, in the scenario-specific fitted samples.

    Much of this gain comes from the longer time baseline. Among the scenario-specific long-period fitted samples, the median $T_{\rm obs}$ increases from $10.18\,{\rm yr}$ for \emph{Gaia} to $17.27\,{\rm yr}$ for the joint scenario, while the median $P_{\rm true}/T_{\rm obs}$ decreases from $2.07$ to $1.20$. The joint data therefore cover a larger fraction of these long-period orbits. Among all long-period fiducial solutions in each scenario, the median absolute relative period error, $|P_{\rm fit}-P_{\rm true}|/P_{\rm true}$, is $40.7\%$ for \emph{Gaia} and $11.1\%$ for the joint fit.

    The fiducial tier requires $\Delta{\rm BIC}\geq10$, $F_{2,12{\rm p}}<25$, and $s_{a_0}>1$, and excludes solutions that converge at the upper eccentricity boundary $e=0.99$. The CSST-only 12p results remain limited by the current survey schedule. Relative to the \emph{Gaia}-only solution, the joint solution increases the 12p model preference and fiducial fractions from $12.57\%$ and $6.76\%$ to $21.38\%$ and $10.46\%$ in the scenario-specific samples, respectively. For the long-period fiducial subsets, the joint solutions have a smaller median absolute relative period error than the \emph{Gaia} solutions.

\section{Effect of the CSST cadence on 12p orbit fitting\label{sect:csst_schedule_opt}}
    CSST will reserve approximately 10\% of its observing time for public proposals during the 10-year initial mission \citep{2026SCPMA..6939501C}. One use of this time is targeted follow-up of sources that are already known or strongly suspected to be binaries. Such observations can improve the binary orbital coverage by scheduling observations at more regular epochs. The number of 12p fits satisfying the selection criteria depends on observational precision, time baseline, and cadence regularity, motivating a test of how a more regular cadence changes this yield.

    To isolate the effect of the observing schedule from the candidate-selection procedure, we randomly selected $500\,000$ binaries from the simulated catalog and did not include single stars in this test. This setup represents a sample of identified binary candidates with no orbital solutions, such as those selected from PMa or overluminosity, for which the main question is whether the epoch data can yield an orbital solution rather than whether the sources can first be distinguished from single stars. For each source and observing schedule, we generated mock observations and fitted both the 5p and 12p models. We then applied the same selection criteria as in Sect.~\ref{subsect:orbfit_selection_tiers}.

    We compared the current CSST survey schedule with three regular schedules. The current schedule is the CSST schedule described in Sect.~\ref{subsect:csst_obs_schedule}. In the regular schedules, we kept the same CSST sky coverage but replaced the schedule with uniformly sampled epochs over the 10-year CSST baseline. We tested three schedules with $25$, $50$, and $100$ CSST observations, denoted U25, U50, and U100, respectively. The U25 schedule is the most direct comparison with the current schedule, because the current schedule provides $27.6$ CSST observations on average. Thus, U25 is approximately a redistribution of the current observations with a similar observational budget. For the joint solutions, each CSST schedule was combined with the same \emph{Gaia} scanning law.

    \begin{figure*}[!h]
        \centering
        \includegraphics[width=0.8\textwidth]{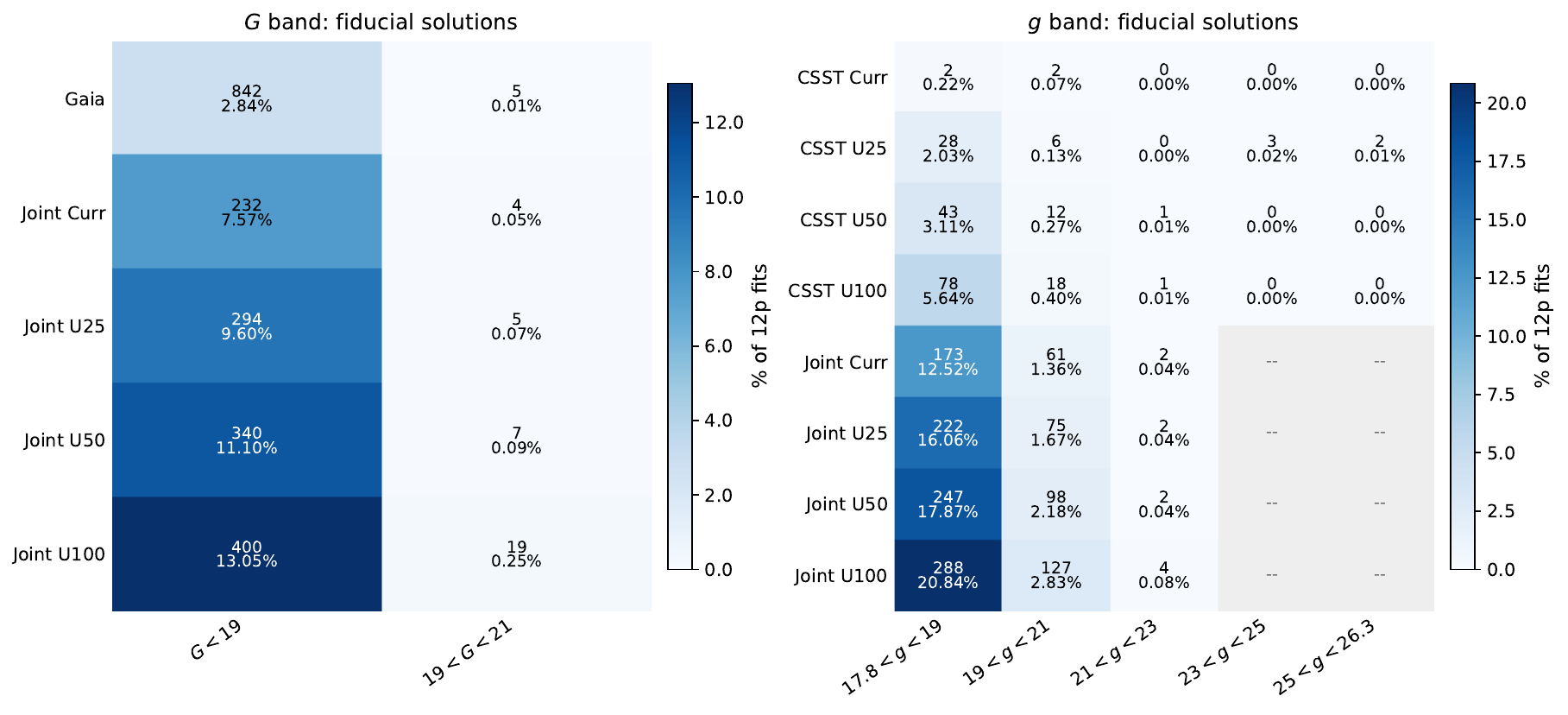}
        \caption{Fiducial 12p solution yields for known mock binaries, binned by Gaia $G$ magnitude on the \emph{left} and CSST $g$ magnitude on the \emph{right}. The joint rows use the same \emph{Gaia} scanning law combined with different CSST schedules. Curr, U25, U50, and U100 denote the current schedule and uniform 25-, 50-, and 100-epoch CSST schedules, respectively. Each cell gives the number of fiducial solutions and their percentage among all 12p-fitted binaries in that row and magnitude bin. The joint solution can still contain sources at $g>21$, because some sources satisfying $G<21$ are included in the \emph{Gaia} mock observations but have $g>21$ (see Fig.~\ref{fig:csst_gaia_obs_err}).}
        \label{fig:csst_optimize_schedule}
    \end{figure*}

    Fig.~\ref{fig:csst_optimize_schedule} compares the \emph{Gaia} and joint solutions in the $G$ band. At $G<19$, the \emph{Gaia} fit yields $842$ fiducial solutions. The joint fit with the current CSST schedule yields $232$, because it is restricted to sources with both \emph{Gaia} and CSST observations. Replacing the current schedule with the U25 schedule increases the joint count from $232$ to $294$ fiducial solutions. The denser U50 and U100 schedules further increase the yield to $340$ and $400$, respectively. At $19<G<21$, the joint fiducial count increases from $4$ for the current schedule to $5$, $7$, and $19$ for U25, U50, and U100.

    For CSST, the fiducial solutions are concentrated at the bright end. Within $17.8<g<19$, the current schedule yields $2$ fiducial solutions. U25 increases this to $28$ solutions, while U50 and U100 further increase the yield to $43$ and $78$, respectively. At $19<g<21$, the corresponding counts for the current, U25, U50, and U100 schedules are $2$, $6$, $12$, and $18$. CSST fiducial solutions remain rare at $g>21$ for all schedules.

    Comparing within the CSST $g$-band, the joint fit yields more fiducial solutions than the CSST-only fit. At $17.8<g<19$, the joint fit with the current CSST schedule yields $173$ fiducial solutions. U25 increases this number to $222$, while U50 and U100 further increase it to $247$ and $288$. At $19<g<21$, the joint fiducial solution count increases from $61$ for the current schedule to $75$, $98$, and $127$ for U25, U50, and U100. At $g>21$, even the joint fit yields very few fiducial solutions. The small number of joint solutions in this regime is expected because some sources with $G<21$ are included in the \emph{Gaia} mock observations but have $g>21$ (see Fig.~\ref{fig:csst_gaia_obs_err}).

    These results show that a more regular CSST cadence can increase the number of fits satisfying the fiducial criteria. U25 increases the yield while using nearly the same number of CSST observations as the current schedule. Additional observations in U50 and U100 continue to increase the yield, but the increase is concentrated in relatively bright sources with $g\lesssim 21$. The simulated schedules suggest that proposed CSST observations, or adjustments to the survey schedule, could be used to reobserve bright binary candidates more regularly and provide better temporal sampling for subsequent orbit fitting.

\section{Conclusion and discussion\label{sect:concl_and_disc}}
    We have presented an assessment of the detectability of unresolved binaries using mock observations for CSST, \emph{Gaia}, and their joint epoch astrometry. We sampled the input catalog from MWMSC, which represents the underlying Galactic population, evolved binaries with \texttt{COSMIC}, calculated stellar parameters and \emph{Gaia} photometry with \texttt{isochrones}, and constructed the CSST photometry using the adopted response curves, blackbody colors, and extinction in the CSST $g$ band.

    The hierarchical candidate classifier identifies candidates for the computationally expensive 12p orbital fit. The classifier uses 5p diagnostics, PMa features, and auxiliary features based on observational cadence. On the independent test set, it selected $72\,554$ candidates among sources eligible for at least one stage, including $58\,199$ true binaries and $14\,355$ single stars. The resulting precision is $0.802$, and the recall among eligible binaries is $0.181$ (Table~\ref{table:candidate_selection_summary}). Candidate selection mainly depends on the photocentric semi-major axis and the orbital period, with only a weak dependence on eccentricity. This trend is consistent with larger photocentric orbits producing stronger 5p residuals, PMa, and related astrometric diagnostics.

    In the candidate set, the 12p model preference criterion $\Delta{\rm BIC}\geq10$ excludes all single stars with formally converged but spurious 12p fits in this simulation. Only a subset of the 12p fits satisfies the fiducial criteria. The fiducial tier requires $\Delta{\rm BIC}\geq10$, $F_{2,12{\rm p}}<25$, $s_{a_0}>1$, and no convergence at the upper eccentricity boundary $e=0.99$. The \emph{Gaia} fit yields $3\,930$ fiducial solutions, while the joint fit yields $984$ from its smaller fitted sample (Table~\ref{tab:orbfit_results}). Relative to the \emph{Gaia}-only fit, the joint fit increases the fraction of fitted binaries satisfying the 12p model preference criterion from $12.57\%$ to $21.38\%$ and the fiducial fraction from $6.76\%$ to $10.46\%$ in the scenario-specific samples.

    Under the current survey schedule, few CSST 12p fits satisfy the fiducial criteria. In the candidate sample, CSST yields only $35$ solutions in the 12p model preference tier and $12$ in the fiducial tier. The current schedule often provides uneven temporal sampling and therefore incomplete orbital-phase coverage. This uneven sampling is associated with the low fiducial yield, despite CSST having single-epoch precision comparable to that of a \emph{Gaia} FoV transit.

    A separate single-stage classifier using only the CSST 5p diagnostics and CSST observational-sampling features was trained, tuned, and applied to the $124\,491$ CSST-eligible test sources, including $41\,302$ true binaries. The CSST classifier selected $1\,736$ candidates, comprising $1\,379$ true binaries and $357$ single stars, with a precision of $0.794$ and a recall of $0.0334$. This low recall and the small CSST 12p yield are consistent with the limited orbital-phase coverage produced by the current uneven survey schedule.

    The idealized cadence experiments motivate investigating operationally feasible ways to obtain more regular follow-up observations of bright binary candidates. Redistributing a comparable number of CSST observations ($27.6$ on average) into a uniformly sampled 25-observation schedule (U25) increases both CSST and joint yields. U50 and U100 further increase the number of fits satisfying the fiducial criteria. The gain is concentrated at the bright end, primarily for sources with $g\lesssim 21$. This result suggests that CSST public proposals or future schedule adjustments could target bright binary candidates with a more regular cadence for subsequent orbit fitting.

    Several limitations remain.
    First, the adopted CSST single-epoch observation errors are predictions with an added attitude-noise term. The in-orbit single-epoch observation errors can only be measured after CSST has operated for a sufficient period.
    Second, the U25, U50, and U100 cases are idealized cadence experiments rather than operational schedule designs. They do not account for operational constraints such as target visibility windows, spacecraft attitude constraints, calibration requirements, or competition with other science proposals.
    Third, the joint 12p fit uses a single photocentric semi-major axis for both missions, whereas $a_{0,\rm G}$ and $a_{0,\rm C}$ are passband-dependent. At the population level, the mismatch associated with this approximation is smaller than the typical reported formal $a_0$ uncertainty for the fiducial joint fits in this simulation.
    Fourth, the simulation does not include photocenter variability induced by stellar activity, crowding, higher-order multiplicity, window truncation, or catalog calibration systematics.
    Finally, the population-level assessment in this work focuses on unresolved binaries. The resolved-binary toy experiment in Appx.~\ref{appx:resolved_binary} shows a sharp decrease in the relative-error distribution between $a=70\,{\rm mas}$ and $a=80\,{\rm mas}$. At $a=80\,{\rm mas}$, all $100$ injected binaries return 13p (12p + mass ratio) solutions and $99$ satisfy the additional quality criteria. Across $90\leq a\leq500\,{\rm mas}$, all $4\,200$ injected binaries return solutions and $4\,199$ satisfy the criteria under the adopted 48-epoch sampling.

    The main conclusion is that, in our simulation, joint CSST and \emph{Gaia} epoch astrometry raises the fractions of fitted unresolved binaries satisfying the 12p model preference and fiducial criteria relative to the \emph{Gaia}-only solution. For fitted true binaries with $P_{\rm true}>15\,{\rm yr}$, joint fitting increases the fiducial fraction from $2.37\%$ to $6.78\%$ in the scenario-specific samples. The longer joint baseline covers a larger fraction of these orbits, and the joint long-period fiducial sample has a smaller median relative period error. Under the present wide-field schedule, CSST alone yields few fiducial 12p fits for unresolved binaries. Binary candidates can be selected using 5p diagnostics and astrometric anomaly features such as PMa. More regular CSST observations of binary candidates can then improve temporal coverage for subsequent 12p orbit fitting and application of the selection criteria. Within the overlapping magnitude range, joint CSST and \emph{Gaia} epoch astrometry can also extend the search to binaries that are fainter than most sources in the \emph{Gaia} non-single-star catalog while retaining sensitivity to longer periods. Such an expanded sample can provide additional dynamical mass measurements of compact companions and help test whether the observed neutron-star--black-hole mass gap traces the underlying compact-object population or observational selection.

\begin{acknowledgements}
    We sincerely thank the expert referee, Fr{\'e}d{\'e}ric Arenou, for his careful, detailed, and insightful comments, which substantially improved the manuscript. This work was supported by the National Key R\&D Program of China (Grant No. 2023YFA1607901), the Youth Innovation Promotion Association CAS, the grants from the Natural Science Foundation of Shanghai through grant 21ZR1474100, National Natural Science Foundation of China (NSFC) through grants 12173069, and the Talent Plan of Shanghai Branch, Chinese Academy of Sciences with No.CASSHB-QNPD-2023-016. We acknowledge the science research grants from the China Manned Space Project with NO.CMS-CSST-2021-A12 and NO.CMS-CSST-2021-B10.

    We thank Yang Chen and Yifan Xuan for their assistance in accessing the Milky Way stellar mock catalog. We are grateful to the editor, Thierry Forveille, for his constructive comment that helped to improve the manuscript. This work made use of data from the \emph{Gaia} Observation Forecast Tool (\href{https://gaia.esac.esa.int/gost/}{https://gaia.esac.esa.int/gost/}) and the High Performance Computing Resource in the Core Facility for Advanced Research Computing at Shanghai Astronomical Observatory.

    We thank the developers and contributors of the following software packages and tools: \texttt{astropy} \citep{2022ApJ...935..167A}, \texttt{COSMIC} \citep{2020ApJ...898...71B}, \texttt{healpy} \citep{2019JOSS....4.1298Z}, \texttt{isochrones} \citep{2015ascl.soft03010M}, \texttt{jplephem} \citep{2011ascl.soft12014R}, \texttt{matplotlib} \citep{2007CSE.....9...90H}, \texttt{numpy} \citep{harris2020array}, \texttt{optuna} \citep{optuna}, \texttt{pandas} \citep{mckinney2010data}, \texttt{TOPCAT} \citep{2005ASPC..347...29T}, \texttt{scikit-learn} \citep{2011JMLR...12.2825P}, and \texttt{scipy} \citep{2020NatMe..17..261V}.
\end{acknowledgements}

\bibliographystyle{aa} 
\bibliography{ref} 

@ARTICLE{2026SCPMA..6939501C,
       author = {{CSST Collaboration} and {Gong}, Yan and {Miao}, Haitao and {Zhan}, Hu and {Li}, Zhao-Yu and {Shangguan}, Jinyi and {Li}, Haining and {Liu}, Chao and {Chen}, Xuefei and {Yuan}, Haibo and {Zhou}, Jilin and {Liu}, Hui-Gen and {Yu}, Cong and {Ji}, Jianghui and {Qi}, Zhaoxiang and {Liu}, Jiacheng and {Dai}, Zigao and {Wang}, Xiaofeng and {Zheng}, Zhenya and {Hao}, Lei and {Dou}, Jiangpei and {Ao}, Yiping and {Lin}, Zhenhui and {Zhang}, Kun and {Wang}, Wei and {Sun}, Guotong and {Li}, Ran and {Li}, Guoliang and {Xu}, Youhua and {Li}, Xinfeng and {Li}, Shengyang and {Wu}, Peng and {Zhang}, Jiuxing and {Wang}, Bo and {Bai}, Jinming and {Cai}, Yi-Fu and {Cai}, Zheng and {Cao}, Jie and {Chan}, Kwan Chuen and {Chang}, Jin and {Chen}, Xiaodian and {Chen}, Xuelei and {Chen}, Yuqin and {Chen}, Yun and {Cui}, Wei and {Dong}, Subo and {Du}, Pu and {Duan}, Wenying and {Fan}, Junhui and {Fan}, LuLu and {Fan}, Zhou and {Fan}, Zuhui and {Fang}, Taotao and {Fu}, Jianning and {Fu}, Liping and {Fu}, Zhensen and {Gao}, Jian and {Gu}, Shenghong and {Gu}, Yidong and {Guo}, Qi and {Han}, Zhanwen and {Hu}, Bin and {Huang}, Zhiqi and {Ho}, Luis C. and {Jiang}, Linhua and {Jiang}, Ning and {Jing}, Yipeng and {Kang}, Xi and {Kong}, Xu and {Li}, Cheng and {Li}, Chengyuan and {Li}, Di and {Li}, Jing and {Li}, Nan and {Li}, Yang A. and {Liao}, Shilong and {Lin}, Weipeng and {Liu}, Fengshan and {Liu}, Jifeng and {Liu}, Xiangkun and {Liu}, Zhuokai and {Mao}, Ruiqing and {Mao}, Shude and {Meng}, Xianmin and {Pang}, Xiaoying and {Peng}, Xiyan and {Peng}, Yingjie and {Shan}, Huanyuan and {Shen}, Juntai and {Shen}, Shiyin and {Shen}, Zhiqiang and {Shi}, Sheng-Cai and {Shi}, Yong and {Tan}, Siyuan and {Tian}, Hao and {Wang}, Jianmin and {Wang}, Jun-Xian and {Wang}, Xin and {Wang}, Yuting and {Wu}, Hong and {Wu}, Jingwen and {Wu}, Xuebing and {Xu}, Chun and {Xue}, Xiang-Xiang and {Xue}, Yongquan and {Yang}, Ji and {Yang}, Xiaohu and {Yao}, Qijun and {Yuan}, Fangting and {Yuan}, Zhen and {Zhang}, Jun and {Zhang}, Pengjie and {Zhang}, Tianmeng and {Zhang}, Wei and {Zhang}, Xin and {Zhao}, Gang and {Zhao}, Gongbo and {Zhong}, Hongen and {Zhong}, Jing and {Zhou}, Liyong and {Zhu}, Wei and {Zu}, Ying},
        title = "{Introduction to the Chinese Space Station Survey Telescope (CSST)}",
      journal = {Science China Physics, Mechanics, and Astronomy},
         year = 2026,
        month = jan,
       volume = {69},
       number = {3},
          eid = {239501},
        pages = {239501},
          doi = {10.1007/s11433-025-2809-0},
archivePrefix = {arXiv},
       eprint = {2507.04618},
 primaryClass = {astro-ph.IM},
       adsurl = {https://ui.adsabs.harvard.edu/abs/2026SCPMA..6939501C}
}

@ARTICLE{2009ApJS..182..205W,
       author = {{Wright}, J.~T. and {Howard}, A.~W.},
        title = "{Efficient Fitting of Multiplanet Keplerian Models to Radial Velocity and Astrometry Data}",
      journal = {\apjs},
         year = 2009,
        month = may,
       volume = {182},
       number = {1},
        pages = {205-215},
          doi = {10.1088/0067-0049/182/1/205},
archivePrefix = {arXiv},
       eprint = {0904.3725},
 primaryClass = {astro-ph.SR},
       adsurl = {https://ui.adsabs.harvard.edu/abs/2009ApJS..182..205W}
}

@ARTICLE{2012Sci...337..444S,
       author = {{Sana}, H. and {de Mink}, S.~E. and {de Koter}, A. and {Langer}, N. and {Evans}, C.~J. and {Gieles}, M. and {Gosset}, E. and {Izzard}, R.~G. and {Le Bouquin}, J. -B. and {Schneider}, F.~R.~N.},
        title = "{Binary Interaction Dominates the Evolution of Massive Stars}",
      journal = {Science},
         year = 2012,
        month = jul,
       volume = {337},
       number = {6093},
        pages = {444},
          doi = {10.1126/science.1223344},
archivePrefix = {arXiv},
       eprint = {1207.6397},
 primaryClass = {astro-ph.SR},
       adsurl = {https://ui.adsabs.harvard.edu/abs/2012Sci...337..444S}
}

@ARTICLE{2013ARA&A..51..269D,
       author = {{Duch{\^e}ne}, Gaspard and {Kraus}, Adam},
        title = "{Stellar Multiplicity}",
      journal = {\araa},
         year = 2013,
        month = aug,
       volume = {51},
       number = {1},
        pages = {269-310},
          doi = {10.1146/annurev-astro-081710-102602},
archivePrefix = {arXiv},
       eprint = {1303.3028},
 primaryClass = {astro-ph.SR},
       adsurl = {https://ui.adsabs.harvard.edu/abs/2013ARA&A..51..269D}
}

@ARTICLE{2012A&A...538A..78L,
       author = {{Lindegren}, L. and {Lammers}, U. and {Hobbs}, D. and {O'Mullane}, W. and {Bastian}, U. and {Hern{\'a}ndez}, J.},
        title = "{The astrometric core solution for the Gaia mission. Overview of models, algorithms, and software implementation}",
      journal = {\aap},
         year = 2012,
        month = feb,
       volume = {538},
          eid = {A78},
        pages = {A78},
          doi = {10.1051/0004-6361/201117905},
archivePrefix = {arXiv},
       eprint = {1112.4139},
 primaryClass = {astro-ph.IM},
       adsurl = {https://ui.adsabs.harvard.edu/abs/2012A&A...538A..78L}
}

@ARTICLE{2016arXiv161205560C,
       author = {{Chambers}, K.~C. and {Magnier}, E.~A. and {Metcalfe}, N. and {Flewelling}, H.~A. and {Huber}, M.~E. and {Waters}, C.~Z. and {Denneau}, L. and {Draper}, P.~W. and {Farrow}, D. and {Finkbeiner}, D.~P. and {Holmberg}, C. and {Koppenhoefer}, J. and {Price}, P.~A. and {Rest}, A. and {Saglia}, R.~P. and {Schlafly}, E.~F. and {Smartt}, S.~J. and {Sweeney}, W. and {Wainscoat}, R.~J. and {Burgett}, W.~S. and {Chastel}, S. and {Grav}, T. and {Heasley}, J.~N. and {Hodapp}, K.~W. and {Jedicke}, R. and {Kaiser}, N. and {Kudritzki}, R.-P. and {Luppino}, G.~A. and {Lupton}, R.~H. and {Monet}, D.~G. and {Morgan}, J.~S. and {Onaka}, P.~M. and {Shiao}, B. and {Stubbs}, C.~W. and {Tonry}, J.~L. and {White}, R. and {Ba{\~n}ados}, E. and {Bell}, E.~F. and {Bender}, R. and {Bernard}, E.~J. and {Boegner}, M. and {Boffi}, F. and {Botticella}, M.~T. and {Calamida}, A. and {Casertano}, S. and {Chen}, W.-P. and {Chen}, X. and {Cole}, S. and {Deacon}, N. and {Frenk}, C. and {Fitzsimmons}, A. and {Gezari}, S. and {Gibbs}, V. and {Goessl}, C. and {Goggia}, T. and {Gourgue}, R. and {Goldman}, B. and {Grant}, P. and {Grebel}, E.~K. and {Hambly}, N.~C. and {Hasinger}, G. and {Heavens}, A.~F. and {Heckman}, T.~M. and {Henderson}, R. and {Henning}, T. and {Holman}, M. and {Hopp}, U. and {Ip}, W.-H. and {Isani}, S. and {Jackson}, M. and {Keyes}, C.~D. and {Koekemoer}, A.~M. and {Kotak}, R. and {Le}, D. and {Liska}, D. and {Long}, K.~S. and {Lucey}, J.~R. and {Liu}, M. and {Martin}, N.~F. and {Masci}, G. and {McLean}, B. and {Mindel}, E. and {Misra}, P. and {Morganson}, E. and {Murphy}, D.~N.~A. and {Obaika}, A. and {Narayan}, G. and {Nieto-Santisteban}, M.~A. and {Norberg}, P. and {Peacock}, J.~A. and {Pier}, E.~A. and {Postman}, M. and {Primak}, N. and {Rae}, C. and {Rai}, A. and {Riess}, A. and {Riffeser}, A. and {Rix}, H.~W. and {R{\"o}ser}, S. and {Russel}, R. and {Rutz}, L. and {Schilbach}, E. and {Schultz}, A.~S.~B. and {Scolnic}, D. and {Strolger}, L. and {Szalay}, A. and {Seitz}, S. and {Small}, E. and {Smith}, K.~W. and {Soderblom}, D.~R. and {Taylor}, P. and {Thomson}, R. and {Taylor}, A.~N. and {Thakar}, A.~R. and {Thiel}, J. and {Thilker}, D. and {Unger}, D. and {Urata}, Y. and {Valenti}, J. and {Wagner}, J. and {Walder}, T. and {Walter}, F. and {Watters}, S.~P. and {Werner}, S. and {Wood-Vasey}, W.~M. and {Wyse}, R.},
        title = "{The Pan-STARRS1 Surveys}",
      journal = {arXiv e-prints},
         year = 2016,
        month = dec,
          eid = {arXiv:1612.05560},
        pages = {arXiv:1612.05560},
          doi = {10.48550/arXiv.1612.05560},
archivePrefix = {arXiv},
       eprint = {1612.05560},
 primaryClass = {astro-ph.IM},
       adsurl = {https://ui.adsabs.harvard.edu/abs/2016arXiv161205560C}
}

@ARTICLE{2025arXiv250510574O,
       author = {{Roman Observations Time Allocation Committee} and {Core Community Survey Definition Committees}},
        title = "{Roman Observations Time Allocation Committee: Final Report and Recommendations}",
      journal = {arXiv e-prints},
         year = 2025,
        month = may,
          eid = {arXiv:2505.10574},
        pages = {arXiv:2505.10574},
          doi = {10.48550/arXiv.2505.10574},
archivePrefix = {arXiv},
       eprint = {2505.10574},
 primaryClass = {astro-ph.IM},
       adsurl = {https://ui.adsabs.harvard.edu/abs/2025arXiv250510574O}
}

@ARTICLE{2022A&A...657A...7K,
       author = {{Kervella}, Pierre and {Arenou}, Fr{\'e}d{\'e}ric and {Th{\'e}venin}, Fr{\'e}d{\'e}ric},
        title = "{Stellar and substellar companions from Gaia EDR3. Proper-motion anomaly and resolved common proper-motion pairs}",
      journal = {\aap},
         year = 2022,
        month = jan,
       volume = {657},
          eid = {A7},
        pages = {A7},
          doi = {10.1051/0004-6361/202142146},
archivePrefix = {arXiv},
       eprint = {2109.10912},
 primaryClass = {astro-ph.SR},
       adsurl = {https://ui.adsabs.harvard.edu/abs/2022A&A...657A...7K}
}

@ARTICLE{2019MNRAS.482.4570G,
       author = {{Gentile Fusillo}, Nicola Pietro and {Tremblay}, Pier-Emmanuel and {G{\"a}nsicke}, Boris T. and {Manser}, Christopher J. and {Cunningham}, Tim and {Cukanovaite}, Elena and {Hollands}, Mark and {Marsh}, Thomas and {Raddi}, Roberto and {Jordan}, Stefan and {Toonen}, Silvia and {Geier}, Stephan and {Barstow}, Martin and {Cummings}, Jeffrey D.},
        title = "{A Gaia Data Release 2 catalogue of white dwarfs and a comparison with SDSS}",
      journal = {\mnras},
         year = 2019,
        month = feb,
       volume = {482},
       number = {4},
        pages = {4570-4591},
          doi = {10.1093/mnras/sty3016},
archivePrefix = {arXiv},
       eprint = {1807.03315},
 primaryClass = {astro-ph.SR},
       adsurl = {https://ui.adsabs.harvard.edu/abs/2019MNRAS.482.4570G}
}

@ARTICLE{1991A&A...248..485D,
       author = {{Duquennoy}, A. and {Mayor}, M.},
        title = "{Multiplicity among Solar Type Stars in the Solar Neighbourhood - Part Two - Distribution of the Orbital Elements in an Unbiased Sample}",
      journal = {\aap},
         year = 1991,
        month = aug,
       volume = {248},
        pages = {485},
       adsurl = {https://ui.adsabs.harvard.edu/abs/1991A&A...248..485D}
}

@ARTICLE{2014A&A...563A.126L,
       author = {{Lucy}, L.~B.},
        title = "{Mass estimates for visual binaries with incomplete orbits}",
      journal = {\aap},
         year = 2014,
        month = mar,
       volume = {563},
          eid = {A126},
        pages = {A126},
          doi = {10.1051/0004-6361/201322649},
archivePrefix = {arXiv},
       eprint = {1309.2868},
 primaryClass = {astro-ph.SR},
       adsurl = {https://ui.adsabs.harvard.edu/abs/2014A&A...563A.126L}
}

@ARTICLE{2005Natur.434..192F,
       author = {{Figer}, Donald F.},
        title = "{An upper limit to the masses of stars}",
      journal = {\nat},
         year = 2005,
        month = mar,
       volume = {434},
       number = {7030},
        pages = {192-194},
          doi = {10.1038/nature03293},
archivePrefix = {arXiv},
       eprint = {astro-ph/0503193},
 primaryClass = {astro-ph},
       adsurl = {https://ui.adsabs.harvard.edu/abs/2005Natur.434..192F}
}

@ARTICLE{2023FrASS..1046603F,
       author = {{Fu}, Zhen-Sen and {Qi}, Zhao-Xiang and {Liao}, Shi-Long and {Peng}, Xi-Yan and {Yu}, Yong and {Wu}, Qi-Qi and {Shao}, Li and {Xu}, You-Hua},
        title = "{Simulation of CSST's astrometric capability}",
      journal = {Frontiers in Astronomy and Space Sciences},
         year = 2023,
        month = jun,
       volume = {10},
          eid = {1146603},
        pages = {1146603},
          doi = {10.3389/fspas.2023.1146603},
archivePrefix = {arXiv},
       eprint = {2304.02196},
 primaryClass = {astro-ph.IM},
       adsurl = {https://ui.adsabs.harvard.edu/abs/2023FrASS..1046603F}
}

@ARTICLE{2022A&A...665A.111W,
       author = {{Wang}, Yilun and {Liao}, Shilong and {Giacobbo}, Nicola and {Olejak}, Aleksandra and {Gao}, Jian and {Liu}, Jifeng},
        title = "{Astrometric mass measurement of compact companions in binary systems with Gaia}",
      journal = {\aap},
         year = 2022,
        month = sep,
       volume = {665},
          eid = {A111},
        pages = {A111},
          doi = {10.1051/0004-6361/202243684},
archivePrefix = {arXiv},
       eprint = {2307.12645},
 primaryClass = {astro-ph.GA},
       adsurl = {https://ui.adsabs.harvard.edu/abs/2022A&A...665A.111W}
}

@ARTICLE{wilson1931distribution,
  title="{The distribution of chi-square}",
  author={{Wilson}, Edwin B. and {Hilferty}, Margaret M.},
  journal={Proceedings of the National Academy of Sciences},
  volume={17},
  number={12},
  pages={684--688},
  year={1931},
  publisher={National Acad Sciences}
}

@ARTICLE{2023A&A...674A...9H,
       author = {{Halbwachs}, Jean-Louis and {Pourbaix}, Dimitri and {Arenou}, Fr{\'e}d{\'e}ric and {Galluccio}, Laurent and {Guillout}, Patrick and {Bauchet}, Nathalie and {Marchal}, Olivier and {Sadowski}, Gilles and {Teyssier}, David},
        title = "{Gaia Data Release 3. Astrometric binary star processing}",
      journal = {\aap},
         year = 2023,
        month = jun,
       volume = {674},
          eid = {A9},
        pages = {A9},
          doi = {10.1051/0004-6361/202243969},
archivePrefix = {arXiv},
       eprint = {2206.05726},
 primaryClass = {astro-ph.SR},
       adsurl = {https://ui.adsabs.harvard.edu/abs/2023A&A...674A...9H}
}

@ARTICLE{2023A&A...674A..10H,
       author = {{Holl}, B. and {Sozzetti}, A. and {Sahlmann}, J. and {Giacobbe}, P. and {S{\'e}gransan}, D. and {Unger}, N. and {Delisle}, J. -B. and {Barbato}, D. and {Lattanzi}, M.~G. and {Morbidelli}, R. and {Sosnowska}, D.},
        title = "{Gaia Data Release 3. Astrometric orbit determination with Markov chain Monte Carlo and genetic algorithms: Systems with stellar, sub-stellar, and planetary mass companions}",
      journal = {\aap},
         year = 2023,
        month = jun,
       volume = {674},
          eid = {A10},
        pages = {A10},
          doi = {10.1051/0004-6361/202244161},
archivePrefix = {arXiv},
       eprint = {2206.05439},
 primaryClass = {astro-ph.EP},
       adsurl = {https://ui.adsabs.harvard.edu/abs/2023A&A...674A..10H}
}

@book{The_Hipparcos_and_Tycho_catalogues,
    author = {{Perryman}, M.},
    publisher = {{ESA SP-1200}},
    title = "{The Hipparcos and Tycho catalogues}",
    year = {1997},
    issn = {0379-6566}
}

@ARTICLE{2023A&A...674A...1G,
       author = {{Gaia Collaboration} and {Vallenari}, A. and {Brown}, A.~G.~A. and {Prusti}, T. and {de Bruijne}, J.~H.~J. and {Arenou}, F. and {Babusiaux}, C. and {Biermann}, M. and {Creevey}, O.~L. and {Ducourant}, C. and {Evans}, D.~W. and {Eyer}, L. and {Guerra}, R. and {Hutton}, A. and {Jordi}, C. and {Klioner}, S.~A. and {Lammers}, U.~L. and {Lindegren}, L. and {Luri}, X. and {Mignard}, F. and {Panem}, C. and {Pourbaix}, D. and {Randich}, S. and {Sartoretti}, P. and {Soubiran}, C. and {Tanga}, P. and {Walton}, N.~A. and {Bailer-Jones}, C.~A.~L. and {Bastian}, U. and {Drimmel}, R. and {Jansen}, F. and {Katz}, D. and {Lattanzi}, M.~G. and {van Leeuwen}, F. and {Bakker}, J. and {Cacciari}, C. and {Casta{\~n}eda}, J. and {De Angeli}, F. and {Fabricius}, C. and {Fouesneau}, M. and {Fr{\'e}mat}, Y. and {Galluccio}, L. and {Guerrier}, A. and {Heiter}, U. and {Masana}, E. and {Messineo}, R. and {Mowlavi}, N. and {Nicolas}, C. and {Nienartowicz}, K. and {Pailler}, F. and {Panuzzo}, P. and {Riclet}, F. and {Roux}, W. and {Seabroke}, G.~M. and {Sordo}, R. and {Th{\'e}venin}, F. and {Gracia-Abril}, G. and {Portell}, J. and {Teyssier}, D. and {Altmann}, M. and {Andrae}, R. and {Audard}, M. and {Bellas-Velidis}, I. and {Benson}, K. and {Berthier}, J. and {Blomme}, R. and {Burgess}, P.~W. and {Busonero}, D. and {Busso}, G. and {C{\'a}novas}, H. and {Carry}, B. and {Cellino}, A. and {Cheek}, N. and {Clementini}, G. and {Damerdji}, Y. and {Davidson}, M. and {de Teodoro}, P. and {Nu{\~n}ez Campos}, M. and {Delchambre}, L. and {Dell'Oro}, A. and {Esquej}, P. and {Fern{\'a}ndez-Hern{\'a}ndez}, J. and {Fraile}, E. and {Garabato}, D. and {Garc{\'\i}a-Lario}, P. and {Gosset}, E. and {Haigron}, R. and {Halbwachs}, J. -L. and {Hambly}, N.~C. and {Harrison}, D.~L. and {Hern{\'a}ndez}, J. and {Hestroffer}, D. and {Hodgkin}, S.~T. and {Holl}, B. and {Jan{\ss}en}, K. and {Jevardat de Fombelle}, G. and {Jordan}, S. and {Krone-Martins}, A. and {Lanzafame}, A.~C. and {L{\"o}ffler}, W. and {Marchal}, O. and {Marrese}, P.~M. and {Moitinho}, A. and {Muinonen}, K. and {Osborne}, P. and {Pancino}, E. and {Pauwels}, T. and {Recio-Blanco}, A. and {Reyl{\'e}}, C. and {Riello}, M. and {Rimoldini}, L. and {Roegiers}, T. and {Rybizki}, J. and {Sarro}, L.~M. and {Siopis}, C. and {Smith}, M. and {Sozzetti}, A. and {Utrilla}, E. and {van Leeuwen}, M. and {Abbas}, U. and {{\'A}brah{\'a}m}, P. and {Abreu Aramburu}, A. and {Aerts}, C. and {Aguado}, J.~J. and {Ajaj}, M. and {Aldea-Montero}, F. and {Altavilla}, G. and {{\'A}lvarez}, M.~A. and {Alves}, J. and {Anders}, F. and {Anderson}, R.~I. and {Anglada Varela}, E. and {Antoja}, T. and {Baines}, D. and {Baker}, S.~G. and {Balaguer-N{\'u}{\~n}ez}, L. and {Balbinot}, E. and {Balog}, Z. and {Barache}, C. and {Barbato}, D. and {Barros}, M. and {Barstow}, M.~A. and {Bartolom{\'e}}, S. and {Bassilana}, J. -L. and {Bauchet}, N. and {Becciani}, U. and {Bellazzini}, M. and {Berihuete}, A. and {Bernet}, M. and {Bertone}, S. and {Bianchi}, L. and {Binnenfeld}, A. and {Blanco-Cuaresma}, S. and {Blazere}, A. and {Boch}, T. and {Bombrun}, A. and {Bossini}, D. and {Bouquillon}, S. and {Bragaglia}, A. and {Bramante}, L. and {Breedt}, E. and {Bressan}, A. and {Brouillet}, N. and {Brugaletta}, E. and {Bucciarelli}, B. and {Burlacu}, A. and {Butkevich}, A.~G. and {Buzzi}, R. and {Caffau}, E. and {Cancelliere}, R. and {Cantat-Gaudin}, T. and {Carballo}, R. and {Carlucci}, T. and {Carnerero}, M.~I. and {Carrasco}, J.~M. and {Casamiquela}, L. and {Castellani}, M. and {Castro-Ginard}, A. and {Chaoul}, L. and {Charlot}, P. and {Chemin}, L. and {Chiaramida}, V. and {Chiavassa}, A. and {Chornay}, N. and {Comoretto}, G. and {Contursi}, G. and {Cooper}, W.~J. and {Cornez}, T. and {Cowell}, S. and {Crifo}, F. and {Cropper}, M. and {Crosta}, M. and {Crowley}, C. and {Dafonte}, C. and {Dapergolas}, A. and {David}, M. and {David}, P. and {de Laverny}, P. and {De Luise}, F. and {De March}, R.},
        title = "{Gaia Data Release 3. Summary of the content and survey properties}",
      journal = {\aap},
         year = 2023,
        month = jun,
       volume = {674},
          eid = {A1},
        pages = {A1},
          doi = {10.1051/0004-6361/202243940},
archivePrefix = {arXiv},
       eprint = {2208.00211},
 primaryClass = {astro-ph.GA},
       adsurl = {https://ui.adsabs.harvard.edu/abs/2023A&A...674A...1G}
}

@ARTICLE{2011SSPMA..41.1441Z,
       author = {{Zhan}, Hu},
        title = "{Consideration for a large-scale multi-color imaging and slitless spectroscopy survey on the Chinese space station and its application in dark energy research}",
      journal = {Scientia Sinica Physica, Mechanica \& Astronomica},
         year = 2011,
        month = jan,
       volume = {41},
       number = {12},
        pages = {1441},
          doi = {10.1360/132011-961},
       adsurl = {https://ui.adsabs.harvard.edu/abs/2011SSPMA..41.1441Z}
}

@ARTICLE{2014RAA....14.1055S,
       author = {{Su}, Ding-Qiang and {Cui}, Xiang-Qun},
        title = "{Two suggested configurations for the Chinese space telescope}",
      journal = {Research in Astronomy and Astrophysics},
         year = 2014,
        month = sep,
       volume = {14},
       number = {9},
          eid = {1055-1060},
        pages = {1055-1060},
          doi = {10.1088/1674-4527/14/9/001},
       adsurl = {https://ui.adsabs.harvard.edu/abs/2014RAA....14.1055S}
}

@ARTICLE{zhan2021wide,
  title="{The wide-field multiband imaging and slitless spectroscopy survey to be carried out by the Survey Space Telescope of China Manned Space Program}",
  author={{Zhan}, Hu},
  journal={Chinese Science Bulletin (Chinese Version)},
  volume={66},
  number={11},
  pages={1290--1298},
  year={2021},
  publisher={SCIENCE IN CHINA PRESS}
}

@ARTICLE{2023A&A...672A..82L,
       author = {{Leclerc}, A. and {Babusiaux}, C. and {Arenou}, F. and {van Leeuwen}, F. and {Bonnefoy}, M. and {Delfosse}, X. and {Forveille}, T. and {Le Bouquin}, J. -B. and {Rodet}, L.},
        title = "{Combining HIPPARCOS and Gaia data for the study of binaries: The BINARYS tool}",
      journal = {\aap},
         year = 2023,
        month = apr,
       volume = {672},
          eid = {A82},
        pages = {A82},
          doi = {10.1051/0004-6361/202244144},
archivePrefix = {arXiv},
       eprint = {2209.04210},
 primaryClass = {astro-ph.SR},
       adsurl = {https://ui.adsabs.harvard.edu/abs/2023A&A...672A..82L}
}

@UNPUBLISHED{LL:LL-136,
  author = {{Lindegren}, L.},
  title="{Expected astrometric properties of binaries in (E)DR3}",
  institution={Lund Observatory},
  year={2022},
  month={May},
  url={https://dms.cosmos.esa.int/COSMOS/doc_fetch.php?id=1566327},
  note={{Gaia Data Processing and Analysis Consortium (DPAC) Technical Note GAIA-C3-TN-LU-LL-136-01}},
  type={Technical Note}
}

@ARTICLE{2021A&ARv..29....4S,
       author = {{Serenelli}, Aldo and {Weiss}, Achim and {Aerts}, Conny and {Angelou}, George C. and {Baroch}, David and {Bastian}, Nate and {Beck}, Paul G. and {Bergemann}, Maria and {Bestenlehner}, Joachim M. and {Czekala}, Ian and {Elias-Rosa}, Nancy and {Escorza}, Ana and {Van Eylen}, Vincent and {Feuillet}, Diane K. and {Gandolfi}, Davide and {Gieles}, Mark and {Girardi}, L{\'e}o and {Lebreton}, Yveline and {Lodieu}, Nicolas and {Martig}, Marie and {Miller Bertolami}, Marcelo M. and {Mombarg}, Joey S.~G. and {Morales}, Juan Carlos and {Moya}, Andr{\'e}s and {Nsamba}, Benard and {Pavlovski}, Kre{\v{s}}imir and {Pedersen}, May G. and {Ribas}, Ignasi and {Schneider}, Fabian R.~N. and {Silva Aguirre}, Victor and {Stassun}, Keivan G. and {Tolstoy}, Eline and {Tremblay}, Pier-Emmanuel and {Zwintz}, Konstanze},
        title = "{Weighing stars from birth to death: mass determination methods across the HRD}",
      journal = {\aapr},
         year = 2021,
        month = dec,
       volume = {29},
       number = {1},
          eid = {4},
        pages = {4},
          doi = {10.1007/s00159-021-00132-9},
archivePrefix = {arXiv},
       eprint = {2006.10868},
 primaryClass = {astro-ph.SR},
       adsurl = {https://ui.adsabs.harvard.edu/abs/2021A&ARv..29....4S}
}

@ARTICLE{2019AJ....157..168D,
       author = {{Dey}, Arjun and {Schlegel}, David J. and {Lang}, Dustin and {Blum}, Robert and {Burleigh}, Kaylan and {Fan}, Xiaohui and {Findlay}, Joseph R. and {Finkbeiner}, Doug and {Herrera}, David and {Juneau}, St{\'e}phanie and {Landriau}, Martin and {Levi}, Michael and {McGreer}, Ian and {Meisner}, Aaron and {Myers}, Adam D. and {Moustakas}, John and {Nugent}, Peter and {Patej}, Anna and {Schlafly}, Edward F. and {Walker}, Alistair R. and {Valdes}, Francisco and {Weaver}, Benjamin A. and {Y{\`e}che}, Christophe and {Zou}, Hu and {Zhou}, Xu and {Abareshi}, Behzad and {Abbott}, T.~M.~C. and {Abolfathi}, Bela and {Aguilera}, C. and {Alam}, Shadab and {Allen}, Lori and {Alvarez}, A. and {Annis}, James and {Ansarinejad}, Behzad and {Aubert}, Marie and {Beechert}, Jacqueline and {Bell}, Eric F. and {BenZvi}, Segev Y. and {Beutler}, Florian and {Bielby}, Richard M. and {Bolton}, Adam S. and {Brice{\~n}o}, C{\'e}sar and {Buckley-Geer}, Elizabeth J. and {Butler}, Karen and {Calamida}, Annalisa and {Carlberg}, Raymond G. and {Carter}, Paul and {Casas}, Ricard and {Castander}, Francisco J. and {Choi}, Yumi and {Comparat}, Johan and {Cukanovaite}, Elena and {Delubac}, Timoth{\'e}e and {DeVries}, Kaitlin and {Dey}, Sharmila and {Dhungana}, Govinda and {Dickinson}, Mark and {Ding}, Zhejie and {Donaldson}, John B. and {Duan}, Yutong and {Duckworth}, Christopher J. and {Eftekharzadeh}, Sarah and {Eisenstein}, Daniel J. and {Etourneau}, Thomas and {Fagrelius}, Parker A. and {Farihi}, Jay and {Fitzpatrick}, Mike and {Font-Ribera}, Andreu and {Fulmer}, Leah and {G{\"a}nsicke}, Boris T. and {Gaztanaga}, Enrique and {George}, Koshy and {Gerdes}, David W. and {Gontcho}, Satya Gontcho A. and {Gorgoni}, Claudio and {Green}, Gregory and {Guy}, Julien and {Harmer}, Diane and {Hernandez}, M. and {Honscheid}, Klaus and {Huang}, Lijuan Wendy and {James}, David J. and {Jannuzi}, Buell T. and {Jiang}, Linhua and {Joyce}, Richard and {Karcher}, Armin and {Karkar}, Sonia and {Kehoe}, Robert and {Kneib}, Jean-Paul and {Kueter-Young}, Andrea and {Lan}, Ting-Wen and {Lauer}, Tod R. and {Le Guillou}, Laurent and {Le Van Suu}, Auguste and {Lee}, Jae Hyeon and {Lesser}, Michael and {Perreault Levasseur}, Laurence and {Li}, Ting S. and {Mann}, Justin L. and {Marshall}, Robert and {Mart{\'\i}nez-V{\'a}zquez}, C.~E. and {Martini}, Paul and {du Mas des Bourboux}, H{\'e}lion and {McManus}, Sean and {Meier}, Tobias Gabriel and {M{\'e}nard}, Brice and {Metcalfe}, Nigel and {Mu{\~n}oz-Guti{\'e}rrez}, Andrea and {Najita}, Joan and {Napier}, Kevin and {Narayan}, Gautham and {Newman}, Jeffrey A. and {Nie}, Jundan and {Nord}, Brian and {Norman}, Dara J. and {Olsen}, Knut A.~G. and {Paat}, Anthony and {Palanque-Delabrouille}, Nathalie and {Peng}, Xiyan and {Poppett}, Claire L. and {Poremba}, Megan R. and {Prakash}, Abhishek and {Rabinowitz}, David and {Raichoor}, Anand and {Rezaie}, Mehdi and {Robertson}, A.~N. and {Roe}, Natalie A. and {Ross}, Ashley J. and {Ross}, Nicholas P. and {Rudnick}, Gregory and {Safonova}, Sasha and {Saha}, Abhijit and {S{\'a}nchez}, F. Javier and {Savary}, Elodie and {Schweiker}, Heidi and {Scott}, Adam and {Seo}, Hee-Jong and {Shan}, Huanyuan and {Silva}, David R. and {Slepian}, Zachary and {Soto}, Christian and {Sprayberry}, David and {Staten}, Ryan and {Stillman}, Coley M. and {Stupak}, Robert J. and {Summers}, David L. and {Sien Tie}, Suk and {Tirado}, H. and {Vargas-Maga{\~n}a}, Mariana and {Vivas}, A. Katherina and {Wechsler}, Risa H. and {Williams}, Doug and {Yang}, Jinyi and {Yang}, Qian and {Yapici}, Tolga and {Zaritsky}, Dennis and {Zenteno}, A. and {Zhang}, Kai and {Zhang}, Tianmeng and {Zhou}, Rongpu and {Zhou}, Zhimin},
        title = "{Overview of the DESI Legacy Imaging Surveys}",
      journal = {\aj},
         year = 2019,
        month = may,
       volume = {157},
       number = {5},
          eid = {168},
        pages = {168},
          doi = {10.3847/1538-3881/ab089d},
archivePrefix = {arXiv},
       eprint = {1804.08657},
 primaryClass = {astro-ph.IM},
       adsurl = {https://ui.adsabs.harvard.edu/abs/2019AJ....157..168D}
}

@ARTICLE{2024OJAp....7E.100E,
       author = {{El-Badry}, Kareem and {Lam}, Casey and {Holl}, Berry and {Halbwachs}, Jean-Louis and {Rix}, Hans-Walter and {Mazeh}, Tsevi and {Shahaf}, Sahar},
        title = "{A generative model for Gaia astrometric orbit catalogs: selection functions for binary stars, giant planets, and compact object companions}",
      journal = {The Open Journal of Astrophysics},
         year = 2024,
        month = nov,
       volume = {7},
          eid = {100},
        pages = {100},
          doi = {10.33232/001c.125461},
archivePrefix = {arXiv},
       eprint = {2411.00088},
 primaryClass = {astro-ph.SR},
       adsurl = {https://ui.adsabs.harvard.edu/abs/2024OJAp....7E.100E}
}

@ARTICLE{2019ApJ...873..111I,
       author = {{Ivezi{\'c}}, {\v{Z}}eljko and {Kahn}, Steven M. and {Tyson}, J. Anthony and {Abel}, Bob and {Acosta}, Emily and {Allsman}, Robyn and {Alonso}, David and {AlSayyad}, Yusra and {Anderson}, Scott F. and {Andrew}, John and {Angel}, James Roger P. and {Angeli}, George Z. and {Ansari}, Reza and {Antilogus}, Pierre and {Araujo}, Constanza and {Armstrong}, Robert and {Arndt}, Kirk T. and {Astier}, Pierre and {Aubourg}, {\'E}ric and {Auza}, Nicole and {Axelrod}, Tim S. and {Bard}, Deborah J. and {Barr}, Jeff D. and {Barrau}, Aurelian and {Bartlett}, James G. and {Bauer}, Amanda E. and {Bauman}, Brian J. and {Baumont}, Sylvain and {Bechtol}, Ellen and {Bechtol}, Keith and {Becker}, Andrew C. and {Becla}, Jacek and {Beldica}, Cristina and {Bellavia}, Steve and {Bianco}, Federica B. and {Biswas}, Rahul and {Blanc}, Guillaume and {Blazek}, Jonathan and {Blandford}, Roger D. and {Bloom}, Josh S. and {Bogart}, Joanne and {Bond}, Tim W. and {Booth}, Michael T. and {Borgland}, Anders W. and {Borne}, Kirk and {Bosch}, James F. and {Boutigny}, Dominique and {Brackett}, Craig A. and {Bradshaw}, Andrew and {Brandt}, William Nielsen and {Brown}, Michael E. and {Bullock}, James S. and {Burchat}, Patricia and {Burke}, David L. and {Cagnoli}, Gianpietro and {Calabrese}, Daniel and {Callahan}, Shawn and {Callen}, Alice L. and {Carlin}, Jeffrey L. and {Carlson}, Erin L. and {Chandrasekharan}, Srinivasan and {Charles-Emerson}, Glenaver and {Chesley}, Steve and {Cheu}, Elliott C. and {Chiang}, Hsin-Fang and {Chiang}, James and {Chirino}, Carol and {Chow}, Derek and {Ciardi}, David R. and {Claver}, Charles F. and {Cohen-Tanugi}, Johann and {Cockrum}, Joseph J. and {Coles}, Rebecca and {Connolly}, Andrew J. and {Cook}, Kem H. and {Cooray}, Asantha and {Covey}, Kevin R. and {Cribbs}, Chris and {Cui}, Wei and {Cutri}, Roc and {Daly}, Philip N. and {Daniel}, Scott F. and {Daruich}, Felipe and {Daubard}, Guillaume and {Daues}, Greg and {Dawson}, William and {Delgado}, Francisco and {Dellapenna}, Alfred and {de Peyster}, Robert and {de Val-Borro}, Miguel and {Digel}, Seth W. and {Doherty}, Peter and {Dubois}, Richard and {Dubois-Felsmann}, Gregory P. and {Durech}, Josef and {Economou}, Frossie and {Eifler}, Tim and {Eracleous}, Michael and {Emmons}, Benjamin L. and {Fausti Neto}, Angelo and {Ferguson}, Henry and {Figueroa}, Enrique and {Fisher-Levine}, Merlin and {Focke}, Warren and {Foss}, Michael D. and {Frank}, James and {Freemon}, Michael D. and {Gangler}, Emmanuel and {Gawiser}, Eric and {Geary}, John C. and {Gee}, Perry and {Geha}, Marla and {Gessner}, Charles J.~B. and {Gibson}, Robert R. and {Gilmore}, D. Kirk and {Glanzman}, Thomas and {Glick}, William and {Goldina}, Tatiana and {Goldstein}, Daniel A. and {Goodenow}, Iain and {Graham}, Melissa L. and {Gressler}, William J. and {Gris}, Philippe and {Guy}, Leanne P. and {Guyonnet}, Augustin and {Haller}, Gunther and {Harris}, Ron and {Hascall}, Patrick A. and {Haupt}, Justine and {Hernandez}, Fabio and {Herrmann}, Sven and {Hileman}, Edward and {Hoblitt}, Joshua and {Hodgson}, John A. and {Hogan}, Craig and {Howard}, James D. and {Huang}, Dajun and {Huffer}, Michael E. and {Ingraham}, Patrick and {Innes}, Walter R. and {Jacoby}, Suzanne H. and {Jain}, Bhuvnesh and {Jammes}, Fabrice and {Jee}, M. James and {Jenness}, Tim and {Jernigan}, Garrett and {Jevremovi{\'c}}, Darko and {Johns}, Kenneth and {Johnson}, Anthony S. and {Johnson}, Margaret W.~G. and {Jones}, R. Lynne and {Juramy-Gilles}, Claire and {Juri{\'c}}, Mario and {Kalirai}, Jason S. and {Kallivayalil}, Nitya J. and {Kalmbach}, Bryce and {Kantor}, Jeffrey P. and {Karst}, Pierre and {Kasliwal}, Mansi M. and {Kelly}, Heather and {Kessler}, Richard and {Kinnison}, Veronica and {Kirkby}, David and {Knox}, Lloyd and {Kotov}, Ivan V. and {Krabbendam}, Victor L. and {Krughoff}, K. Simon and {Kub{\'a}nek}, Petr and {Kuczewski}, John and {Kulkarni}, Shri and {Ku}, John and {Kurita}, Nadine R. and {Lage}, Craig S. and {Lambert}, Ron and {Lange}, Travis and {Langton}, J. Brian and {Le Guillou}, Laurent and {Levine}, Deborah and {Liang}, Ming and {Lim}, Kian-Tat and {Lintott}, Chris J. and {Long}, Kevin E. and {Lopez}, Margaux and {Lotz}, Paul J. and {Lupton}, Robert H. and {Lust}, Nate B. and {MacArthur}, Lauren A. and {Mahabal}, Ashish and {Mandelbaum}, Rachel and {Markiewicz}, Thomas W. and {Marsh}, Darren S. and {Marshall}, Philip J. and {Marshall}, Stuart and {May}, Morgan and {McKercher}, Robert and {McQueen}, Michelle and {Meyers}, Joshua and {Migliore}, Myriam and {Miller}, Michelle and {Mills}, David J.},
        title = "{LSST: From Science Drivers to Reference Design and Anticipated Data Products}",
      journal = {\apj},
         year = 2019,
        month = mar,
       volume = {873},
       number = {2},
          eid = {111},
        pages = {111},
          doi = {10.3847/1538-4357/ab042c},
archivePrefix = {arXiv},
       eprint = {0805.2366},
 primaryClass = {astro-ph},
       adsurl = {https://ui.adsabs.harvard.edu/abs/2019ApJ...873..111I}
}

@ARTICLE{2019NatAs...3..482K,
       author = {{Kroupa}, Pavel and {Jerabkova}, Tereza},
        title = "{The Salpeter IMF and its descendants}",
      journal = {Nature Astronomy},
         year = 2019,
        month = jun,
       volume = {3},
        pages = {482-484},
          doi = {10.1038/s41550-019-0793-0},
archivePrefix = {arXiv},
       eprint = {1910.01126},
 primaryClass = {astro-ph.SR},
       adsurl = {https://ui.adsabs.harvard.edu/abs/2019NatAs...3..482K}
}

@ARTICLE{2022A&A...662A.112E,
       author = {{Euclid Collaboration} and {Scaramella}, R. and {Amiaux}, J. and {Mellier}, Y. and {Burigana}, C. and {Carvalho}, C.~S. and {Cuillandre}, J. -C. and {Da Silva}, A. and {Derosa}, A. and {Dinis}, J. and {Maiorano}, E. and {Maris}, M. and {Tereno}, I. and {Laureijs}, R. and {Boenke}, T. and {Buenadicha}, G. and {Dupac}, X. and {Gaspar Venancio}, L.~M. and {G{\'o}mez-{\'A}lvarez}, P. and {Hoar}, J. and {Lorenzo Alvarez}, J. and {Racca}, G.~D. and {Saavedra-Criado}, G. and {Schwartz}, J. and {Vavrek}, R. and {Schirmer}, M. and {Aussel}, H. and {Azzollini}, R. and {Cardone}, V.~F. and {Cropper}, M. and {Ealet}, A. and {Garilli}, B. and {Gillard}, W. and {Granett}, B.~R. and {Guzzo}, L. and {Hoekstra}, H. and {Jahnke}, K. and {Kitching}, T. and {Maciaszek}, T. and {Meneghetti}, M. and {Miller}, L. and {Nakajima}, R. and {Niemi}, S.~M. and {Pasian}, F. and {Percival}, W.~J. and {Pottinger}, S. and {Sauvage}, M. and {Scodeggio}, M. and {Wachter}, S. and {Zacchei}, A. and {Aghanim}, N. and {Amara}, A. and {Auphan}, T. and {Auricchio}, N. and {Awan}, S. and {Balestra}, A. and {Bender}, R. and {Bodendorf}, C. and {Bonino}, D. and {Branchini}, E. and {Brau-Nogue}, S. and {Brescia}, M. and {Candini}, G.~P. and {Capobianco}, V. and {Carbone}, C. and {Carlberg}, R.~G. and {Carretero}, J. and {Casas}, R. and {Castander}, F.~J. and {Castellano}, M. and {Cavuoti}, S. and {Cimatti}, A. and {Cledassou}, R. and {Congedo}, G. and {Conselice}, C.~J. and {Conversi}, L. and {Copin}, Y. and {Corcione}, L. and {Costille}, A. and {Courbin}, F. and {Degaudenzi}, H. and {Douspis}, M. and {Dubath}, F. and {Duncan}, C.~A.~J. and {Dusini}, S. and {Farrens}, S. and {Ferriol}, S. and {Fosalba}, P. and {Fourmanoit}, N. and {Frailis}, M. and {Franceschi}, E. and {Franzetti}, P. and {Fumana}, M. and {Gillis}, B. and {Giocoli}, C. and {Grazian}, A. and {Grupp}, F. and {Haugan}, S.~V.~H. and {Holmes}, W. and {Hormuth}, F. and {Hudelot}, P. and {Kermiche}, S. and {Kiessling}, A. and {Kilbinger}, M. and {Kohley}, R. and {Kubik}, B. and {K{\"u}mmel}, M. and {Kunz}, M. and {Kurki-Suonio}, H. and {Lahav}, O. and {Ligori}, S. and {Lilje}, P.~B. and {Lloro}, I. and {Mansutti}, O. and {Marggraf}, O. and {Markovic}, K. and {Marulli}, F. and {Massey}, R. and {Maurogordato}, S. and {Melchior}, M. and {Merlin}, E. and {Meylan}, G. and {Mohr}, J.~J. and {Moresco}, M. and {Morin}, B. and {Moscardini}, L. and {Munari}, E. and {Nichol}, R.~C. and {Padilla}, C. and {Paltani}, S. and {Peacock}, J. and {Pedersen}, K. and {Pettorino}, V. and {Pires}, S. and {Poncet}, M. and {Popa}, L. and {Pozzetti}, L. and {Raison}, F. and {Rebolo}, R. and {Rhodes}, J. and {Rix}, H. -W. and {Roncarelli}, M. and {Rossetti}, E. and {Saglia}, R. and {Schneider}, P. and {Schrabback}, T. and {Secroun}, A. and {Seidel}, G. and {Serrano}, S. and {Sirignano}, C. and {Sirri}, G. and {Skottfelt}, J. and {Stanco}, L. and {Starck}, J.~L. and {Tallada-Cresp{\'\i}}, P. and {Tavagnacco}, D. and {Taylor}, A.~N. and {Teplitz}, H.~I. and {Toledo-Moreo}, R. and {Torradeflot}, F. and {Trifoglio}, M. and {Valentijn}, E.~A. and {Valenziano}, L. and {Verdoes Kleijn}, G.~A. and {Wang}, Y. and {Welikala}, N. and {Weller}, J. and {Wetzstein}, M. and {Zamorani}, G. and {Zoubian}, J. and {Andreon}, S. and {Baldi}, M. and {Bardelli}, S. and {Boucaud}, A. and {Camera}, S. and {Di Ferdinando}, D. and {Fabbian}, G. and {Farinelli}, R. and {Galeotta}, S. and {Graci{\'a}-Carpio}, J. and {Maino}, D. and {Medinaceli}, E. and {Mei}, S. and {Neissner}, C. and {Polenta}, G. and {Renzi}, A. and {Romelli}, E. and {Rosset}, C. and {Sureau}, F. and {Tenti}, M. and {Vassallo}, T. and {Zucca}, E. and {Baccigalupi}, C. and {Balaguera-Antol{\'\i}nez}, A. and {Battaglia}, P. and {Biviano}, A. and {Borgani}, S. and {Bozzo}, E. and {Cabanac}, R. and {Cappi}, A.},
        title = "{Euclid preparation. I. The Euclid Wide Survey}",
      journal = {\aap},
         year = 2022,
        month = jun,
       volume = {662},
          eid = {A112},
        pages = {A112},
          doi = {10.1051/0004-6361/202141938},
archivePrefix = {arXiv},
       eprint = {2108.01201},
 primaryClass = {astro-ph.CO},
       adsurl = {https://ui.adsabs.harvard.edu/abs/2022A&A...662A.112E}
}

@ARTICLE{2024A&A...691A..81D,
       author = {{Ding}, Ye and {Liao}, Shilong and {Wu}, Qiqi and {Qi}, Zhaoxiang and {Tang}, Zhenghong},
        title = "{Analysis of the Gaia Data Release 3 parallax bias in the Galactic plane}",
      journal = {\aap},
         year = 2024,
        month = nov,
       volume = {691},
          eid = {A81},
        pages = {A81},
          doi = {10.1051/0004-6361/202450967},
archivePrefix = {arXiv},
       eprint = {2409.15694},
 primaryClass = {astro-ph.IM},
       adsurl = {https://ui.adsabs.harvard.edu/abs/2024A&A...691A..81D}
}

@ARTICLE{2025AJ....169..211D,
       author = {{Ding}, Ye and {Liao}, Shilong and {Wen}, Shangyu and {Qi}, Zhaoxiang},
        title = "{Analysis of the Gaia Data Release 3 Parallax Bias at Bright Magnitudes}",
      journal = {\aj},
         year = 2025,
        month = apr,
       volume = {169},
       number = {4},
          eid = {211},
        pages = {211},
          doi = {10.3847/1538-3881/adba44},
archivePrefix = {arXiv},
       eprint = {2502.08068},
 primaryClass = {astro-ph.SR},
       adsurl = {https://ui.adsabs.harvard.edu/abs/2025AJ....169..211D}
}

@ARTICLE{2019A&A...623A..72K,
       author = {{Kervella}, Pierre and {Arenou}, Fr{\'e}d{\'e}ric and {Mignard}, Fran{\c{c}}ois and {Th{\'e}venin}, Fr{\'e}d{\'e}ric},
        title = "{Stellar and substellar companions of nearby stars from Gaia DR2. Binarity from proper motion anomaly}",
      journal = {\aap},
         year = 2019,
        month = mar,
       volume = {623},
          eid = {A72},
        pages = {A72},
          doi = {10.1051/0004-6361/201834371},
archivePrefix = {arXiv},
       eprint = {1811.08902},
 primaryClass = {astro-ph.SR},
       adsurl = {https://ui.adsabs.harvard.edu/abs/2019A&A...623A..72K}
}

@ARTICLE{2016A&A...587A..61C,
       author = {{Corral-Santana}, J.~M. and {Casares}, J. and {Mu{\~n}oz-Darias}, T. and {Bauer}, F.~E. and {Mart{\'\i}nez-Pais}, I.~G. and {Russell}, D.~M.},
        title = "{BlackCAT: A catalogue of stellar-mass black holes in X-ray transients}",
      journal = {Astronomy \& Astrophysics},
         year = 2016,
        month = mar,
       volume = {587},
          eid = {A61},
        pages = {A61},
          doi = {10.1051/0004-6361/201527130},
archivePrefix = {arXiv},
       eprint = {1510.08869},
 primaryClass = {astro-ph.HE},
       adsurl = {https://ui.adsabs.harvard.edu/abs/2016A&A...587A..61C}
}

@ARTICLE{2012ApJ...757...36K,
       author = {{Kreidberg}, Laura and {Bailyn}, Charles D. and {Farr}, Will M. and {Kalogera}, Vicky},
        title = "{Mass Measurements of Black Holes in X-Ray Transients: Is There a Mass Gap?}",
      journal = {\apj},
         year = 2012,
        month = sep,
       volume = {757},
       number = {1},
          eid = {36},
        pages = {36},
          doi = {10.1088/0004-637X/757/1/36},
archivePrefix = {arXiv},
       eprint = {1205.1805},
 primaryClass = {astro-ph.HE},
       adsurl = {https://ui.adsabs.harvard.edu/abs/2012ApJ...757...36K}
}

@ARTICLE{2021A&A...649A...2L,
       author = {{Lindegren}, L. and {Klioner}, S.~A. and {Hern{\'a}ndez}, J. and {Bombrun}, A. and {Ramos-Lerate}, M. and {Steidelm{\"u}ller}, H. and {Bastian}, U. and {Biermann}, M. and {de Torres}, A. and {Gerlach}, E. and {Geyer}, R. and {Hilger}, T. and {Hobbs}, D. and {Lammers}, U. and {McMillan}, P.~J. and {Stephenson}, C.~A. and {Casta{\~n}eda}, J. and {Davidson}, M. and {Fabricius}, C. and {Gracia-Abril}, G. and {Portell}, J. and {Rowell}, N. and {Teyssier}, D. and {Torra}, F. and {Bartolom{\'e}}, S. and {Clotet}, M. and {Garralda}, N. and {Gonz{\'a}lez-Vidal}, J.~J. and {Torra}, J. and {Abbas}, U. and {Altmann}, M. and {Anglada Varela}, E. and {Balaguer-N{\'u}{\~n}ez}, L. and {Balog}, Z. and {Barache}, C. and {Becciani}, U. and {Bernet}, M. and {Bertone}, S. and {Bianchi}, L. and {Bouquillon}, S. and {Brown}, A.~G.~A. and {Bucciarelli}, B. and {Busonero}, D. and {Butkevich}, A.~G. and {Buzzi}, R. and {Cancelliere}, R. and {Carlucci}, T. and {Charlot}, P. and {Cioni}, M.-R.~L. and {Crosta}, M. and {Crowley}, C. and {del Peloso}, E.~F. and {del Pozo}, E. and {Drimmel}, R. and {Esquej}, P. and {Fienga}, A. and {Fraile}, E. and {Gai}, M. and {Garcia-Reinaldos}, M. and {Guerra}, R. and {Hambly}, N.~C. and {Hauser}, M. and {Jan{\ss}en}, K. and {Jordan}, S. and {Kostrzewa-Rutkowska}, Z. and {Lattanzi}, M.~G. and {Liao}, S. and {Licata}, E. and {Lister}, T.~A. and {L{\"o}ffler}, W. and {Marchant}, J.~M. and {Masip}, A. and {Mignard}, F. and {Mints}, A. and {Molina}, D. and {Mora}, A. and {Morbidelli}, R. and {Murphy}, C.~P. and {Pagani}, C. and {Panuzzo}, P. and {Pe{\~n}alosa Esteller}, X. and {Poggio}, E. and {Re Fiorentin}, P. and {Riva}, A. and {Sagrist{\`a} Sell{\'e}s}, A. and {Sanchez Gimenez}, V. and {Sarasso}, M. and {Sciacca}, E. and {Siddiqui}, H.~I. and {Smart}, R.~L. and {Souami}, D. and {Spagna}, A. and {Steele}, I.~A. and {Taris}, F. and {Utrilla}, E. and {van Reeven}, W. and {Vecchiato}, A.},
        title = "{Gaia Early Data Release 3. The astrometric solution}",
      journal = {\aap},
         year = 2021,
        month = may,
       volume = {649},
          eid = {A2},
        pages = {A2},
          doi = {10.1051/0004-6361/202039709},
archivePrefix = {arXiv},
       eprint = {2012.03380},
 primaryClass = {astro-ph.IM},
       adsurl = {https://ui.adsabs.harvard.edu/abs/2021A&A...649A...2L}
}

@ARTICLE{2021A&A...649A...4L,
       author = {{Lindegren}, L. and {Bastian}, U. and {Biermann}, M. and {Bombrun}, A. and {de Torres}, A. and {Gerlach}, E. and {Geyer}, R. and {Hern{\'a}ndez}, J. and {Hilger}, T. and {Hobbs}, D. and {Klioner}, S.~A. and {Lammers}, U. and {McMillan}, P.~J. and {Ramos-Lerate}, M. and {Steidelm{\"u}ller}, H. and {Stephenson}, C.~A. and {van Leeuwen}, F.},
        title = "{Gaia Early Data Release 3. Parallax bias versus magnitude, colour, and position}",
      journal = {\aap},
         year = 2021,
        month = may,
       volume = {649},
          eid = {A4},
        pages = {A4},
          doi = {10.1051/0004-6361/202039653},
archivePrefix = {arXiv},
       eprint = {2012.01742},
 primaryClass = {astro-ph.IM},
       adsurl = {https://ui.adsabs.harvard.edu/abs/2021A&A...649A...4L}
}

@ARTICLE{2026AJ....171..121X,
       author = {{Xuan}, Yifan and {Feng}, Fabo and {Fu}, Zhensen and {Liao}, Shilong and {Qi}, Zhaoxiang and {Chen}, Yang},
        title = "{Predicting the Detection Yields of Giant Planets and Brown Dwarfs with CSST Astrometry}",
      journal = {\aj},
         year = 2026,
        month = mar,
       volume = {171},
       number = {3},
          eid = {121},
        pages = {121},
          doi = {10.3847/1538-3881/ae3176},
archivePrefix = {arXiv},
       eprint = {2512.23378},
 primaryClass = {astro-ph.EP},
       adsurl = {https://ui.adsabs.harvard.edu/abs/2026AJ....171..121X}
}

@ARTICLE{2026A&A...710A.398L,
       author = {{Liao}, Shilong and {Ding}, Ye and {Wen}, Shangyu and {Qi}, Zhaoxiang and {Wu}, Qiqi},
        title = "{Constraining the inclination of binary system orbits with the astrometric excess noise from Gaia DR3}",
      journal = {\aap},
         year = 2026,
        month = jul,
       volume = {710},
          eid = {A398},
        pages = {A398},
          doi = {10.1051/0004-6361/202452921},
archivePrefix = {arXiv},
       eprint = {2605.25482},
 primaryClass = {astro-ph.IM},
       adsurl = {https://ui.adsabs.harvard.edu/abs/2026A&A...710A.398L}
}

@ARTICLE{2023A&A...674A..25H,
       author = {{Holl}, B. and {Fabricius}, C. and {Portell}, J. and {Lindegren}, L. and {Panuzzo}, P. and {Bernet}, M. and {Casta{\~n}eda}, J. and {Jevardat de Fombelle}, G. and {Audard}, M. and {Ducourant}, C. and {Harrison}, D.~L. and {Evans}, D.~W. and {Busso}, G. and {Sozzetti}, A. and {Gosset}, E. and {Arenou}, F. and {De Angeli}, F. and {Riello}, M. and {Eyer}, L. and {Rimoldini}, L. and {Gavras}, P. and {Mowlavi}, N. and {Nienartowicz}, K. and {Lecoeur-Ta{\"\i}bi}, I. and {Garc{\'\i}a-Lario}, P. and {Pourbaix}, D.},
        title = "{Gaia Data Release 3. Gaia scan-angle-dependent signals and spurious periods}",
      journal = {\aap},
         year = 2023,
        month = jun,
       volume = {674},
          eid = {A25},
        pages = {A25},
          doi = {10.1051/0004-6361/202245353},
archivePrefix = {arXiv},
       eprint = {2212.11971},
 primaryClass = {astro-ph.IM},
       adsurl = {https://ui.adsabs.harvard.edu/abs/2023A&A...674A..25H}
}

@ARTICLE{mwmsc,
       author = {{Chen}, Yang and {Fu}, Xiaoting and {Liu}, Chao and {Dal Tio}, Piero and {Girardi}, L{\'e}o and  {Pastorelli}, Giada and {Mazzi}, Alessandro and {Trabucchi}, Michele and {Tian}, Hao and {Fan}, Dongwei and {Marigo}, Paola and {Bressan}, Alessandro},
        title = "{The First Comprehensive Milky Way Stellar Mock Catalogue for the Chinese Space Station Telescope Survey Camera}",
      journal = {Science China Physics, Mechanics and Astronomy},
         year = 2023,
        month = nov,
       volume = {66},
       number = {11},
          eid = {119511},
        pages = {119511},
          doi = {10.1007/s11433-023-2181-x}
}

@ARTICLE{2016AN....337..871G,
       author = {{Girardi}, L.},
        title = "{Milky Way populations with TRILEGAL}",
      journal = {Astronomische Nachrichten},
         year = 2016,
        month = sep,
       volume = {337},
       number = {8-9},
        pages = {871},
          doi = {10.1002/asna.201612388},
       adsurl = {https://ui.adsabs.harvard.edu/abs/2016AN....337..871G}
}

@ARTICLE{2005A&A...436..895G,
       author = {{Girardi}, L. and {Groenewegen}, M.~A.~T. and {Hatziminaoglou}, E. and {da Costa}, L.},
        title = "{Star counts in the Galaxy. Simulating from very deep to very shallow photometric surveys with the TRILEGAL code}",
      journal = {\aap},
         year = 2005,
        month = jun,
       volume = {436},
       number = {3},
        pages = {895-915},
          doi = {10.1051/0004-6361:20042352},
archivePrefix = {arXiv},
       eprint = {astro-ph/0504047},
 primaryClass = {astro-ph},
       adsurl = {https://ui.adsabs.harvard.edu/abs/2005A&A...436..895G}
}

@ARTICLE{2016ApJS..222....8D,
       author = {{Dotter}, Aaron},
        title = "{MESA Isochrones and Stellar Tracks (MIST) 0: Methods for the Construction of Stellar Isochrones}",
      journal = {\apjs},
         year = 2016,
        month = jan,
       volume = {222},
       number = {1},
          eid = {8},
        pages = {8},
          doi = {10.3847/0067-0049/222/1/8},
archivePrefix = {arXiv},
       eprint = {1601.05144},
 primaryClass = {astro-ph.SR},
       adsurl = {https://ui.adsabs.harvard.edu/abs/2016ApJS..222....8D}
}

@ARTICLE{2016ApJ...823..102C,
       author = {{Choi}, Jieun and {Dotter}, Aaron and {Conroy}, Charlie and {Cantiello}, Matteo and {Paxton}, Bill and {Johnson}, Benjamin D.},
        title = "{Mesa Isochrones and Stellar Tracks (MIST). I. Solar-scaled Models}",
      journal = {\apj},
         year = 2016,
        month = jun,
       volume = {823},
       number = {2},
          eid = {102},
        pages = {102},
          doi = {10.3847/0004-637X/823/2/102},
archivePrefix = {arXiv},
       eprint = {1604.08592},
 primaryClass = {astro-ph.SR},
       adsurl = {https://ui.adsabs.harvard.edu/abs/2016ApJ...823..102C}
}

@ARTICLE{2021A&A...649A...3R,
       author = {{Riello}, M. and {De Angeli}, F. and {Evans}, D.~W. and {Montegriffo}, P. and {Carrasco}, J.~M. and {Busso}, G. and {Palaversa}, L. and {Burgess}, P.~W. and {Diener}, C. and {Davidson}, M. and {Rowell}, N. and {Fabricius}, C. and {Jordi}, C. and {Bellazzini}, M. and {Pancino}, E. and {Harrison}, D.~L. and {Cacciari}, C. and {van Leeuwen}, F. and {Hambly}, N.~C. and {Hodgkin}, S.~T. and {Osborne}, P.~J. and {Altavilla}, G. and {Barstow}, M.~A. and {Brown}, A.~G.~A. and {Castellani}, M. and {Cowell}, S. and {De Luise}, F. and {Gilmore}, G. and {Giuffrida}, G. and {Hidalgo}, S. and {Holland}, G. and {Marinoni}, S. and {Pagani}, C. and {Piersimoni}, A.~M. and {Pulone}, L. and {Ragaini}, S. and {Rainer}, M. and {Richards}, P.~J. and {Sanna}, N. and {Walton}, N.~A. and {Weiler}, M. and {Yoldas}, A.},
        title = "{Gaia Early Data Release 3. Photometric content and validation}",
      journal = {\aap},
         year = 2021,
        month = may,
       volume = {649},
          eid = {A3},
        pages = {A3},
          doi = {10.1051/0004-6361/202039587},
archivePrefix = {arXiv},
       eprint = {2012.01916},
 primaryClass = {astro-ph.IM},
       adsurl = {https://ui.adsabs.harvard.edu/abs/2021A&A...649A...3R}
}

@ARTICLE{2018A&A...616A...2L,
       author = {{Lindegren}, L. and {Hern{\'a}ndez}, J. and {Bombrun}, A. and {Klioner}, S. and {Bastian}, U. and {Ramos-Lerate}, M. and {de Torres}, A. and {Steidelm{\"u}ller}, H. and {Stephenson}, C. and {Hobbs}, D. and {Lammers}, U. and {Biermann}, M. and {Geyer}, R. and {Hilger}, T. and {Michalik}, D. and {Stampa}, U. and {McMillan}, P.~J. and {Casta{\~n}eda}, J. and {Clotet}, M. and {Comoretto}, G. and {Davidson}, M. and {Fabricius}, C. and {Gracia}, G. and {Hambly}, N.~C. and {Hutton}, A. and {Mora}, A. and {Portell}, J. and {van Leeuwen}, F. and {Abbas}, U. and {Abreu}, A. and {Altmann}, M. and {Andrei}, A. and {Anglada}, E. and {Balaguer-N{\'u}{\~n}ez}, L. and {Barache}, C. and {Becciani}, U. and {Bertone}, S. and {Bianchi}, L. and {Bouquillon}, S. and {Bourda}, G. and {Br{\"u}semeister}, T. and {Bucciarelli}, B. and {Busonero}, D. and {Buzzi}, R. and {Cancelliere}, R. and {Carlucci}, T. and {Charlot}, P. and {Cheek}, N. and {Crosta}, M. and {Crowley}, C. and {de Bruijne}, J. and {de Felice}, F. and {Drimmel}, R. and {Esquej}, P. and {Fienga}, A. and {Fraile}, E. and {Gai}, M. and {Garralda}, N. and {Gonz{\'a}lez-Vidal}, J.~J. and {Guerra}, R. and {Hauser}, M. and {Hofmann}, W. and {Holl}, B. and {Jordan}, S. and {Lattanzi}, M.~G. and {Lenhardt}, H. and {Liao}, S. and {Licata}, E. and {Lister}, T. and {L{\"o}ffler}, W. and {Marchant}, J. and {Martin-Fleitas}, J.-M. and {Messineo}, R. and {Mignard}, F. and {Morbidelli}, R. and {Poggio}, E. and {Riva}, A. and {Rowell}, N. and {Salguero}, E. and {Sarasso}, M. and {Sciacca}, E. and {Siddiqui}, H. and {Smart}, R.~L. and {Spagna}, A. and {Steele}, I. and {Taris}, F. and {Torra}, J. and {van Elteren}, A. and {van Reeven}, W. and {Vecchiato}, A.},
        title = "{Gaia Data Release 2. The astrometric solution}",
      journal = {\aap},
         year = 2018,
        month = aug,
       volume = {616},
          eid = {A2},
        pages = {A2},
          doi = {10.1051/0004-6361/201832727},
archivePrefix = {arXiv},
       eprint = {1804.09366},
 primaryClass = {astro-ph.IM},
       adsurl = {https://ui.adsabs.harvard.edu/abs/2018A&A...616A...2L}
}

@ARTICLE{2022MNRAS.510.3885G,
       author = {{Gandhi}, P. and {Buckley}, D.~A.~H. and {Charles}, P.~A. and {Hodgkin}, S. and {Scaringi}, S. and {Knigge}, C. and {Rao}, A. and {Paice}, J.~A. and {Zhao}, Y.},
        title = "{Astrometric excess noise in Gaia EDR3 and the search for X-ray binaries}",
      journal = {\mnras},
         year = 2022,
        month = mar,
       volume = {510},
       number = {3},
        pages = {3885-3895},
          doi = {10.1093/mnras/stab3771},
archivePrefix = {arXiv},
       eprint = {2201.00833},
 primaryClass = {astro-ph.HE},
       adsurl = {https://ui.adsabs.harvard.edu/abs/2022MNRAS.510.3885G}
}

@ARTICLE{2019A&A...632L...9K,
       author = {{Kiefer}, Flavien},
        title = "{Determining the mass of the planetary candidate HD 114762 b using Gaia}",
      journal = {\aap},
         year = 2019,
        month = dec,
       volume = {632},
          eid = {L9},
        pages = {L9},
          doi = {10.1051/0004-6361/201936942},
archivePrefix = {arXiv},
       eprint = {1910.07835},
 primaryClass = {astro-ph.EP},
       adsurl = {https://ui.adsabs.harvard.edu/abs/2019A&A...632L...9K}
}

@ARTICLE{2021A&A...645A...7K,
       author = {{Kiefer}, F. and {H{\'e}brard}, G. and {Lecavelier des Etangs}, A. and {Martioli}, E. and {Dalal}, S. and {Vidal-Madjar}, A.},
        title = "{Determining the true mass of radial-velocity exoplanets with Gaia. Nine planet candidates in the brown dwarf or stellar regime and 27 confirmed planets}",
      journal = {\aap},
         year = 2021,
        month = jan,
       volume = {645},
          eid = {A7},
        pages = {A7},
          doi = {10.1051/0004-6361/202039168},
archivePrefix = {arXiv},
       eprint = {2009.14164},
 primaryClass = {astro-ph.EP},
       adsurl = {https://ui.adsabs.harvard.edu/abs/2021A&A...645A...7K}
}

@ARTICLE{2020MNRAS.495..321P,
       author = {{Penoyre}, Zephyr and {Belokurov}, Vasily and {Wyn Evans}, N. and {Everall}, A. and {Koposov}, S.~E.},
        title = "{Binary deviations from single object astrometry}",
      journal = {\mnras},
         year = 2020,
        month = jun,
       volume = {495},
       number = {1},
        pages = {321-337},
          doi = {10.1093/mnras/staa1148},
archivePrefix = {arXiv},
       eprint = {2003.05456},
 primaryClass = {astro-ph.GA},
       adsurl = {https://ui.adsabs.harvard.edu/abs/2020MNRAS.495..321P}
}

@ARTICLE{mahalanobis,
       author = {{Mahalanobis}, P.~C.},
        title = "{On the Generalised Distance in Statistics}",
      journal = {Proceedings of the National Institute of Sciences of India},
         year = 1936,
       volume = {2},
       number = {1},
        pages = {49--55}
}

@ARTICLE{2024A&A...688A...1C,
       author = {{Castro-Ginard}, Alfred and {Penoyre}, Zephyr and {Casey}, Andrew R. and {Brown}, Anthony G.~A. and {Belokurov}, Vasily and {Cantat-Gaudin}, Tristan and {Drimmel}, Ronald and {Fouesneau}, Morgan and {Khanna}, Shourya and {Kurbatov}, Evgeny P. and {Price-Whelan}, Adrian M. and {Rix}, Hans-Walter and {Smart}, Richard L.},
        title = "{Gaia DR3 detectability of unresolved binary systems}",
      journal = {\aap},
         year = 2024,
        month = aug,
       volume = {688},
          eid = {A1},
        pages = {A1},
          doi = {10.1051/0004-6361/202450172},
archivePrefix = {arXiv},
       eprint = {2404.14127},
 primaryClass = {astro-ph.GA},
       adsurl = {https://ui.adsabs.harvard.edu/abs/2024A&A...688A...1C}
}

@ARTICLE{2021ApJS..254...42B,
       author = {{Brandt}, Timothy D.},
        title = "{The Hipparcos-Gaia Catalog of Accelerations: Gaia EDR3 Edition}",
      journal = {\apjs},
         year = 2021,
        month = jun,
       volume = {254},
       number = {2},
          eid = {42},
        pages = {42},
          doi = {10.3847/1538-4365/abf93c},
archivePrefix = {arXiv},
       eprint = {2105.11662},
 primaryClass = {astro-ph.GA},
       adsurl = {https://ui.adsabs.harvard.edu/abs/2021ApJS..254...42B}
}

@ARTICLE{2025A&A...704A.150A,
       author = {{Abreu}, A. and {Lillo-Box}, J. and {Perez-Garcia}, A.~M. and {Sahlmann}, J. and {de Bruijne}, J.~H.~J. and {Cifuentes}, C.},
        title = "{ExoDNN: Boosting exoplanet detection with artificial intelligence: Application to Gaia Data Release 3}",
      journal = {\aap},
         year = 2025,
        month = dec,
       volume = {704},
          eid = {A150},
        pages = {A150},
          doi = {10.1051/0004-6361/202555598},
archivePrefix = {arXiv},
       eprint = {2602.02910},
 primaryClass = {astro-ph.EP},
       adsurl = {https://ui.adsabs.harvard.edu/abs/2025A&A...704A.150A}
}

@ARTICLE{2007MNRAS.377L..74L,
       author = {{Liddle}, Andrew R.},
        title = "{Information criteria for astrophysical model selection}",
      journal = {\mnras},
         year = 2007,
        month = may,
       volume = {377},
       number = {1},
        pages = {L74-L78},
          doi = {10.1111/j.1745-3933.2007.00306.x},
archivePrefix = {arXiv},
       eprint = {astro-ph/0701113},
 primaryClass = {astro-ph},
       adsurl = {https://ui.adsabs.harvard.edu/abs/2007MNRAS.377L..74L}
}

@ARTICLE{2019MNRAS.489..738H,
       author = {{Hara}, Nathan C. and {Bou{\'e}}, G. and {Laskar}, J. and {Delisle}, J.-B. and {Unger}, N.},
        title = "{Bias and robustness of eccentricity estimates from radial velocity data}",
      journal = {\mnras},
         year = 2019,
        month = oct,
       volume = {489},
       number = {1},
        pages = {738-762},
          doi = {10.1093/mnras/stz1849},
archivePrefix = {arXiv},
       eprint = {1907.02048},
 primaryClass = {astro-ph.EP},
       adsurl = {https://ui.adsabs.harvard.edu/abs/2019MNRAS.489..738H}
}

@ARTICLE{2026AJ....172...53T,
       author = {{Thompson}, William and {Blakely}, Dori and {Xuan}, Jerry W. and {Blouin}, Simon and {Zhang}, Jingwen and {Johnstone}, Doug and {Ruffio}, Jean-Baptiste and {Nielsen}, Eric and {Speedie}, Jessica and {Bowler}, Brendan P. and {Bouchard-C{\^o}t{\'e}}, Alexandre and {Franson}, Kyle and {Roberson}, William and {Cloutier}, Ryan and {Marois}, Christian and {Rochon}, Alexandra and others},
        title = "{Detecting and Characterizing Companions with a Calibrated Gaia DR2, DR3, and Hipparcos Catalog (G23H)}",
      journal = {\aj},
         year = 2026,
        month = jul,
       volume = {172},
       number = {1},
          eid = {53},
        pages = {53},
          doi = {10.3847/1538-3881/ae64ed},
archivePrefix = {arXiv},
       eprint = {2606.16777},
 primaryClass = {astro-ph.EP},
       adsurl = {https://ui.adsabs.harvard.edu/abs/2026AJ....172...53T}
}

@ARTICLE{2002A&A...391..647G,
       author = {{Gontcharov}, G.~A. and {Kiyaeva}, O.~V.},
        title = "{Photocentric orbits from a direct combination of ground-based astrometry with Hipparcos. I. Comparison with known orbits}",
      journal = {\aap},
         year = 2002,
        month = aug,
       volume = {391},
        pages = {647-657},
          doi = {10.1051/0004-6361:20020896},
       adsurl = {https://ui.adsabs.harvard.edu/abs/2002A&A...391..647G}
}

@ARTICLE{2021MNRAS.506.2269E,
       author = {{El-Badry}, Kareem and {Rix}, Hans-Walter and {Heintz}, Tyler M.},
        title = "{A million binaries from Gaia eDR3: sample selection and validation of Gaia parallax uncertainties}",
      journal = {\mnras},
         year = 2021,
        month = sep,
       volume = {506},
       number = {2},
        pages = {2269-2295},
          doi = {10.1093/mnras/stab323},
archivePrefix = {arXiv},
       eprint = {2101.05282},
 primaryClass = {astro-ph.SR},
       adsurl = {https://ui.adsabs.harvard.edu/abs/2021MNRAS.506.2269E}
}

@INPROCEEDINGS{2005ASPC..347...29T,
       author = {{Taylor}, M.~B.},
        title = "{TOPCAT \& STIL: Starlink Table/VOTable Processing Software}",
    booktitle = {Astronomical Data Analysis Software and Systems XIV},
         year = 2005,
       editor = {{Shopbell}, P. and {Britton}, M. and {Ebert}, R.},
       series = {Astronomical Society of the Pacific Conference Series},
       volume = {347},
        month = dec,
        pages = {29},
       adsurl = {https://ui.adsabs.harvard.edu/abs/2005ASPC..347...29T}
}

@ARTICLE{2022ApJ...935..167A,
       author = {{Astropy Collaboration} and {Price-Whelan}, Adrian M. and {Lim}, Pey Lian and {Earl}, Nicholas and {Starkman}, Nathaniel and {Bradley}, Larry and {Shupe}, David L. and {Patil}, Aarya A. and {Corrales}, Lia and {Brasseur}, C.~E. and {N{\"o}the}, Maximilian and {Donath}, Axel and {Tollerud}, Erik and {Morris}, Brett M. and {Ginsburg}, Adam and {Vaher}, Eero and {Weaver}, Benjamin A. and {Tocknell}, James and {Jamieson}, William and {van Kerkwijk}, Marten H. and {Robitaille}, Thomas P. and {Merry}, Bruce and {Bachetti}, Matteo and {G{\"u}nther}, H. Moritz and {Aldcroft}, Thomas L. and {Alvarado-Montes}, Jaime A. and {Archibald}, Anne M. and {B{\'o}di}, Attila and {Bapat}, Shreyas and {Barentsen}, Geert and {Baz{\'a}n}, Juanjo and {Biswas}, Manish and {Boquien}, M{\'e}d{\'e}ric and {Burke}, D.~J. and {Cara}, Daria and {Cara}, Mihai and {Conroy}, Kyle E. and {Conseil}, Simon and {Craig}, Matthew W. and {Cross}, Robert M. and {Cruz}, Kelle L. and {D'Eugenio}, Francesco and {Dencheva}, Nadia and {Devillepoix}, Hadrien A.~R. and {Dietrich}, J{\"o}rg P. and {Eigenbrot}, Arthur Davis and {Erben}, Thomas and {Ferreira}, Leonardo and {Foreman-Mackey}, Daniel and {Fox}, Ryan and {Freij}, Nabil and {Garg}, Suyog and {Geda}, Robel and {Glattly}, Lauren and {Gondhalekar}, Yash and {Gordon}, Karl D. and {Grant}, David and {Greenfield}, Perry and {Groener}, Austen M. and {Guest}, Steve and {Gurovich}, Sebastian and {Handberg}, Rasmus and {Hart}, Akeem and {Hatfield-Dodds}, Zac and {Homeier}, Derek and {Hosseinzadeh}, Griffin and {Jenness}, Tim and {Jones}, Craig K. and {Joseph}, Prajwel and {Kalmbach}, J. Bryce and {Karamehmetoglu}, Emir and {Ka{\l}uszy{\'n}ski}, Miko{\l}aj and {Kelley}, Michael S.~P. and {Kern}, Nicholas and {Kerzendorf}, Wolfgang E. and {Koch}, Eric W. and {Kulumani}, Shankar and {Lee}, Antony and {Ly}, Chun and {Ma}, Zhiyuan and {MacBride}, Conor and {Maljaars}, Jakob M. and {Muna}, Demitri and {Murphy}, N.~A. and {Norman}, Henrik and {O'Steen}, Richard and {Oman}, Kyle A. and {Pacifici}, Camilla and {Pascual}, Sergio and {Pascual-Granado}, J. and {Patil}, Rohit R. and {Perren}, Gabriel I. and {Pickering}, Timothy E. and {Rastogi}, Tanuj and {Roulston}, Benjamin R. and {Ryan}, Daniel F. and {Rykoff}, Eli S. and {Sabater}, Jose and {Sakurikar}, Parikshit and {Salgado}, Jes{\'u}s and {Sanghi}, Aniket and {Saunders}, Nicholas and {Savchenko}, Volodymyr and {Schwardt}, Ludwig and {Seifert-Eckert}, Michael and {Shih}, Albert Y. and {Jain}, Anany Shrey and {Shukla}, Gyanendra and {Sick}, Jonathan and {Simpson}, Chris and {Singanamalla}, Sudheesh and {Singer}, Leo P. and {Singhal}, Jaladh and {Sinha}, Manodeep and {Sip{\H{o}}cz}, Brigitta M. and {Spitler}, Lee R. and {Stansby}, David and {Streicher}, Ole and {{\v{S}}umak}, Jani and {Swinbank}, John D. and {Taranu}, Dan S. and {Tewary}, Nikita and {Tremblay}, Grant R. and {de Val-Borro}, Miguel and {Van Kooten}, Samuel J. and {Vasovi{\'c}}, Zlatan and {Verma}, Shresth and {de Miranda Cardoso}, Jos{\'e} Vin{\'\i}cius and {Williams}, Peter K.~G. and {Wilson}, Tom J. and {Winkel}, Benjamin and {Wood-Vasey}, W.~M. and {Xue}, Rui and {Yoachim}, Peter and {Zhang}, Chen and {Zonca}, Andrea and {Astropy Project Contributors}},
        title = "{The Astropy Project: Sustaining and Growing a Community-oriented Open-source Project and the Latest Major Release (v5.0) of the Core Package}",
      journal = {\apj},
         year = 2022,
        month = aug,
       volume = {935},
       number = {2},
          eid = {167},
        pages = {167},
          doi = {10.3847/1538-4357/ac7c74},
archivePrefix = {arXiv},
       eprint = {2206.14220},
 primaryClass = {astro-ph.IM},
       adsurl = {https://ui.adsabs.harvard.edu/abs/2022ApJ...935..167A}
}

@ARTICLE{2007CSE.....9...90H,
       author = {{Hunter}, John D.},
        title = "{Matplotlib: A 2D Graphics Environment}",
      journal = {Computing in Science and Engineering},
         year = 2007,
        month = may,
       volume = {9},
       number = {3},
        pages = {90-95},
          doi = {10.1109/MCSE.2007.55},
       adsurl = {https://ui.adsabs.harvard.edu/abs/2007CSE.....9...90H}
}

@ARTICLE{2020NatMe..17..261V,
       author = {{Virtanen}, Pauli and {Gommers}, Ralf and {Oliphant}, Travis E. and {Haberland}, Matt and {Reddy}, Tyler and {Cournapeau}, David and {Burovski}, Evgeni and {Peterson}, Pearu and {Weckesser}, Warren and {Bright}, Jonathan and {van der Walt}, St{\'e}fan J. and {Brett}, Matthew and {Wilson}, Joshua and {Millman}, K. Jarrod and {Mayorov}, Nikolay and {Nelson}, Andrew R.~J. and {Jones}, Eric and {Kern}, Robert and {Larson}, Eric and {Carey}, C.~J. and {Polat}, {\.I}lhan and {Feng}, Yu and {Moore}, Eric W. and {VanderPlas}, Jake and {Laxalde}, Denis and {Perktold}, Josef and {Cimrman}, Robert and {Henriksen}, Ian and {Quintero}, E.~A. and {Harris}, Charles R. and {Archibald}, Anne M. and {Ribeiro}, Ant{\^o}nio H. and {Pedregosa}, Fabian and {van Mulbregt}, Paul and {SciPy 1. 0 Contributors}},
        title = "{SciPy 1.0: fundamental algorithms for scientific computing in Python}",
      journal = {Nature Methods},
         year = 2020,
        month = feb,
       volume = {17},
        pages = {261-272},
          doi = {10.1038/s41592-019-0686-2},
archivePrefix = {arXiv},
       eprint = {1907.10121},
 primaryClass = {cs.MS},
       adsurl = {https://ui.adsabs.harvard.edu/abs/2020NatMe..17..261V}
}

@ARTICLE{harris2020array,
 title         = {Array programming with {NumPy}},
 author        = {Charles R. Harris and K. Jarrod Millman and St{\'{e}}fan J.
                 van der Walt and Ralf Gommers and Pauli Virtanen and David
                 Cournapeau and Eric Wieser and Julian Taylor and Sebastian
                 Berg and Nathaniel J. Smith and Robert Kern and Matti Picus
                 and Stephan Hoyer and Marten H. van Kerkwijk and Matthew
                 Brett and Allan Haldane and Jaime Fern{\'{a}}ndez del
                 R{\'{i}}o and Mark Wiebe and Pearu Peterson and Pierre
                 G{\'{e}}rard-Marchant and Kevin Sheppard and Tyler Reddy and
                 Warren Weckesser and Hameer Abbasi and Christoph Gohlke and
                 Travis E. Oliphant},
 year          = {2020},
 month         = sep,
 journal       = {Nature},
 volume        = {585},
 number        = {7825},
 pages         = {357--362},
 doi           = {10.1038/s41586-020-2649-2},
 publisher     = {Springer Science and Business Media {LLC}},
 url           = {https://doi.org/10.1038/s41586-020-2649-2}
}

@inproceedings{mckinney2010data,
  author = {McKinney, Wes},
  title = {Data Structures for Statistical Computing in Python},
  pages = {56--61},
  bookpagination = {page},
  publisher = {SciPy},
  series = {Proceedings of the Python in Science Conference},
  editor = {{van der Walt}, St{\'e}fan and Millman, Jarrod},
  booktitle = {Proceedings of the 9th Python in Science Conference},
  year = {2010},
  doi = {10.25080/Majora-92bf1922-00a},
  booksubtitle = {SciPy 2010},
  eventtitle = {Python in Science Conference},
  venue = {Austin, Texas},
  eventdate = {June 28 - July 3 2010}
}

@ARTICLE{2019JOSS....4.1298Z,
       author = {{Zonca}, Andrea and {Singer}, Leo and {Lenz}, Daniel and {Reinecke}, Martin and {Rosset}, Cyrille and {Hivon}, Eric and {Gorski}, Krzysztof},
        title = "{healpy: equal area pixelization and spherical harmonics transforms for data on the sphere in Python}",
      journal = {The Journal of Open Source Software},
         year = 2019,
        month = mar,
       volume = {4},
       number = {35},
          eid = {1298},
        pages = {1298},
          doi = {10.21105/joss.01298},
       adsurl = {https://ui.adsabs.harvard.edu/abs/2019JOSS....4.1298Z}
}

@ARTICLE{2020ApJ...898...71B,
       author = {{Breivik}, Katelyn and {Coughlin}, Scott and {Zevin}, Michael and {Rodriguez}, Carl L. and {Kremer}, Kyle and {Ye}, Claire S. and {Andrews}, Jeff J. and {Kurkowski}, Michael and {Digman}, Matthew C. and {Larson}, Shane L. and {Rasio}, Frederic A.},
        title = "{COSMIC Variance in Binary Population Synthesis}",
      journal = {\apj},
         year = 2020,
        month = jul,
       volume = {898},
       number = {1},
          eid = {71},
        pages = {71},
          doi = {10.3847/1538-4357/ab9d85},
archivePrefix = {arXiv},
       eprint = {1911.00903},
 primaryClass = {astro-ph.HE},
       adsurl = {https://ui.adsabs.harvard.edu/abs/2020ApJ...898...71B}
}

@software{2015ascl.soft03010M,
       author = {{Morton}, Timothy D.},
        title = "{isochrones: Stellar model grid package}",
 howpublished = {Astrophysics Source Code Library, record ascl:1503.010},
         year = 2015,
        month = mar,
          eid = {ascl:1503.010},
archivePrefix = {ascl},
       eprint = {1503.010},
       adsurl = {https://ui.adsabs.harvard.edu/abs/2015ascl.soft03010M}
}

@software{2011ascl.soft12014R,
       author = {{Rhodes}, Brandon Craig},
        title = "{PyEphem: Astronomical Ephemeris for Python}",
 howpublished = {Astrophysics Source Code Library, record ascl:1112.014},
         year = 2011,
        month = dec,
          eid = {ascl:1112.014},
archivePrefix = {ascl},
       eprint = {1112.014},
       adsurl = {https://ui.adsabs.harvard.edu/abs/2011ascl.soft12014R}
}

@INPROCEEDINGS{optuna,
       author = {{Akiba}, Takuya and {Sano}, Shotaro and {Yanase}, Toshihiko and
                 {Ohta}, Takeru and {Koyama}, Masanori},
        title = "{Optuna: A Next-generation Hyperparameter Optimization Framework}",
    booktitle = {Proceedings of the 25th ACM SIGKDD International Conference on
                 Knowledge Discovery and Data Mining},
         year = 2019,
        month = jul,
        pages = {2623-2631},
          doi = {10.1145/3292500.3330701},
}

@ARTICLE{2011JMLR...12.2825P,
       author = {{Pedregosa}, Fabian and {Varoquaux}, Ga{\"e}l and {Gramfort}, Alexandre and {Michel}, Vincent and {Thirion}, Bertrand and {Grisel}, Olivier and {Blondel}, Mathieu and {M{\"u}ller}, Andreas and {Nothman}, Joel and {Louppe}, Gilles and {Prettenhofer}, Peter and {Weiss}, Ron and {Dubourg}, Vincent and {Vanderplas}, Jake and {Passos}, Alexandre and {Cournapeau}, David and {Brucher}, Matthieu and {Perrot}, Matthieu and {Duchesnay}, {\'E}douard},
        title = "{Scikit-learn: Machine Learning in Python}",
      journal = {Journal of Machine Learning Research},
         year = 2011,
        month = oct,
       volume = {12},
        pages = {2825-2830},
          doi = {10.48550/arXiv.1201.0490},
archivePrefix = {arXiv},
       eprint = {1201.0490},
 primaryClass = {cs.LG},
       adsurl = {https://ui.adsabs.harvard.edu/abs/2011JMLR...12.2825P}
}

\begin{appendix}
\section{Detailed astrometric models\label{appx:detailed_ast_models}}
    For each source, we define a tangent plane at the catalog position $(\alpha_0,\delta_0)$ at the catalog epoch $t_0={\rm J}2026.0$. The local east, north, and line-of-sight unit vectors are
    \begin{align}
        \begin{split}
            \hat{\vec{p}}&=\left[-\sin\alpha_{0},\ \cos\alpha_{0},\ 0\right]^{\rm T},\\
            \hat{\vec{q}}&=\left[-\sin\delta_{0}\cos\alpha_{0},\ -\sin\delta_{0}\sin\alpha_{0},\ \cos\delta_{0}\right]^{\rm T},\\
            \hat{\vec{r}}&=\left[\cos\delta_{0}\cos\alpha_{0},\ \cos\delta_{0}\sin\alpha_{0},\ \sin\delta_{0}\right]^{\rm T},
        \end{split}
    \end{align}
    where the hats denote unit vectors. At epoch $t_{j}$, the tangent-plane coordinates along $\hat{\vec{p}}$ and $\hat{\vec{q}}$ are denoted by $\xi_{j}$ and $\zeta_{j}$, respectively, and are measured in mas. For a telescope barycentric position $\vec{b}_j$ at epoch $t_j$, the parallax factors are
    \begin{align}
        \Pi_{\xi,j}=-\hat{\vec{p}}\cdot\frac{\vec{b}_j}{\rm AU},\qquad \Pi_{\zeta,j}=-\hat{\vec{q}}\cdot\frac{\vec{b}_j}{\rm AU}.
    \end{align}

    We describe the astrometric motion on the local tangent plane. \emph{Gaia} records the projection of this two-dimensional motion onto its AL direction, giving \citep{2023A&A...674A...9H}
    \begin{align}
        \eta_{{\rm s},j}=
        \left(\Delta\alpha^\ast+\mu_{\alpha^\ast}\Delta t_j\right)\sin\psi_j
        +\left(\Delta\delta+\mu_\delta\Delta t_j\right)\cos\psi_j
        +\varpi\Pi_{\eta,j},
        \label{eq:gaia_single}
    \end{align}
    where the subscript s denotes a single star, $\Pi_{\eta,j}$ is the AL parallax factor, $\psi_j$ is the scan angle, $\Delta t_j=t_j-t_{\rm ref}$, and $t_{\rm ref}$ is the reference epoch of the corresponding astrometric solution. $\Pi_{\eta,j}$, $\psi_j$, and $t_j$ were provided by GOST. The offsets $\Delta\alpha^\ast\equiv\Delta\alpha\cos\delta$ and $\Delta\delta$ are zero in the input mock catalog, but are retained as free parameters in the fitted models.

    CSST records two-dimensional astrometric motion. Following \citet{2023FrASS..1046603F}, which adopted Eq. (1.2.26) of \citet{The_Hipparcos_and_Tycho_catalogues}, we write
    \begin{align}
        \xi_{{\rm s},j}&=\Delta\alpha^\ast+
        \frac{\mu_{\alpha^\ast}\Delta t_j+\Pi_{\xi,j}\varpi}{1-\hat{\vec{r}}\cdot\frac{\vec{b}_j}{\rm AU}\varpi},\label{eq:csst_single_xi}\\
        \zeta_{{\rm s},j}&=\Delta\delta+
        \frac{\mu_{\delta}\Delta t_j+\Pi_{\zeta,j}\varpi}{1-\hat{\vec{r}}\cdot\frac{\vec{b}_j}{\rm AU}\varpi}.\label{eq:csst_single_zeta}
    \end{align}
    The denominator is retained when generating the CSST mock observations. In the fitting procedure, we set this denominator to unity because it remains very close to unity.

    For binaries, the eccentric anomaly $E_j$ is obtained from
    \begin{align}
        M_j=\frac{2\pi\left(t_{j}-t_{0}\right)}{P}+M_{0}=E_j-e\sin E_j,
    \end{align}
    where $P$ is the orbital period, $e$ is the eccentricity, and $M_0$ is the mean anomaly at $t_0={\rm J}2026.0$. The relative semi-major axis $a_{\rm rel}$ is computed from Kepler's third law. Using the Thiele--Innes elements \citep{2009ApJS..182..205W} of the relative orbit, $A_{\rm rel},B_{\rm rel},F_{\rm rel},G_{\rm rel}$, the instantaneous separation of the secondary relative to the primary in the tangent plane is
    \begin{align}
        \Delta\xi_j&=-B_{\rm rel}X_j-G_{\rm rel}Y_j,\\
        \Delta\zeta_j&=-A_{\rm rel}X_j-F_{\rm rel}Y_j.
    \end{align}
    Here $X_{j}$ and $Y_{j}$ are the elliptical rectangular coordinates at epoch $t_{j}$. We denote this separation vector and its length by
    \begin{align}
        \vec{\rho}_j=\left[\Delta\xi_j,\Delta\zeta_j\right]^{\rm T},\qquad
        \rho_j=\left|\vec{\rho}_j\right|.
    \end{align}
    With this convention, the primary position relative to the binary barycenter is $-q\vec{\rho}_j/(1+q)$.

    We now model the orbital term relative to the binary barycenter. Although the catalog-level cut in Sect.~\ref{subsect:phot} removes the few binaries with very large photocentric semi-major axes, the projected separation at an individual epoch can still be comparable to the angular resolution. We therefore used the binary observation bias model of \citet{LL:LL-136}, also implemented for generating \emph{Gaia} mock observations by \citet{2024OJAp....7E.100E}, to describe the transition from the unresolved photocenter regime to the marginally resolved regime.

    We introduce the dimensionless bias function $\mathcal{B}(f,h)$, whose value is obtained by iterating
    \begin{align}
        \mathcal{B}(f,h)=
        \frac{fh}{f+\exp\left[h^2/2-h\mathcal{B}(f,h)\right]},
        \label{eq:bias_function}
    \end{align}
    where $f$ is the passband-specific flux ratio and $h$ is the component separation normalized by the angular resolution scale. For \emph{Gaia}, the AL binary separation is
    \begin{align}
        \Delta\eta_{j}=\Delta\xi_{j}\sin\psi_{j}+\Delta\zeta_{j}\cos\psi_{j}.
    \end{align}
    We set $h_{{\rm G}, j}=\Delta\eta_j/u_{\rm G}$ adopting the commonly used scale $u_{\rm G}=90\,{\rm mas}$ \citep{LL:LL-136,2023A&A...674A..25H,2024OJAp....7E.100E}. The \emph{Gaia} AL displacement relative to the binary barycenter is then
    \begin{align}
        \delta\eta_{j}=
        \begin{cases}
            \left(\dfrac{f_{\rm G}}{1+f_{\rm G}}-\dfrac{q}{1+q}\right)\Delta\eta_j, & |h_{{\rm G}, j}|\leq 0.1,\\
            u_{\rm G}{\cal B}(f_{\rm G},h_{{\rm G}, j})-\dfrac{q}{1+q}\Delta\eta_j, & 0.1<|h_{{\rm G}, j}|\leq 3-f_{\rm G},\\
            -\dfrac{q}{1+q}\Delta\eta_j, & |h_{{\rm G}, j}|>3-f_{\rm G}.
        \end{cases}
        \label{eq:gaia_binary_displacement}
    \end{align}

    We applied an analogous binary observation bias model to CSST. Assuming that the PSF is centrally symmetric and similar for the two components, the displacement of the fitted peak is parallel to the instantaneous separation vector $\vec{\rho}_j$. We therefore define $h_{{\rm C}, j}=\rho_j/u_{\rm C}$ and write
    \begin{align}
        \delta\xi_{j}=c_j\Delta\xi_j,\qquad \delta\zeta_{j}=c_j\Delta\zeta_j,\label{eq:csst_binary_displacement}
    \end{align}
    where
    \begin{align}
        c_j=
        \begin{cases}
            \dfrac{f_{\rm C}}{1+f_{\rm C}}-\dfrac{q}{1+q}, & |h_{{\rm C}, j}|\leq 0.1,\\
            \dfrac{u_{\rm C}{\cal B}(f_{\rm C},h_{{\rm C}, j})}{\rho_j}-\dfrac{q}{1+q}, & 0.1<|h_{{\rm C}, j}|\leq 3-f_{\rm C},\\
            -\dfrac{q}{1+q}, & |h_{{\rm C}, j}|>3-f_{\rm C}.
        \end{cases}
    \end{align}

    For luminous star--compact object binaries, the compact object is assigned negligible flux. Eqs. \eqref{eq:gaia_binary_displacement} and \eqref{eq:csst_binary_displacement} then reduce to the motion of the luminous star around the barycenter:
    \begin{align}
        \delta\xi_j&=BX_j+GY_j,\\
        \delta\zeta_j&=AX_j+FY_j,\\
        \delta\eta_j&=\delta\xi_{j}\sin\psi_{j}+\delta\zeta_{j}\cos\psi_{j},
    \end{align}
    where $A,B,F,G$ are the Thiele--Innes elements of the orbit of the luminous component.

    The noiseless mock observables were generated by adding the binary displacement to the corresponding single-star motion in Eqs. (\ref{eq:gaia_single}--\ref{eq:csst_single_zeta}). For \emph{Gaia},
    \begin{align}
        \eta_{{\rm b}, j}=
        \left(\Delta\alpha^\ast+\mu_{\alpha^\ast}\Delta t_j\right)\sin\psi_j
        +\left(\Delta\delta+\mu_\delta\Delta t_j\right)\cos\psi_j
        +\varpi\Pi_{\eta,j}+\delta\eta_{j}.
        \label{eq:gaia_binary}
    \end{align}
    For CSST,
    \begin{align}
        \xi_{{\rm b}, j}&=\Delta\alpha^\ast+
        \frac{\mu_{\alpha^\ast}\Delta t_j+\Pi_{\xi,j}\varpi}{1-\hat{\vec{r}}\cdot\frac{\vec{b}_j}{\rm AU}\varpi}
        +\delta\xi_{j},\label{eq:csst_binary_xi}\\
        \zeta_{{\rm b}, j}&=\Delta\delta+
        \frac{\mu_{\delta}\Delta t_j+\Pi_{\zeta,j}\varpi}{1-\hat{\vec{r}}\cdot\frac{\vec{b}_j}{\rm AU}\varpi}
        +\delta\zeta_{j},
        \label{eq:csst_binary_zeta}
    \end{align}
    where the subscript b denotes a binary. In the binary fitting procedure, we also set the denominator to unity.

\section{Calculation of proper motion anomaly\label{appx:pma_calc}}
    All position offsets are defined on the local tangent plane centered on the input catalog position at the reference epoch $t_0={\rm J}2026.0$. We denote the \emph{Gaia} and CSST reference epochs by $t_{\rm G}={\rm J}2019.75$ and $t_{\rm C}={\rm J}2032.5$, respectively, and define
    \begin{align}
        \Delta t_\mathcal{M} = t_\mathcal{M}-t_0,\qquad
        \Delta t_{\rm CG}=t_{\rm C}-t_{\rm G},
    \end{align}
    where $\mathcal{M}\in\{{\rm G},{\rm C}\}$ labels the mission and $\Delta t_{\rm CG}$ is the reference epoch separation between \emph{Gaia} and CSST.

    Neglecting perspective acceleration, the linearly propagated input position offset at the reference epoch of mission $\mathcal{M}$ is
    \begin{align}
        \vec{P}_{0,\mathcal{M}}=\left[\mu_{\alpha^\ast,0}\Delta t_\mathcal{M},\mu_{\delta,0}\Delta t_\mathcal{M}\right]^{\rm T}.
    \end{align}
    Let $\Delta\vec{P}_\mathcal{M}$ be the fitted position offset at the reference epoch of mission $\mathcal{M}$, measured with respect to the propagated input position. The astrometric position at the mission reference epoch is therefore
    \begin{align}
        \vec{P}_\mathcal{M} = \vec{P}_{0,\mathcal{M}}+\Delta\vec{P}_\mathcal{M},
    \end{align}
    where $\Delta\vec{P}_\mathcal{M}=\left[\Delta\alpha^{\ast}_\mathcal{M},\Delta\delta_\mathcal{M}\right]^{\rm T}$. Since the Milky Way stellar mock catalog \citep{mwmsc} does not provide parameter uncertainties, the input astrometric parameters are treated as exact in this work. Thus, the covariance matrix of $\vec{P}_\mathcal{M}$ is simply the covariance matrix of $\Delta\vec{P}_\mathcal{M}$. We denote
    \begin{align}
        \begin{split}
        \vec{C}_{P,\mathcal{M}}&={\rm Cov}(\Delta\vec{P}_\mathcal{M}),\\
        \vec{C}_{\mu,\mathcal{M}}&={\rm Cov}(\vec{\mu}_\mathcal{M}),\\
        \vec{C}_{\mu P,\mathcal{M}}&={\rm Cov}(\vec{\mu}_\mathcal{M},\Delta\vec{P}_\mathcal{M}),
        \end{split}
    \end{align}
    where $\vec{\mu}_\mathcal{M}$ is the proper motion from the five-parameter solution of mission $\mathcal{M}$, and the cross-covariance block is defined such that $(\vec{C}_{\mu P,\mathcal{M}})_{ij}={\rm Cov}(\mu_{\mathcal{M},i},\Delta P_{\mathcal{M},j})$, where $i$ and $j$ index the two tangent-plane coordinates.

    The long-term \emph{Gaia}--CSST proper motion is defined as the positional difference between the \emph{Gaia} and CSST reference epochs divided by their reference epoch separation \citep{2019A&A...623A..72K,2021ApJS..254...42B}:
    \begin{align}
        \vec{\mu}_{\rm CG}=\frac{\vec{P}_{\rm C}-\vec{P}_{\rm G}}{\Delta t_{\rm CG}}=\vec{\mu}_0+\frac{\Delta\vec{P}_{\rm C}-\Delta\vec{P}_{\rm G}}{\Delta t_{\rm CG}}.\label{eq:pma_def}
    \end{align}
    Since the \emph{Gaia} and CSST astrometric solutions are independent, the covariance of $\vec{\mu}_{\rm CG}$ is
    \begin{align}
        \vec{C}_{\rm CG}={\rm Cov}(\vec{\mu}_{\rm CG})=\frac{\vec{C}_{P,{\rm G}}+\vec{C}_{P,{\rm C}}}{\left(\Delta t_{\rm CG}\right)^2}.
    \end{align}

    The PMa of mission $\mathcal{M}$ is defined as the difference between the mission proper motion and the long-term \emph{Gaia}--CSST proper motion:
    \begin{align}
        \Delta\vec{\mu}_\mathcal{M}=\vec{\mu}_\mathcal{M}-\vec{\mu}_{\rm CG}=\vec{\mu}_\mathcal{M}-\vec{\mu}_0-\frac{\Delta\vec{P}_{\rm C}-\Delta\vec{P}_{\rm G}}{\Delta t_{\rm CG}}.
    \end{align}
    The corresponding covariance matrix can be written as
    \begin{align}
        \vec{C}_{\Delta\mu,\mathcal{M}}=\vec{C}_{\mu,\mathcal{M}}
        +\frac{\vec{C}_{P,{\rm G}}+\vec{C}_{P,{\rm C}}}{\left(\Delta t_{\rm CG}\right)^2}
        +{\rm sign}(\mathcal{M})\frac{\vec{C}_{\mu P,\mathcal{M}}+\vec{C}_{\mu P,\mathcal{M}}^{\rm T}}{\Delta t_{\rm CG}},
        \label{eq:pma_cov_general}
    \end{align}
    where ${\rm sign}({\rm G})=+1$ and ${\rm sign}({\rm C})=-1$, following the coefficients of the position offsets in Eq.~\eqref{eq:pma_def}.

    For each mission, we also define the norm of the PMa vector as $\Delta\mu_\mathcal{M}=\left|\Delta\vec{\mu}_\mathcal{M}\right|$. Its uncertainty is obtained by error propagation:
    \begin{align}
        \sigma^2(\Delta\mu_\mathcal{M})=\frac{\Delta\vec{\mu}_\mathcal{M}^{\rm T}\vec{C}_{\Delta\mu,\mathcal{M}}\Delta\vec{\mu}_\mathcal{M}}{\left(\Delta\mu_\mathcal{M}\right)^{2}}.
    \end{align}

    We quantify the significance of the PMa vector by its Mahalanobis distance \citep{mahalanobis} from the null vector.
    \begin{align}
        s_{\Delta\mu,\mathcal{M}}=\sqrt{\Delta\vec{\mu}_\mathcal{M}^{\rm T}\vec{C}_{\Delta\mu,\mathcal{M}}^{-1}\Delta\vec{\mu}_\mathcal{M}}.
    \end{align}

\section{Binary candidate selection\label{appx:selection}}
    We used the histogram-based gradient-boosting classifier \texttt{HistGradientBoostingClassifier} from \texttt{scikit-learn} \citep{2011JMLR...12.2825P}. The input features encode multiple physical and observational effects whose response to orbital motion is unlikely to be captured by independent threshold cuts or linear combinations of such cuts. Gradient-boosted decision trees can model nonlinear relations among the input features, while the histogram implementation accelerates training on the large data set.

\begin{figure*}[!t]
        \centering
        \resizebox{0.8\textwidth}{!}{
            \begin{tikzpicture}[
                    node distance=5mm and 7mm,
                    every node/.style={font=\footnotesize},
                    data/.style={
                        rounded corners, draw, align=center,
                        minimum width=31mm, minimum height=9mm, fill=gray!8
                    },
                    process/.style={
                        draw, rounded corners=1pt, align=center,
                        minimum width=37mm, minimum height=10mm, fill=blue!4
                    },
                    filter/.style={
                        diamond, draw, align=center, aspect=1.65,
                        inner sep=1pt, minimum width=30mm,
                        minimum height=15mm, fill=orange!6
                    },
                    cand/.style={
                        draw, align=center,
                        minimum width=25mm, minimum height=8mm, fill=green!7
                    },
                    end/.style={
                        rounded corners, draw, align=center,
                        minimum width=25mm, minimum height=9mm, fill=gray!8
                    },
                    classifierbox/.style={
                        draw, rounded corners=2pt, align=left,
                        text width=90mm, minimum height=25mm,
                        fill=orange!8, inner sep=3pt
                    },
                    arrow/.style={-{Latex[length=2mm]}, thick},
                    bus/.style={thick},
                    dashedlink/.style={dashed, thick}
                ]

                \node[data] (input) {
                    5p diagnostics\\
                    PMa features\\
                    auxiliary observational-sampling features
                };

                \node[process, below=of input] (split) {
                    split into\\
                    training, tuning, validation, test\\
                    30\%, 10\%, 10\%, 50\%
                };

                \node[process, below=of split] (tune) {
                    train four classifiers with training set\\
                    tune hyperparameters on tuning set
                };

                \node[process, below=of tune] (retrain) {
                    retrain four classifiers with training + tuning sets\\
                    and \emph{fixed} tuned hyperparameters
                };

                \node[classifierbox, right=8mm of retrain, yshift=8mm] (classifiers) {
                    \textbf{Classifiers}\\[1mm]
                        $C_1$: joint + PMa classifier (joint 5p + PMa + joint obs. features)\\[1mm]
                        $C_2$: joint classifier (joint 5p + joint obs. features)\\[1mm]
                        $C_3$: \emph{Gaia} classifier (\emph{Gaia} 5p + \emph{Gaia} obs. features)\\[1mm]
                        $C_4$: CSST classifier (CSST 5p + CSST obs. features)
                };

                \draw[arrow] (input) -- (split);
                \draw[arrow] (split) -- (tune);
                \draw[arrow] (tune) -- (retrain);

                \node[
                    draw, dashed, rounded corners,
                    fit=(tune)(retrain)(classifiers),
                    inner xsep=2mm,
                    inner ysep=2mm,
                    label={[font=\normalsize\bfseries, fill=white, inner sep=1pt]above:training and tuning of four classifiers}
                ] {};

                \node[process, below=30mm of retrain] (threshold) {
                    choose thresholds $\theta_1,\theta_2,\theta_3,\theta_4$\\
                    on validation set\\
                    with overall precision $\geq 0.8$
                };

                \node[process, below=30mm of threshold] (testeval) {
                    fix classifiers and thresholds\\
                    evaluate on the test set
                };

                \draw[arrow] (retrain) -- (threshold);
                \draw[arrow] (threshold) -- (testeval);

                \node[end, right=20mm of threshold, yshift=20mm] (cascadeinput) {input};

                \node[filter, below=of cascadeinput] (m1) {
                    $C_1$ eligible and\\
                    score $\geq\theta_1$?
                };
                \node[cand, right=14mm of m1] (c1) {candidate\\ set 1};

                \node[filter, below=of m1] (m2) {
                    $C_2$ eligible and\\
                    score $\geq\theta_2$?
                };
                \node[cand, right=14mm of m2] (c2) {candidate\\ set 2};

                \node[filter, below=of m2] (m3) {
                    $C_3$ eligible and\\
                    score $\geq\theta_3$?
                };
                \node[cand, right=14mm of m3] (c3) {candidate\\ set 3};

                \node[filter, below=of m3] (m4) {
                    $C_4$ eligible and\\
                    score $\geq\theta_4$?
                };
                \node[cand, right=14mm of m4] (c4) {candidate\\ set 4};

                \node[end, below=10mm of m4] (reject) {rejected};

                \draw[arrow] (cascadeinput) -- (m1);
                \draw[arrow] (m1) -- node[above] {Y} (c1);
                \draw[arrow] (m1) -- node[left] {N} (m2);
                \draw[arrow] (m2) -- node[above] {Y} (c2);
                \draw[arrow] (m2) -- node[left] {N} (m3);
                \draw[arrow] (m3) -- node[above] {Y} (c3);
                \draw[arrow] (m3) -- node[left] {N} (m4);
                \draw[arrow] (m4) -- node[above] {Y} (c4);
                \draw[arrow] (m4) -- node[left] {N} (reject);

                \coordinate (bus1) at ($(c1.east)+(4mm,0)$);
                \coordinate (bus2) at ($(bus1 |- c2.east)$);
                \coordinate (bus3) at ($(bus1 |- c3.east)$);
                \coordinate (bus4) at ($(bus1 |- c4.east)$);

                \node[end] (merge) at (bus4 |- reject) {binary candidates};

                \draw[bus] (c1.east) -- (bus1);
                \draw[bus] (c2.east) -| (bus2);
                \draw[bus] (c3.east) -| (bus3);
                \draw[bus] (c4.east) -| (bus4);
                \draw[bus] (bus1) -- (bus4);
                \draw[arrow] (bus4) -- (merge.north);

                \node[
                    draw, dashed, rounded corners,
                    fit=(cascadeinput)(m1)(m2)(m3)(m4)(c1)(c2)(c3)(c4)(reject)(merge),
                    inner xsep=6mm,
                    inner ysep=6mm,
                    label={[font=\normalsize\bfseries, fill=white, inner sep=1pt]above:hierarchical cascade classifier}
                ] (cascadeclassifier) {};

                \draw[dashedlink] (threshold.east) -- ($(cascadeclassifier.west |- threshold.east)$);
                \draw[dashedlink] (testeval.east) -- ($(cascadeclassifier.west |- testeval.east)$);

            \end{tikzpicture}
        }
        \caption{Hierarchical binary candidate selection pipeline. Each of the four stage-specific histogram-based gradient-boosting classifiers is trained on the subset of sources eligible for that stage using a stage-specific set of input features. Hyperparameters are tuned on the tuning set, and the classifiers are then retrained on the combined training and tuning sets using the fixed tuned hyperparameters. The score thresholds $\theta_1,\theta_2,\theta_3,\theta_4$ are optimized jointly on the validation set by applying the hierarchical cascade classifier with a required overall precision of 0.8. The final performance is measured once on the independent test set using the fixed classifiers, fixed thresholds, and the same hierarchical cascade classifier. Sources accepted by an earlier stage are not passed to later stages, while sources not accepted by a stage are considered by the next stage if they are eligible for it.}
        \label{fig:hgbt_cascade}
    \end{figure*}
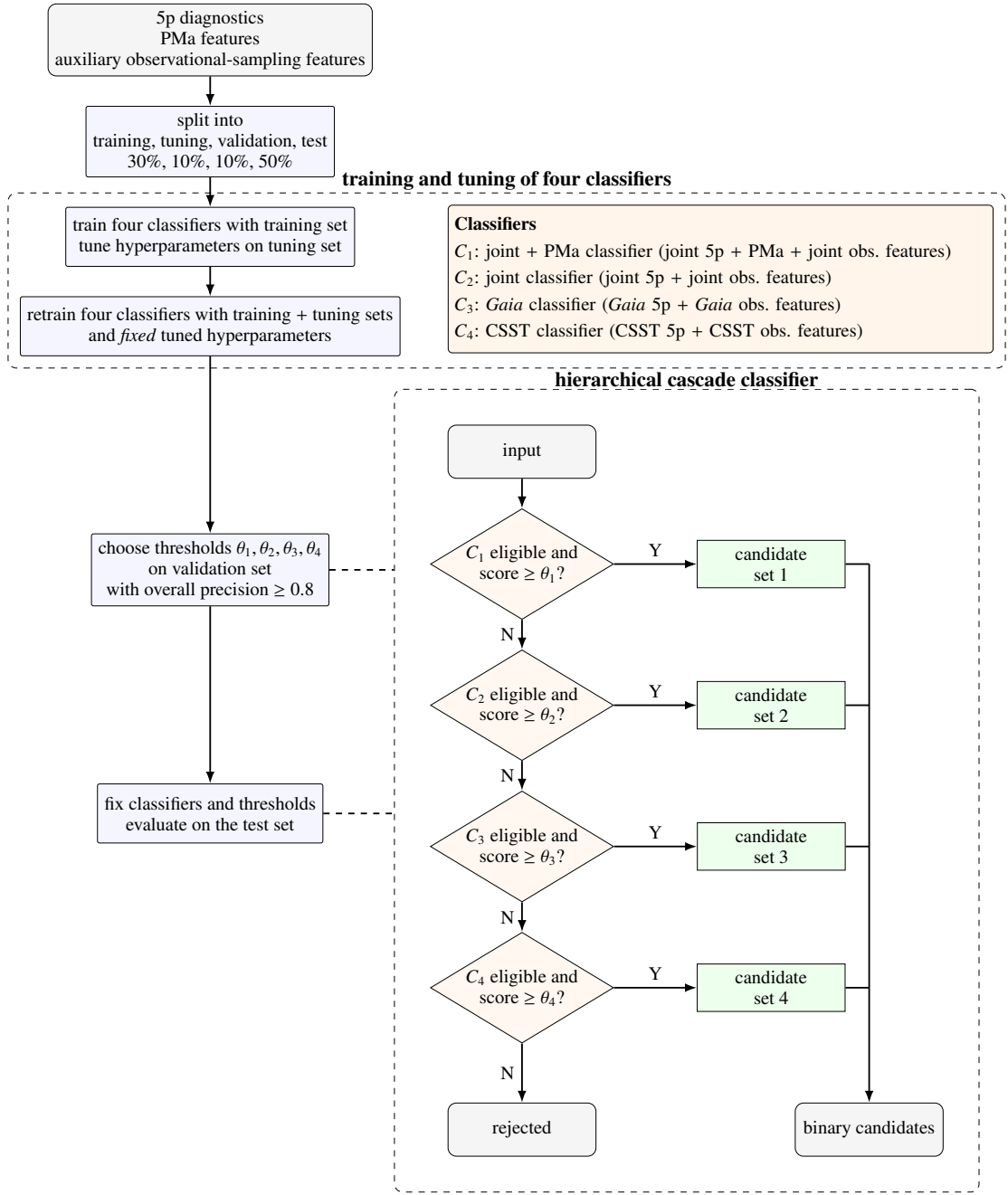

    Since the available feature sets are not identical for all sources, we organized the candidate-selection procedure as a four-stage cascade (Fig.~\ref{fig:hgbt_cascade}). Each stage uses \texttt{HistGradientBoostingClassifier}, but is trained separately on the sources \emph{eligible} for that stage using a feature set specific to that stage. We define a source as \emph{eligible} for a stage if it passes the corresponding data-quality preselection before classification. In practice, this requires $s_{\rm 5p,\varpi}\geq 1$ for the corresponding scenario. The first stage additionally requires PMa significance $s_{\Delta\mu,\rm G}\geq 1$ or $s_{\Delta\mu,\rm C}\geq 1$.

    The training set contains 30\% of the sources and is used to fit the stage classifiers; the tuning set contains 10\% and is used to tune the hyperparameters; the validation set contains 10\% and is used to optimize the score thresholds $(\theta_1,\theta_2,\theta_3,\theta_4)$; and the test set contains 50\% and is used only for the final independent evaluation.
    Since the simulated catalog is large, the 30\% training subset still provides sufficient statistics for training the classifiers. We assigned a larger fraction to the test set because the selected candidates are subsequently passed to the 12p orbital-fitting step, and not all selected binary candidates yield fits satisfying the fiducial criteria. A larger test set therefore provides more candidates and improves the statistical robustness of the final 12p solution assessment.

    We tuned the hyperparameters of each classifier on the tuning set with \texttt{optuna} \citep{optuna}. For each \texttt{optuna} trial, a temporary classifier with the trial hyperparameters was trained on the subset of the training set that was eligible for that stage and evaluated on the corresponding tuning set. The objective was to maximize the recall among eligible binaries while requiring a precision of 0.8 on the tuning set. After the best hyperparameters were selected, the classifier for that stage was retrained on the combined training and tuning sets using the \emph{fixed} tuned hyperparameters. After this retraining step, no classifier was refitted and no hyperparameter was retuned.

    Because the stages interact within the cascade, we optimized the four score thresholds jointly using only the scores obtained from the validation set. Each \texttt{optuna} trial applied the complete four-stage cascade to the validation set and maximized the recall among eligible binaries at a target precision of $0.8$. We adopted this target precision as a practical constraint to retain as many true binaries as possible while maintaining high candidate purity. The selected thresholds were fixed and applied once to the test set. After selecting the four thresholds, we fixed them, computed the scores on the test set, and applied the cascade once for the final independent evaluation.

    In the hierarchical cascade classifier, the four stages are ordered by the amount of information available to the classifier. Stage 1 (joint+PMa) uses the joint 5p diagnostics, both \emph{Gaia} and CSST PMa features, and the joint auxiliary observational-sampling features. Stage 2 (joint) uses the joint 5p diagnostics and joint auxiliary observational-sampling features, but does not require PMa. Stages 3 (\emph{Gaia}) and 4 (CSST) are fallback stages using only the \emph{Gaia} and CSST 5p diagnostics and their corresponding auxiliary observational-sampling features. A rejected source is passed to the next stage if the source is eligible for that stage. Once a source is accepted by a stage, it is not passed to later stages. The final candidate set is the union of the candidates accepted by the four stages.

\section{Orbital fitting algorithm\label{appx:orb_sol}}
    The 12 parameters consist of the five astrometric parameters ($\Delta\alpha^\ast,\Delta\delta,\varpi,\mu_{\alpha^\ast},\mu_{\delta}$), four photocentric Thiele--Innes elements ($A,B,F,G$), and three nonlinear orbital parameters $(P,e,M_{0})$. The fitting procedure follows the implementation in \texttt{gaiamock} \citep{2024OJAp....7E.100E}, which we extended to CSST and joint astrometry.

    The search over the nonlinear parameters $(P,e,M_{0})$ was carried out by minimizing the $\chi^2$ using adaptive simulated annealing over $50\,{\rm d}<P<10^{4}\,{\rm d}$, $0\leq e<0.99$, and $0\leq M_0<2\pi$. For each trial set of nonlinear parameters, we computed the eccentric anomaly $E_j$ and the elliptical rectangular coordinates
    \begin{align}
        X_j&=\cos E_j-e,\qquad Y_j=\sqrt{1-e^2}\sin E_j.
    \end{align}
    Treating $X_j$ and $Y_j$ as coefficients, the model is linear in the remaining nine parameters. For CSST, we set the denominator in Eqs. \eqref{eq:csst_binary_xi} and \eqref{eq:csst_binary_zeta} to unity in fitting. The CSST fitting model is therefore
    \begin{align}
        \begin{split}
            \xi_j^{\rm fit}&=\Delta\alpha^\ast+\mu_{\alpha^\ast}\Delta t_j+\Pi_{\xi,j}\varpi+B X_j+G Y_j,\\
            \zeta_j^{\rm fit}&=\Delta\delta+\mu_\delta\Delta t_j+\Pi_{\zeta,j}\varpi+A X_j+F Y_j,
        \end{split}
        \label{eq:fit_csst}
    \end{align}
    The 12p \emph{Gaia} AL model is \citep{2022A&A...665A.111W}
    \begin{align}
        \begin{split}
            \eta_j^{\rm fit}=&
            \left[\Delta\alpha^\ast+\mu_{\alpha^\ast}\Delta t_j+B X_j+G Y_j\right]\sin\psi_j\\
            &+\left[\Delta\delta+\mu_\delta\Delta t_j+A X_j+F Y_j\right]\cos\psi_j
            +\varpi\Pi_{\eta,j}.
        \end{split}
        \label{eq:fit_gaia_12p}
    \end{align}
    Joint fits use one shared set of photocentric Thiele--Innes elements and hence one fitted $a_0$ for the combined \emph{Gaia} and CSST epoch data. The \emph{Gaia} and CSST true photocentric semi-major axes differ because they are passband-dependent. The shared-$a_0$ approximation therefore introduces a mismatch. Section \ref{sect:bin_orb_sol} shows that, at the population level, this mismatch is smaller than the typical reported formal $a_0$ uncertainty for the fiducial joint fits in this simulation.

    For each trial set of $(P,e,M_0)$, we solved the nine linear parameters by weighted least squares, passed the resulting parameter values to the nonlinear optimizer, and repeated the nonlinear search. After convergence, the best-fit Thiele--Innes elements were converted to the Campbell elements $(a_0,\omega,\Omega,i)$ following Appx. A of \citet{2023A&A...674A...9H}.

\section{Resolved binary toy experiment\label{appx:resolved_binary}}
    We performed an injection--recovery experiment to assess whether CSST epoch astrometry can recover the orbits of resolved binaries. We generated $4\,500$ binaries with relative angular semi-major axes $a$ ranging from $60$ to $500\,{\rm mas}$ in steps of $10\,{\rm mas}$, with $100$ binaries at each grid point. The orbital periods were sampled from $\lg (P/{\rm d}) \sim N(4.8,2.3^2)$, truncated to $400\,{\rm d}<P<1000\,{\rm d}$. The eccentricities, orbital orientations, mean anomalies, and mass ratios were sampled from \citet{LL:LL-136}. The primary masses were sampled from the Salpeter initial mass function \citep{2019NatAs...3..482K}. We fixed the component magnitudes at $g_1=19$ and $g_2=20$. All binaries were placed in a sky region where CSST provides relatively regular temporal sampling, yielding $48$ observations per source.

    At each epoch, observations of the two components were generated only when their instantaneous projected separation satisfied $\rho\geq60\,{\rm mas}$. For each binary system, epochs below this resolution threshold were excluded from the mock observations. For the retained epochs, we generated mock observations for both components around a common barycenter and fitted a 13-parameter model, hereafter 13p, comprising five barycentric astrometric parameters, four relative Thiele--Innes elements, $(P,e,M_0)$, and the mass ratio. The fitting algorithm returned a solution when the retained data provided at least $14$ effective observations. Returned solutions were classified as satisfying the additional quality criteria when $F_{2,13{\rm p}}<25$, $a/\sigma_a>1$, and neither the fitted $P$ nor $e$ lay at or near an adopted boundary. Of the $4\,500$ injected binaries, $4\,476$ returned 13p solutions. The remaining $24$ binaries, all with $a=60\,{\rm mas}$, had too few effective observations.

    \begin{figure}[!h]
        \centering
        \includegraphics[width=0.4\textwidth]{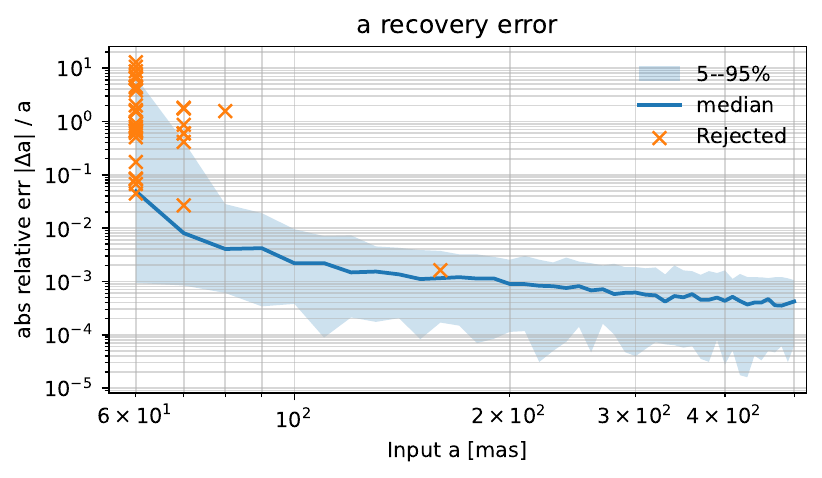}
        \vspace{1mm}
        \includegraphics[width=0.4\textwidth]{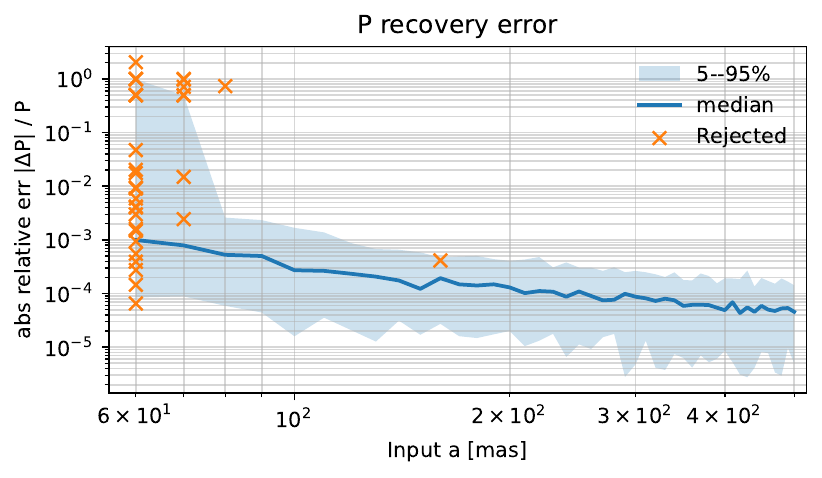}
        \caption{Absolute relative errors in the recovered relative semi-major axis $a$ (\emph{top}) and orbital period $P$ (\emph{bottom}) versus the injected $a$. The blue curves and the shaded region show the median and 5th--95th percentiles of the returned 13p solutions. Orange crosses mark returned solutions that failed to pass the criteria.}
        \label{fig:resolved_binary}
    \end{figure}

    At $a=60\,{\rm mas}$, $76$ of the $100$ binaries returned 13p solutions and $50$ satisfied the additional quality criteria. Their median and 95th-percentile absolute relative errors in $a$ were $4.89\%$ and $675\%$, respectively, indicating a substantial tail of poorly recovered orbits. At $a=70$ and $80\,{\rm mas}$, all $100$ binaries at each grid point returned solutions, with $93$ and $99$ satisfying the criteria, respectively; the 95th-percentile error in $a$ decreased from $42.1\%$ to $2.81\%$. Fig.~\ref{fig:resolved_binary} shows the corresponding decrease in the upper error tails of $a$ and $P$.

    For $90\leq a\leq500\,{\rm mas}$, all $4\,200$ injected binaries returned 13p solutions, and $4\,199$ satisfied the additional quality criteria. These results show that the injected resolved-binary orbits are recovered with high completeness over most of the tested $a$ range in this toy experiment.

\end{appendix}
\end{document}